\documentclass[preprint]{aastex}
\newcommand\br{\mbox{$B\!-\!R$}}
\begin{document}

\title{Analysis of the Interaction Effects in the Southern Galaxy Pair 
Tol1238-364 and ESO381-G009\footnote{Based on data collected at the ESO-MPIA 2.2 m telescope in La Silla,
Chile}}

\author{S. Temporin}
\affil{Institut f\"ur Astrophysik, Leopold-Franzens-Universit\"at Innsbruck, 
Technikerstra{\ss}e 25, A-6020 Innsbruck, Austria}
\email{giovanna.temporin@uibk.ac.at}
\author{S. Ciroi and P. Rafanelli}
\affil{Dipartimento di Astronomia, Vicolo dell'Osservatorio 2, I-35122  
Padova, Italy}
\email{ciroi@pd.astro.it, piraf@pd.astro.it}
\author{M. Radovich}
\affil{Astronomical Observatory of Capodimonte, Via
Moiariello 16, I-80131 Napoli, Italy}
\email{radovich@na.astro.it}
\author{J. Vennik}
\affil{Tartu Observatory, 61602 T\~{o}ravere, Tartumaa, Estonia}
\email{vennik@aai.ee}
\author{G. M. Richter}
\affil{Astrophysikalisches Institut Potsdam, An der Sternwarte 16, D-14482 
Potsdam, Germany}
\email{gmrichter@aip.de}
\and
\author{K. Birkle}
\affil{Max-Planck-Institut f\"ur Astronomie, K\"onigstuhl 17, D-69117 Heidelberg, 
Germany}
\email{birkle@mpia-hd.mpg.de}

\begin{abstract}
In the context of the connection among galaxy-galaxy interaction, starbursts
and nuclear activity, we present and discuss a quantitative morphological
analysis based on $BVR$ images and a detailed spectroscopic investigation
of two interacting galaxies, the Seyfert 2 
Tol1238-364 (IC 3639) and  its companion ESO381-G009, forming a triple system
with ESO381-G006. Broad-band optical photometry is complemented by H$\alpha$
imaging, which provides information about the distribution of star forming regions
across the galaxies.
Long-slit spectroscopic data obtained at  
different position angles of the slit are employed to determine the physical 
conditions of circumnuclear and extranuclear regions. 
A mixture of thermal and non-thermal ionizing radiation is found in the 
surroundings of the nucleus of Tol1238-364 and the energy budget supports the 
presence of a circumnuclear starburst.
Several regions in both the galaxies show anomalous line ratios: 
additional ionization by shock-heating and
low ionization of some extranuclear \ion{H}{2} regions are suggested as possible 
explanations.
An analysis of the emission-line profiles
reveals the presence of a broad H$\alpha$ component in the nuclear region of 
Tol1238-364.
Independent estimates of the star formation rates (SFR) were obtained through 
flux-calibrated H$\alpha$-images and FIR emission in the four IRAS bands.
Overall SFR densities have been compared with the SFR densities derived from 
H$\alpha$ emission in the individual regions of the galaxies sampled by long-slit spectra.
In both the galaxies an enhancement of the star formation activity with respect to
isolated galaxies is revealed.
The prevalence of starburst or nuclear activity has been examined through FIR 
color indices. The interaction scenario is discussed on the basis of the 
observed galaxy properties.

\end{abstract}


\section{INTRODUCTION}

The importance of the role played by interaction in triggering nuclear bursts
of star formation and activity in galaxies has been clearly pointed out in
the last decades. N-body simulations have shown that interactions drive large
gas flows toward the center of galaxies (e.g. Barnes \&\ Hernquist 1991).
Disk instabilities in colliding galaxies lead rapidly to the formation of
strong bars and to gas inflows fueling early starbursts and/or AGN activity 
before the galaxy merging \citep{mh96,bh96,mih99}.
The role of the orbital geometry in triggering inflow and
activity has been considered: prograde encounters favor bar instabilities
\citep{bh96}.

Other clues to the connection between interaction and starburst and AGN 
activity
come from several statistical studies: an excess of both starburst and Seyfert
galaxies among galaxy pairs and an excess of physical companions in samples of
starburst galaxies (SBGs) and Seyferts have been observed \citep[e.g.][]{kvs92,rvb95,rtb97}.

Detailed studies of interacting systems can provide a mean to evaluate the 
validity of this scenario \citep[e.g.][]{aga01}. In this context we performed a 
photometric and spectroscopic analysis of the Seyfert 2 Tol1238-364 and of 
its companion 
ESO381-G009, as well as an evaluation of far-infrared (FIR) emission of each galaxy 
of 
the pair in order to determine the physical properties of circumnuclear and 
extranuclear regions and to obtain independent estimates of the star formation 
rates. 

Tol1238-364 is a Seyfert 2 galaxy morphologically classified as SB(rs)bc; its 
near infrared bar is aligned at P. A. = 150$^{\circ}$ \citep{mrk97,goetal98}. 
After the recent detection of a broad 
H$\alpha$ 
component in polarized light \citep{hlb97}, the galaxy is 
listed
among the Seyfert 2s hosting a hidden broad line region (HBLR).
The column density, as measured by means of the photoelectric absorption cutoff
in the hard X-ray spectrum, assumes values larger than 10$^{25}$ cm$^{-2}$ \citep{rms99}.
It is also classified as Luminous Infrared 
Galaxy (L$_{FIR}$ = 4.36$\times10^{10}$ L$_{\odot}$, Lutz 1992) and is located 
in the ``radio-bright'' side of the  FIR/radio
correlation \citep{hsr85}, as shown by \citet{braetal98}.
It forms a triple system along with the almost face-on SB(r?)
ESO381-G009, located at 1$^{\prime}$.8 arcmin North-East and an edge-on galaxy 2$^{\prime}$.6 
arcmin to North-West \citep[e.g.][]{kk00}. 
Even if the optical images do not reveal strongly distorted
morphologies, traces of plumes and bridges of neutral gas have been detected in the \ion{H}{1} 21 cm line 
\citep{bpj00,bw01}. However, the neutral hydrogen appears still bound to the individual
galaxies. Its mass, based on Australia Telescope Compact Array observations in the 21 cm line \citep{bw01},
is of 3.6, 4.0, and 1.5 $\times$10$^9h_{75}^{-2}$ M$_{\odot}$ for Tol1238-364, ESO381-G009, and ESO381-G006, 
respectively.
The same data show that the radial velocities, listed in Table~\ref{literature} together with the
coordinates and B-band absolute magnitudes of the three galaxies, are concordant.

The Seyfert 2 and its NE-companion are 
characterized by strongly enhanced star formation also in the nuclear region. 
The starburst nature of
Tol1238-364 is pointed out by UV \citep{goetal98,gh99}, FIR
\citep{lu92} and radio \citep{braetal98}
observations. 
ROSAT HRI data show soft X-ray emission extended up to a radius of 6.8 kpc 
best fitted with a two-component model of a power law and thermal emission \citep{lwh01}.
 ESO381-G009 presents enhanced FIR emission, even if its FIR
luminosity is not sufficient to classify it as 
Infrared Luminous Galaxy (L$_{FIR}$ = 7.42$\times10^{9}$ L$_{\odot}$, Lutz 1992).

The present paper is organized as follows:
Observations and data reduction are described in \S~2. 
The main properties of the galaxies are derived through the analysis of the photometric and 
spectroscopic data and the comparison with photoionization models
in \S~3. Comments on the FIR emission of each of the two galaxies 
obtained applying Maximum Entropy procedures to the IRAS data, are given in 
\S~4. The results are summarized and extensively discussed in \S~5, where considerations on
the environment of the galaxies and the evolutionary history of the triplet are expressed, as well.

\section{OBSERVATIONS AND DATA REDUCTION}

Our data were obtained under good seeing conditions 
($\lesssim$ 1$^{\prime\prime}$) 
during an observing run in April 1995 at the ESO-MPIA 2.2 m telescope in La 
Silla, equipped with the Faint Object Spectrograph and Camera EFOSC2.
The detector was a 1k$\times$1k CCD with a pixel size of 19 $\mu$m and a scale 
of 0.336 arcsec/pixel.
Two images were taken during the same nights in 
redshifted, narrow band H$\alpha$ ($\lambda_c$ = 6651.7 \AA, FWHM = 61.3 \AA) and continuum 
($\lambda_c$ = 6521.2 \AA, FWHM = 74.3 \AA) filters. 
The spectrophotometric standard star Kopff 27 was also observed with the same
filters. Additionally broad-band $BVR$ images of both galaxies, together with standard star fields, 
were taken. Observation details concerning both imaging and spectroscopy, including exposure times, 
slit position angles (P.A.), and limit surface brightnesses and magnitudes 
are summarized in Table~\ref{tab1}.

H$\alpha$ images were reduced in a standard way: they were
bias-subtracted, flat-fielded, cleaned from cosmic rays, and background subtracted.
The tabulated spectrum of Kopff 27 was convoluted with the system 
 (filter + CCD + telescope) transmission curve and normalized to the 
 bandwidth of the H$\alpha$ filter. From the comparison with the 
 counts measured on the images, corrected for atmospheric extinction, 
 we computed the zero points to be applied for flux calibration.
 
 The flux calibrated images were corrected for atmospheric extinction. 
 The continuum image was aligned to the H$\alpha$ image using field stars 
 close to the galaxy and then subtracted from it; a check that the scaling 
 factor derived from the photometric calibration was correct was given by the 
 disappearance of non saturated stars in the continuum-subtracted image.
 A further check consisted in comparing the H$\alpha$ fluxes of the nucleus and
 the extranuclear regions extracted from the spectra (see below) with those 
 of the relevant regions in the calibrated image. The fluxes were found in good agreement
 (they differ by less than 20\%, which is within the errors of measurement).
 
\begin{figure}
\plottwo{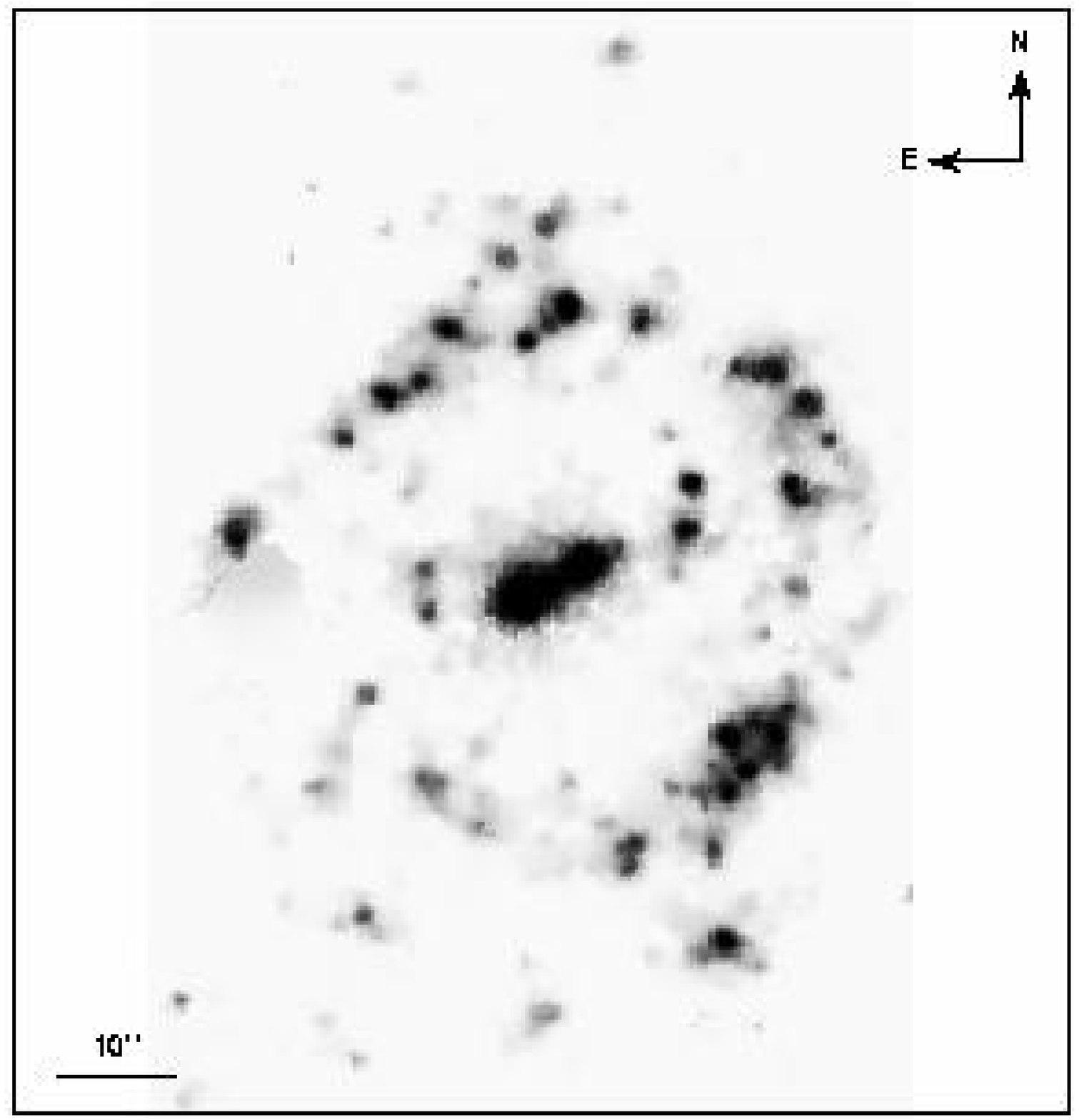}{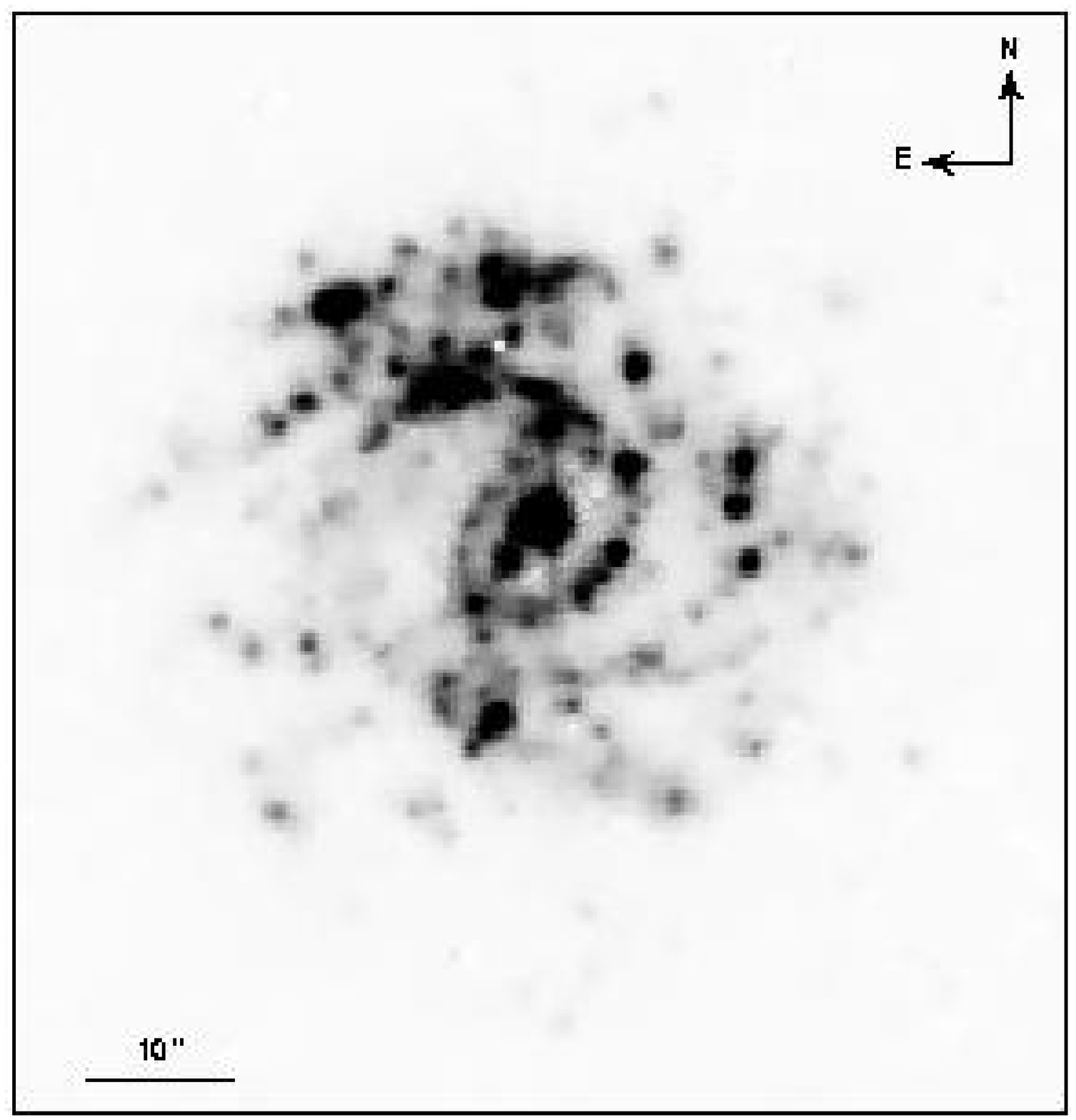}
\caption{H$\alpha$ images of
ESO381-G009  (left) and Tol1238-364 (right), after application of an adaptive 
smooth filter.\label{images}}
\end{figure}

\begin{figure}
\plotone{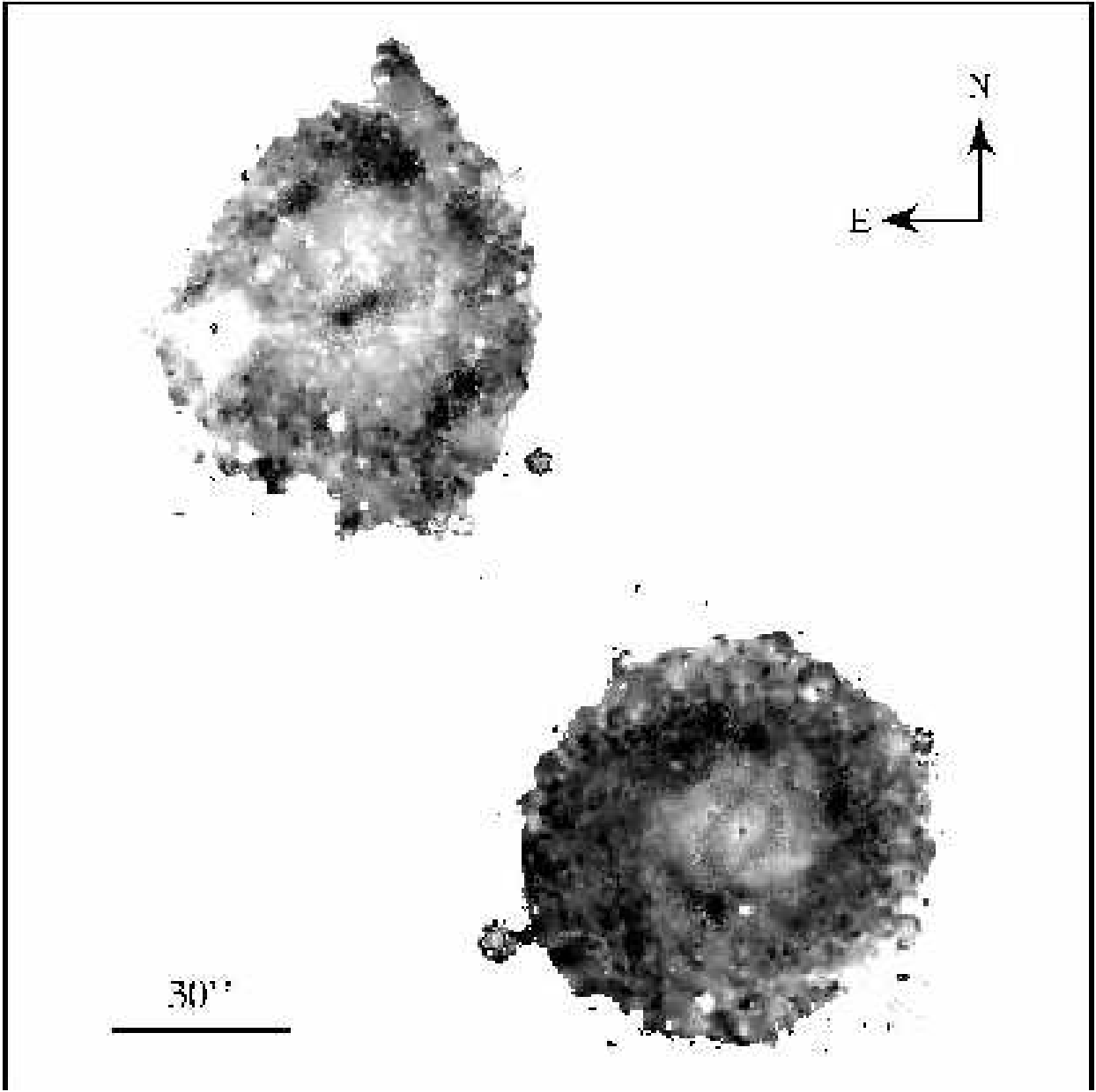}
\caption{\br\ image of ESO381-G009 (upper left)
and Tol1238-364 (lower right). Darker regions are bluer\label{cmap}}
\end{figure}

After the 
subtraction of the  H$\alpha$-continuum component, the images were 
processed by means of
an adaptive smooth filtering procedure developed at the Astrophysikalisches 
Institut
Potsdam \citep{rietal91,loetal93} in order to reduce the noise 
and enhance 
faint details of the internal structure of the galaxies (Fig.~\ref{images}).
$BVR$ images were also reduced in a standard way and registered.
The non-saturated foreground stars were subtracted by fitting them with a model of the
point spread function (PSF) obtained with IRAF\footnote{IRAF is
distributed by the National Optical Astronomy Observatories, which are 
operated by the Association of Universities for Research in Astronomy, 
Inc., under cooperative agreement with the National Science Foundation}/DAOPHOT routines. 
Saturated stars were masked with the task IMEDIT. The photometric calibration constants
have been derived by use of the package PHOTCAL applied to the standard stars, with the
usual transformation equations.
A \br\ color map (Fig.~\ref{cmap}) was obtained to analyze the color gradient across the 
galaxies and to have an insight into the spatial distribution of young and old stellar populations. 

\placefigure{images}
\placefigure{cmap} 

A 2$^{\prime\prime}$-wide slit was used to obtain long-slit spectra covering 
the spectral ranges 3200--6030 \AA\ (grism \#3) and 3850--7950 \AA\ (grism \#6) 
with dispersions 1.9 
\AA\ pixel$^{-1}$
and 2.6 \AA\ pixel$^{-1}$ respectively. 
The resulting spectral resolution estimated from the FWHM of faint
comparison lines is of the order of $\sim$10 \AA, while the scale along 
the slit 
is 0$^{\prime\prime}$.336 pixel$^{-1}$.

\placetable{tab1}
  
\begin{figure}
\epsscale{.9}
\plotone{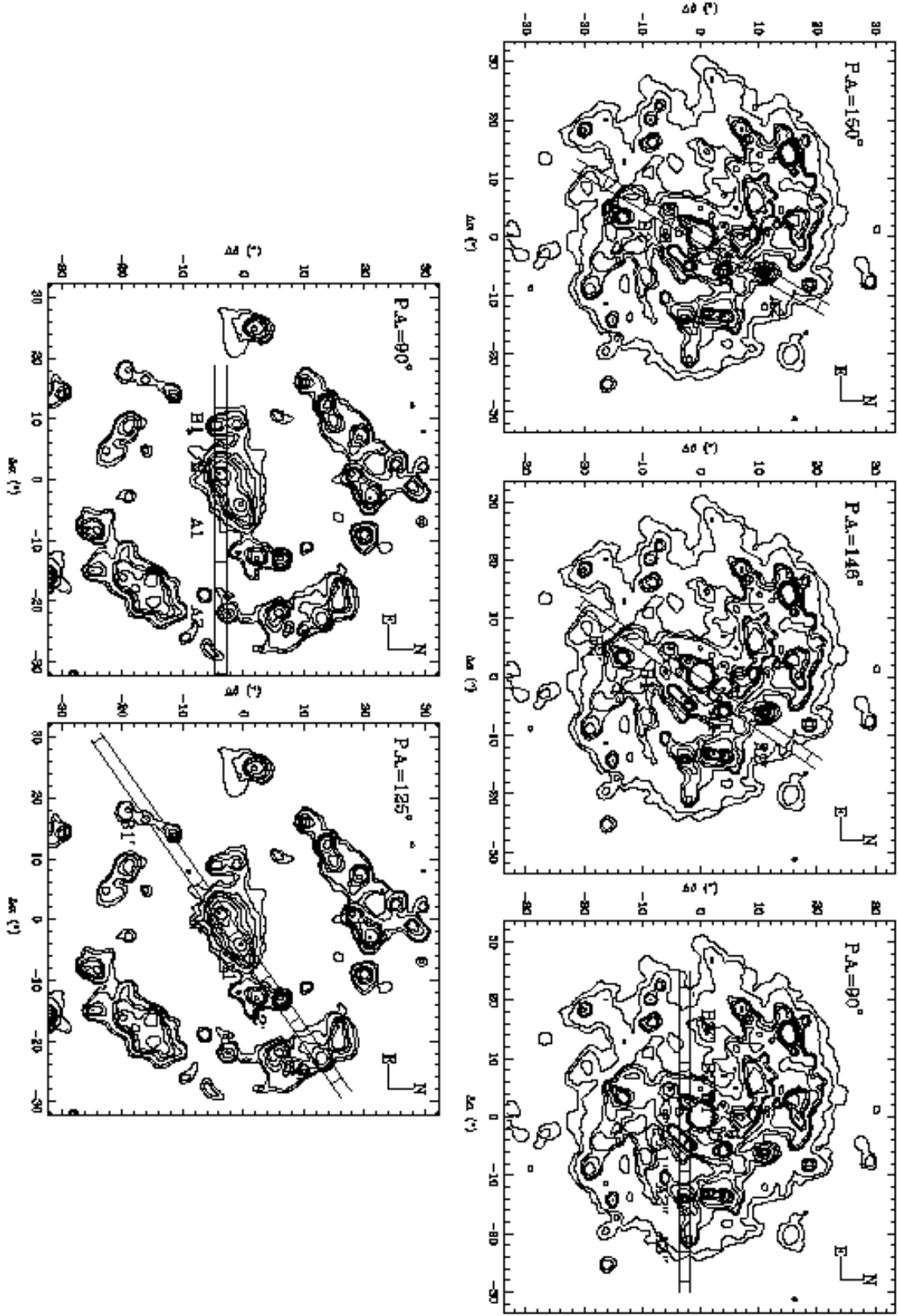}
\caption{Contour maps of the H$\alpha$   
adaptive-smoothed images of 
Tol1238-364 (top) and ESO381-G009 (bottom) with superimposed the traces 
of the slit positions and the regions extracted from the
spectra. Regions are labeled according to the explanations in the text. 
\label{contour}}
\end{figure}

Tol1238-364 was observed at three different slit position 
angles: P.A. = 90$^{\circ}$, 146$^{\circ}$, and 150$^{\circ}$. 
The spectra at P.A. = 150$^{\circ}$ and 146$^{\circ}$
were taken across the center of the galaxy (the first one along the galaxy's bar and
centered on the nucleus), 
while the spectrum at P.A. =  90$^{\circ}$ covers the region $\sim$ 
3$^{\prime\prime}$ South of 
the nucleus, where many knots of emission are visible in the H$\alpha$ image 
(Fig.~\ref{contour}, top). 
The spectra of ESO381-G009 were taken at P.A. 90$^{\circ}$ 
and 125$^{\circ}$ through the nucleus (Fig.~\ref{contour}, bottom). 
This last orientation of the slit allowed to observe both the nucleus and the 
emission knot on its NW side (see Fig.~\ref{contour}, bottom).
Spectra of the standard stars Kopff 27 and Feige 56 were 
taken immediately before or after the spectra of the galaxies, to allow their 
flux calibration.

\placefigure{contour}

The two-dimensional spectra were reduced with standard IRAF
packages: after bias
subtraction, flat-fielding and a careful subtraction of cosmic rays, they were
wavelength calibrated. 
The IRAF tasks IDENTIFY, REIDENTIFY, FITCOORDS, and TRANSFORM were used in sequence 
to determine the dispersion solution and linearize the two-dimensional spectra.
An estimate of the wavelength calibration error was obtained by evaluating the 
$rms$ of the mean difference between measured and predicted sky-line 
wavelengths. This $rms$ value
is $\sim$ 1.3 \AA, equivalent to 80 km s$^{-1}$ at $\lambda$ = 5000 \AA.
The spectra were corrected for atmospheric extinction and calibrated (CALIBRATE) to a flux scale 
using sensitivity functions obtained with the tasks STANDARD and SENSFUNC applied to the observed
spectrophotometric standard stars. 
Finally the sky-background was fitted by using two column samples at both sides of the galaxy 
spectrum and subtracted with the task BACKGROUND.

Contour maps of the extended H$\alpha$ + [\ion{N}{2}] $\lambda$6548,6583 
emission lines and continuum-subtracted H$\alpha$ profiles along the slit were
used to identify and extract the spectra of the nucleus and of several 
extranuclear emitting regions. Such regions are labeled on the slit images 
overlaid to the contour 
maps of the galaxies in Fig.~\ref{contour}: labels without apex
are used for regions at P.A. = 150$^{\circ}$ in Tol1238-364
and P.A. = 90$^{\circ}$ in ESO381-G009, an apostrophe indicates the regions 
at P.A. = 146$^{\circ}$ in Tol1238-364 and P.A. = 125$^{\circ}$ in 
ESO381-G009, while a quotation mark indicates the regions at 
P.A. = 90$^{\circ}$ of Tol1238-364.

\begin{figure}
\epsscale{1.1}
\plotone{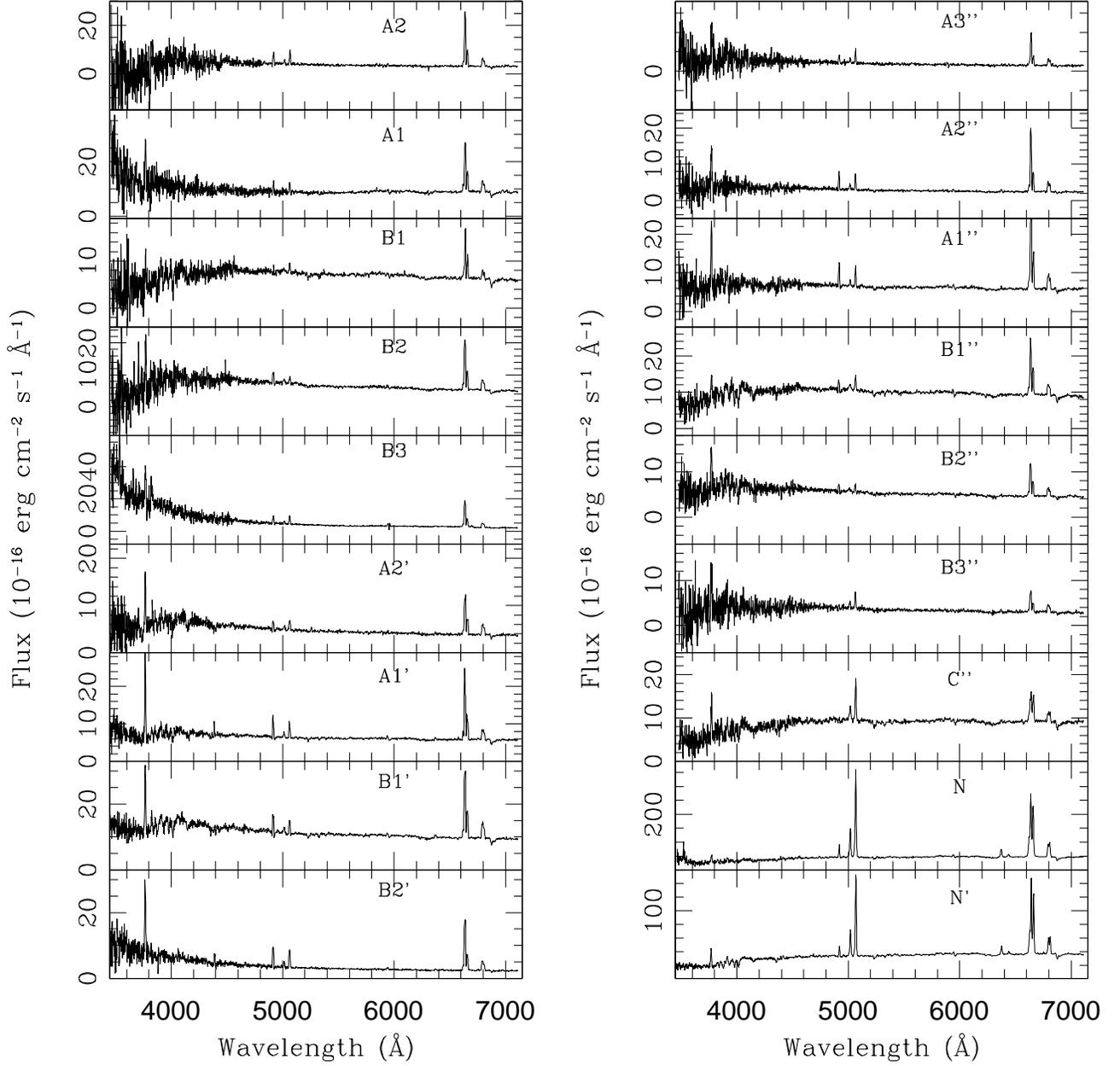}
\caption{Nuclear and extranuclear spectra of Tol1238-364, labeled 
according to Fig.~\ref{contour}. \label{tolspec}}
\end{figure}

\begin{figure}
\epsscale{1.0}
\plotone{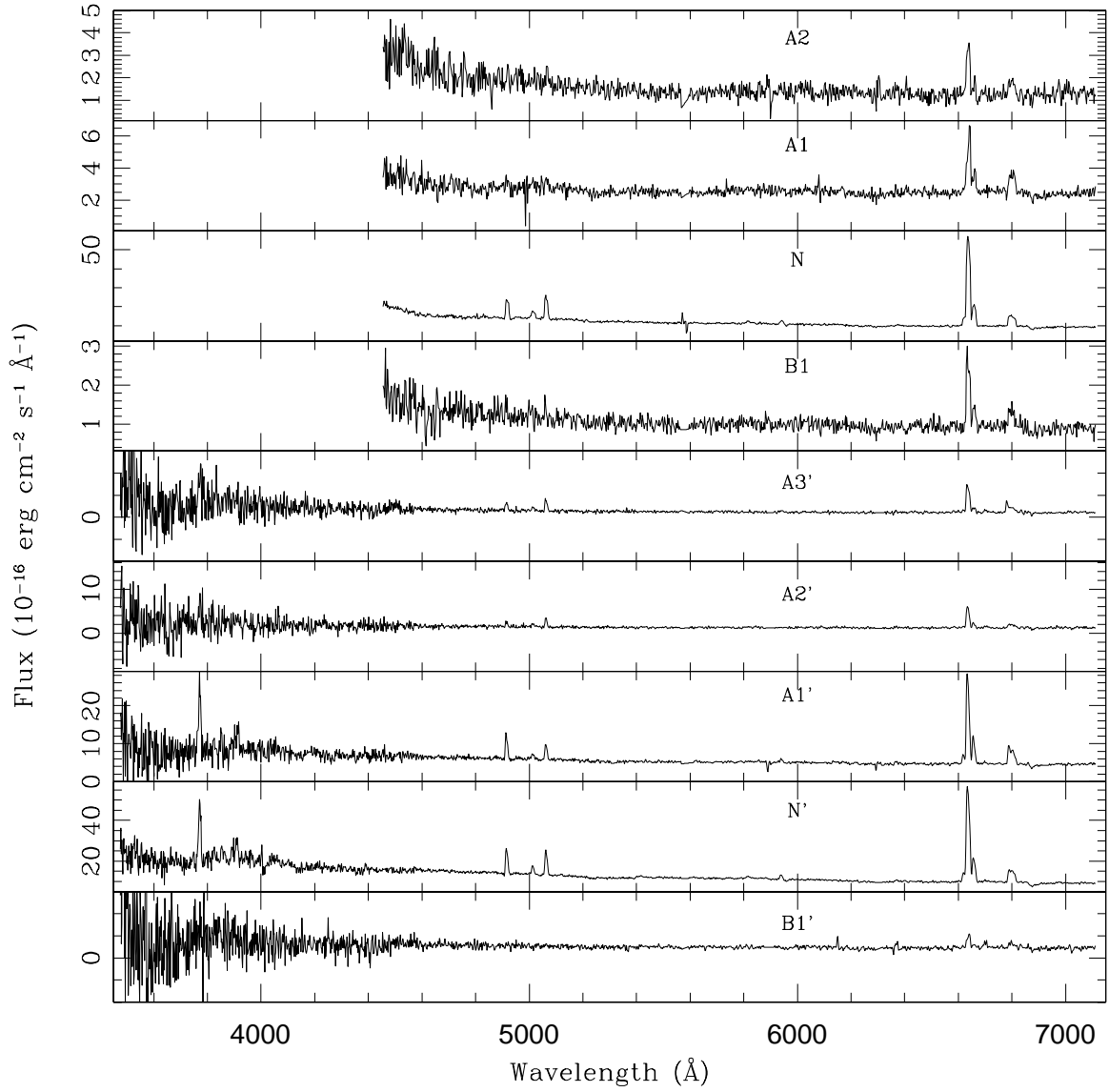}
\caption{Spectra of the regions extracted from ESO381-G009, labeled 
according to Fig.~\ref{contour}. \label{esospec}}
\end{figure}

\begin{figure}
\plotone{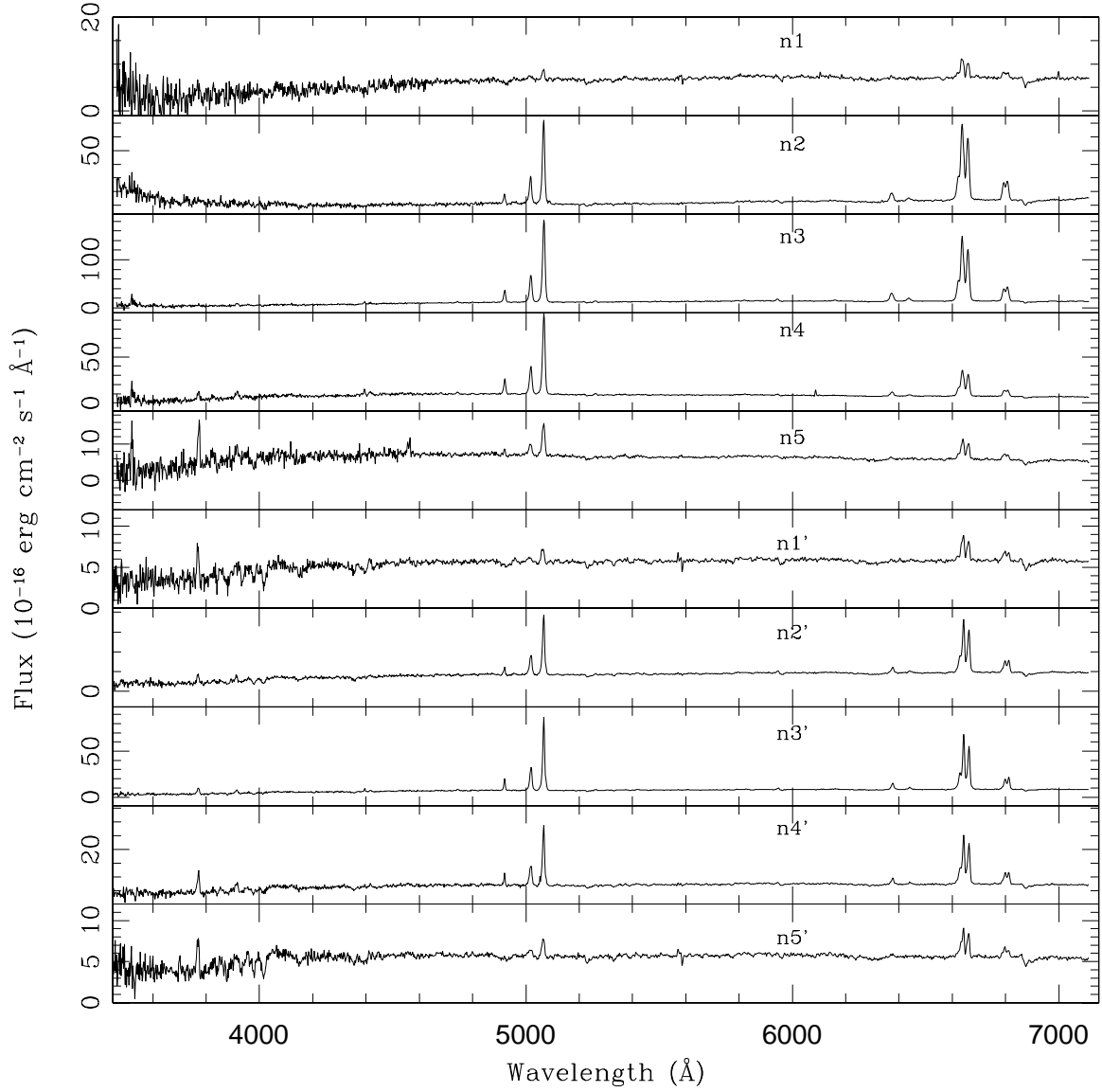}
\caption{Spectra of the nuclear and circumnuclear regions extracted from the central portions N and N'
of Tol1238-364. The regions are labeled with increasing numbers from NW to SE.\label{nucspec}}.
\end{figure}

A comparison with the H$\alpha$ images revealed a clear one-to-one 
correspondence among the so-selected emitting regions and the emission 
knots enhanced by the adaptive smooth filter (see Fig.~\ref{contour}). 1-D
spectra of these regions (Figs.~\ref{tolspec} and \ref{esospec}) were extracted 
and the corresponding blue and red parts were combined together. 
The central regions N and N' of Tol1238-364 were further divided into 
$\gtrsim$ 1.0$^{\prime\prime}$-wide subregions (n1, n2, n3, n4, n5, n1', n2', n3', n4', n5')
in order to analyze the circumnuclear region in more detail. 
Their spectra are shown in Fig.~\ref{nucspec}.
Regions n3 and n3' are centered on the H$\alpha$ emission peaks along the slit.
The signal-to-noise ratio (SNR) of the spectra ranges from 5 to 15 in the blue ($\lambda$ $\sim$ 4000 \AA)
and from 35 to 55 in the red ($\lambda$ $\sim$ 6500 \AA) in the central regions of Tol1238-364 
(n1, ..., n5, n1', ..., n5', C'').
In the outermost regions the SNR in the continuum is significantly lower,
although it is still sufficiently good in the prominent emission lines.
The regions with 
the lowest values are the outermost ones at P.A. = 150\degr\ and have SNR $\sim$ 2 in the blue,
around $\lambda$ 4000 \AA, $\sim$ 5 around $\lambda$ 5000 \AA, and $\sim$ 15 in the red.
For ESO381-G009 the SNR in the center is $\sim$ 8 in the blue and $\sim$ 40 in the red, while
in the outer regions SNR $\sim$ 2 in the blue and $\sim$ 8 in the red. 

A correction for Galactic extinction 
(A$_V$ = 0.17 mag, as derived from the A$_B$ value given by \citet{bh82}
following \citet{ccm89} and assuming a visual selective 
extinction R$_V$ = 3.1) was applied\footnote{The use of the Galactic extinction value derived from the
maps of \citet{sch98}, A$_V$ = 0.229 mag, would imply a 5\% difference in the measured fluxes.}.

\placefigure{tolspec}
\placefigure{esospec}
\placefigure{nucspec}

Before measuring the emission-line fluxes, a template spectrum --conveniently 
diluted to match the absorption 
features following the method outlined by \citet{hfs93}-- was used to correct the
spectra for the effects of 
the underlying stellar population, particularly affecting H$\beta$.
The H$\beta$ absorption was detected in all but one  
of the extracted 1-D spectra of Tol1238-3634 and in the central regions of
ESO381-G009. 

\begin{figure}
\plotone{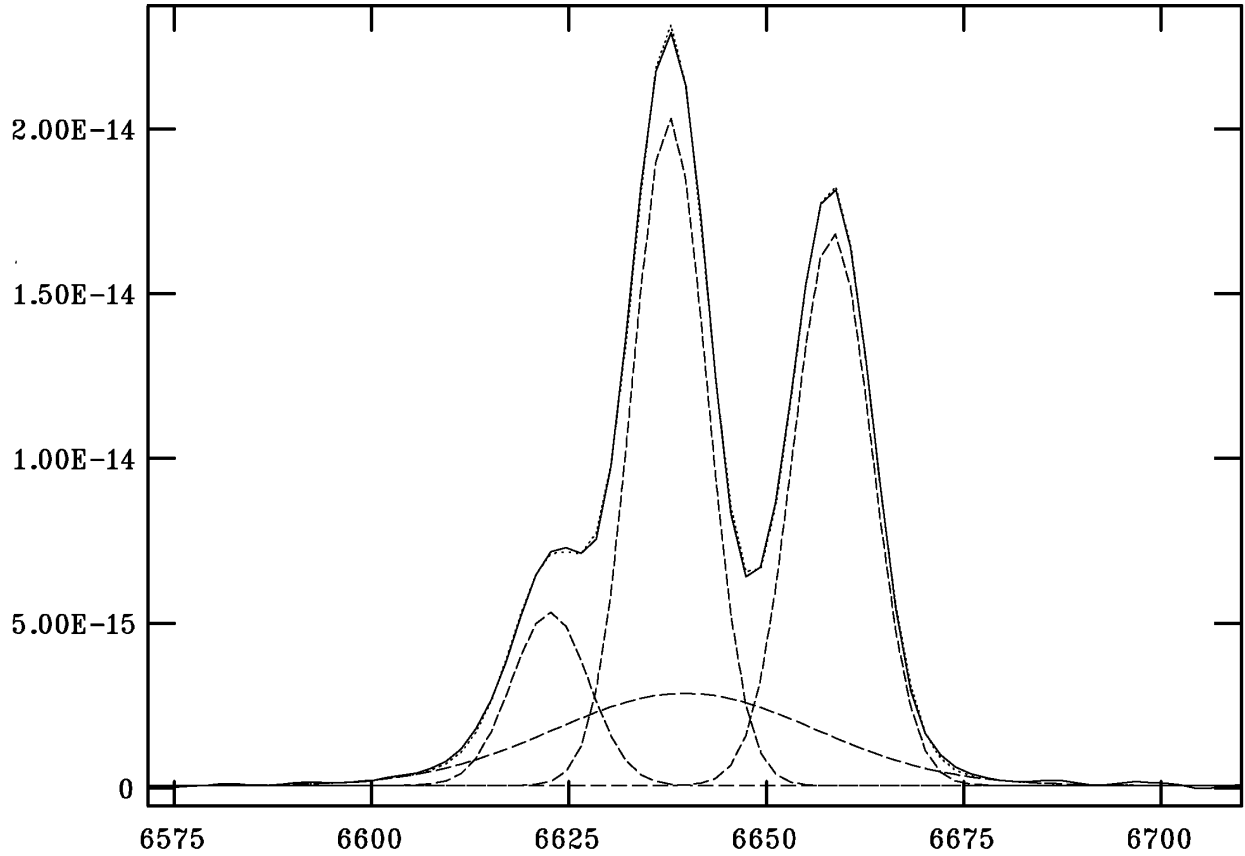}
\caption{Multi-gaussian fit of the blend
H$\alpha$ + [NII] $\lambda$6548,6563 for the
nuclear region of Tol1238-364. Fluxes in erg cm$^{-2}$ s$^{-1}$ \AA$^{-1}$
are plotted versus the wavelength in \AA. The fit of the blend is drawn with a 
solid line,
while the individual fit components are drawn with dashed lines. 
A broad H$\alpha$ component
was required to match the observed profile. \label{multigauss}} 
\end{figure}

A careful deblending of H$\alpha$ and the [NII] satellite-lines
was performed using a multi-gaussian fit procedure.
In the nuclear region of the Seyfert galaxy a better fit of the blend
was obtained assuming the presence of a broad H$\alpha$ component (Fig.~\ref{multigauss}),
possibly indicating that this galaxy is a Seyfert of type 1.9. The mean 
full-width at half-maximum (FWHM) of
the broad component found in the nuclear regions at P.A. = 150$^{\circ}$
and 146$^{\circ}$ is 1735 $\pm$ 60 km s$^{-1}$, in agreement with 
the FWHM $\sim$ 1860 km s$^{-1}$ found in polarized light by Heisler
et al. (1997).
The mean FWHM of the narrow line is $\sim$ 300 km s$^{-1}$.

\placefigure{multigauss}

Measured fluxes of the narrow lines were corrected for internal extinction 
assuming as intrinsic 
value of the Balmer decrement H$\alpha$/H$\beta$=2.85 \citep{ost89}. The 
Galactic interstellar reddening curve parametrized by \citet{mm72} has been 
used. Hereafter we indicate with the term ``intensities'' these extinction-corrected
fluxes.
The internal extinction derived in the above way can be considered an upper 
limit in the case of the nucleus of the Seyfert galaxy, since the
intrinsic ratio H$\alpha$/H$\beta$=3.1 might be more
appropriate according to some authors \citep[e.g.][]{vo87}.
The observed emission-line fluxes and intensities relative to H$\beta$, 
the absolute fluxes of H$\alpha$ and H$\beta$, and the values of 
internal extinction for the individual regions are listed in the
Appendix (Tables A1 through A6). The distances of the extremes of every region to the 
intensity peak along the slit, expressed in arcseconds, are indicated, as well.

\section{MORPHOLOGICAL, PHOTOMETRIC, AND SPECTROSCOPIC PROPERTIES}

\subsection{Morphology}

\begin{figure}
\plotone{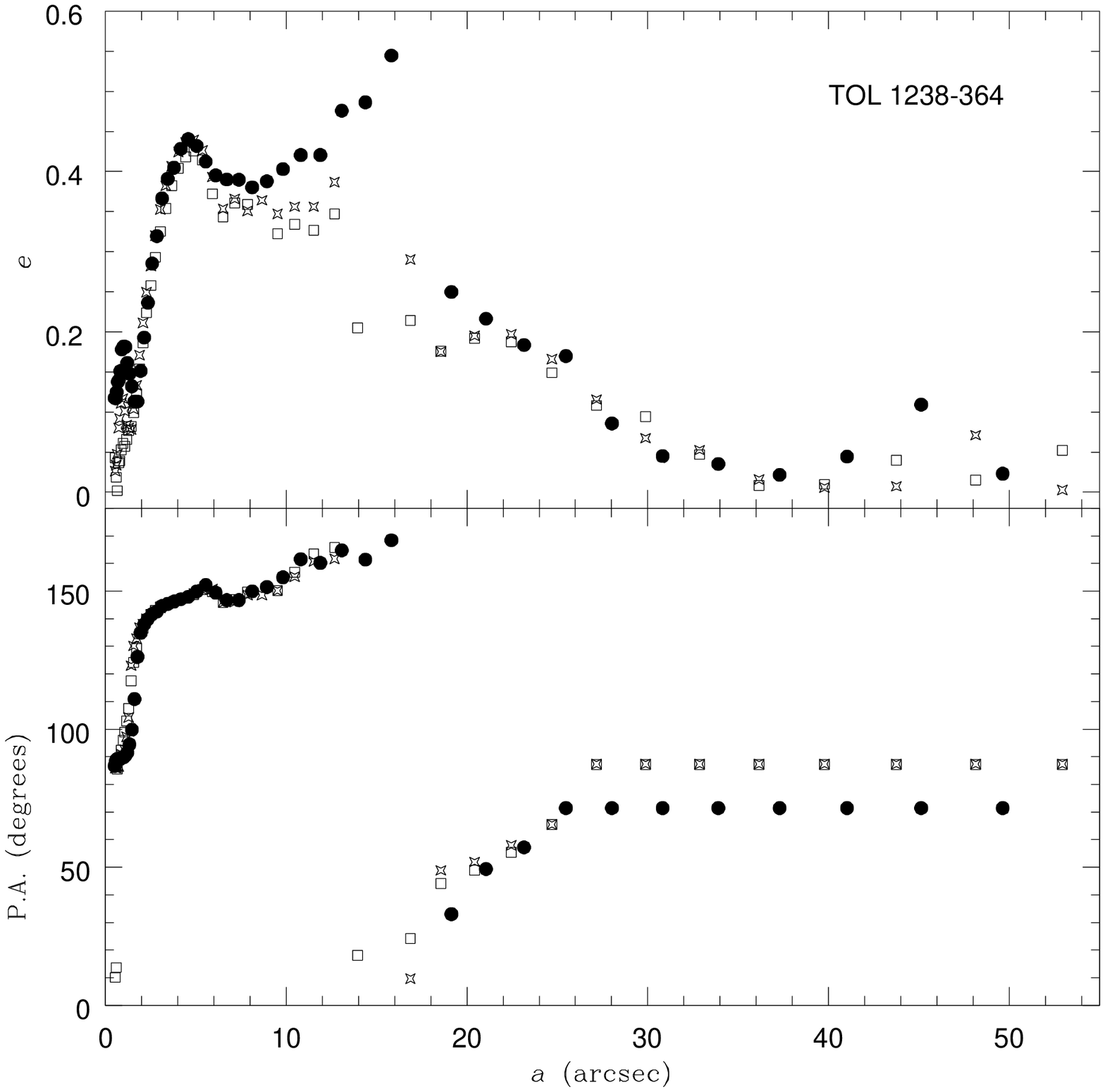}
\caption{Trend of the isophotal ellipticity (top) and position angle (bottom) with 
increasing radius for Tol1238-364 in B (full circles), V (stars), and R (squares).
\label{pa_ell_tol}}
\end{figure}

\begin{figure}
\plotone{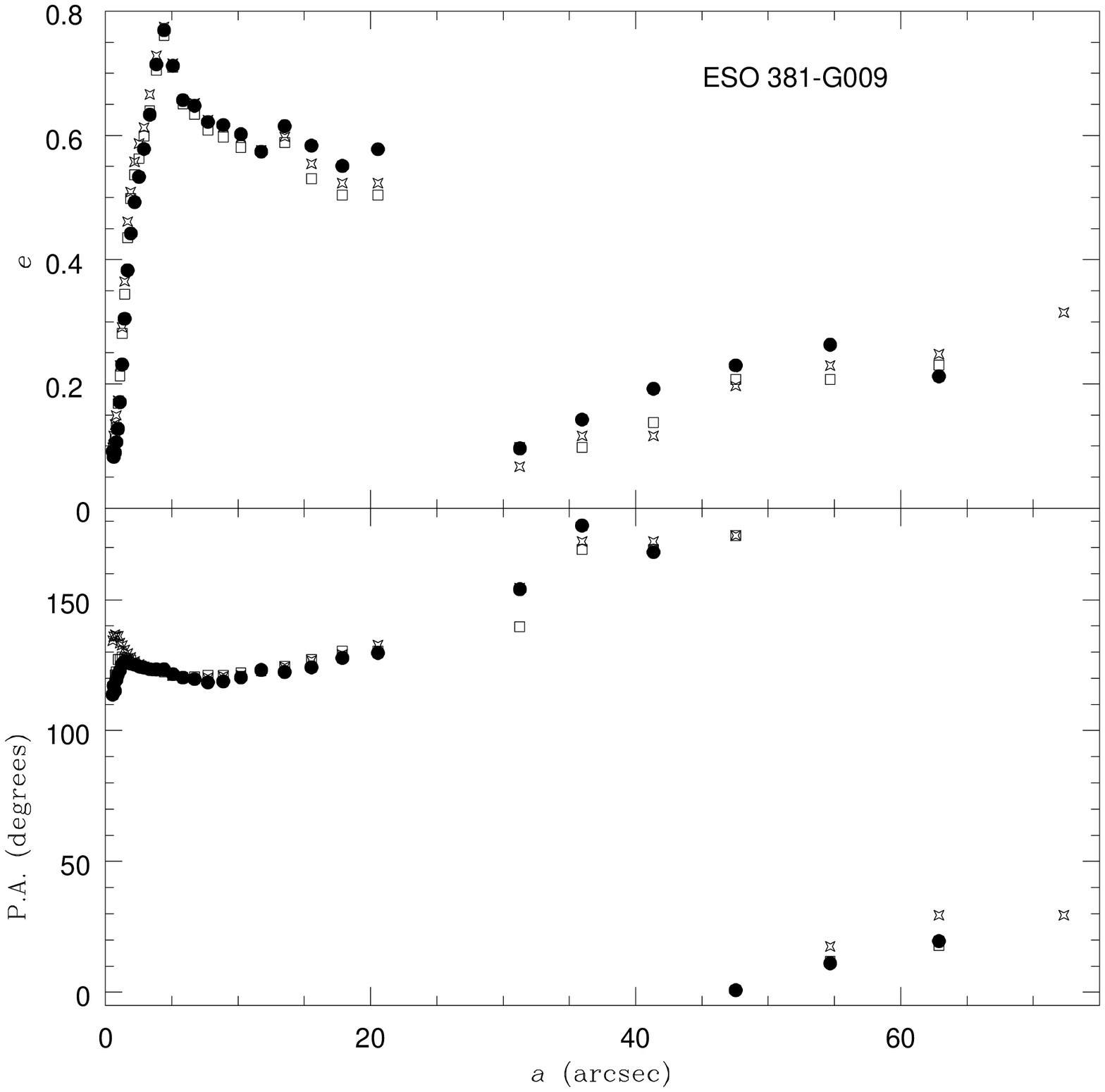}
\caption{Trend of the isophotal ellipticity (top) and position angle (bottom) with 
increasing radius for ESO381-G009 in B (full circles), V (stars), and R (suqares).
\label{pa_ell_eso}}
\end{figure}

The morphology of the two galaxies was investigated through the analysis of the 
broad-band optical images. In order to study the radial trend of the ellipticity 
($e = 1-\frac{b}{a}$ with $a$ and $b$ semiaxes of the ellipse)
and position angle (P.A.), their isophotes were fitted with ellipses with fixed center. 
In the outermost isophotes of Tol1238-364 the P.A. had to be fixed as well.
The $e$ and P.A. versus $a$ plots in Fig.~\ref{pa_ell_tol} reveal in Tol1238-364 
the existence of a bar extended until $a\, \sim\, 5^{\prime\prime}$ with 
$e\, \simeq \, 0.5$ and P.A. $\sim$ 145\degr, spiral arms mostly visible in B and
extended until $a\, \sim\, 16^{\prime\prime}$, and a clear twist of the isophotes 
(P.A.  $\sim$ 80\degr), which become nearly circular beyond $a\, \sim\, 30^{\prime\prime}$.
Also in ESO381-G009 (Fig.~\ref{pa_ell_eso}) we observe a twist of the isophotes, 
whose P.A. changes from 
$\sim$ 120\degr\ to $\sim$ 30\degr\ and whose ellipticity increases beyond $a\, 
\sim\, 30^{\prime\prime}$, giving the impression that they are stretching towards 
Tol1238-364.
A strongly elliptical bar ($e\, \sim\, 0.8$) is located at P.A. $\sim$ 125\degr\
and seems to extend its isophotes up to $a\, \sim \, 20^{\prime\prime}$, after which we 
observe a sudden transition to the spiral arms arranged into a ring-like shape.

We repeated the isophote fitting with free ellipses in order to investigate the change in the
position of the isophote center. We estimated a decentering degree, as defined by
\citet{mm99}, of 7.5\% for Tol1238-364 and 3.4\% for ESO381-G009. 
This parameter gives an indication of the degree of asymmetry of the galaxy disks.

\placefigure{pa_ell_tol}
\placefigure{pa_ell_eso}
  
\begin{figure}
\plotone{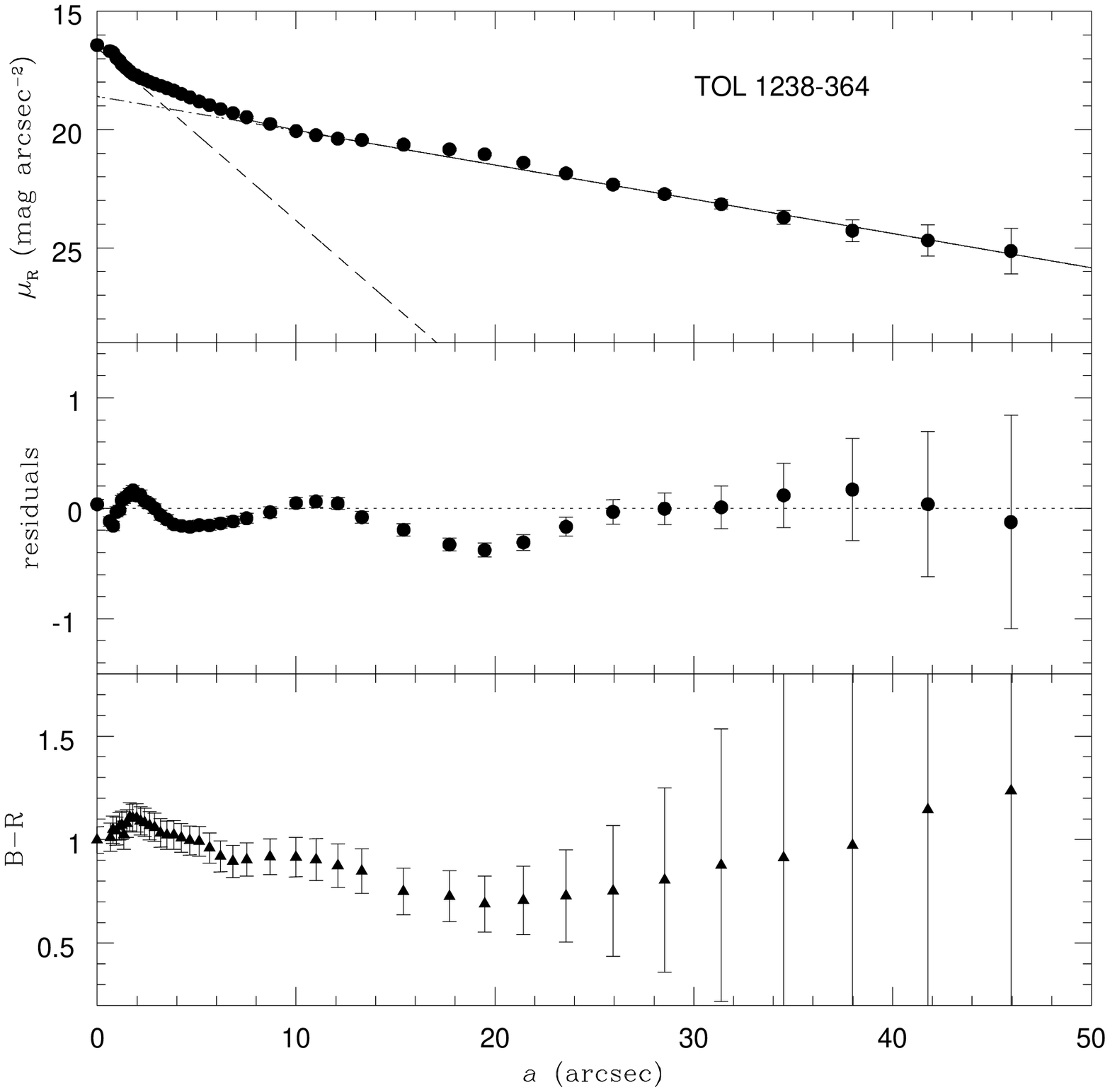}
\caption{From top to bottom: fit to the $R$-band surface brightness profile of Tol1238-364 
with an exponential bulge (dashed line) and an exponential disk (dash-dotted line) component; 
fit residuals, and \br\ radial profile. \label{tol_col}}
\end{figure}

\begin{figure}
\plotone{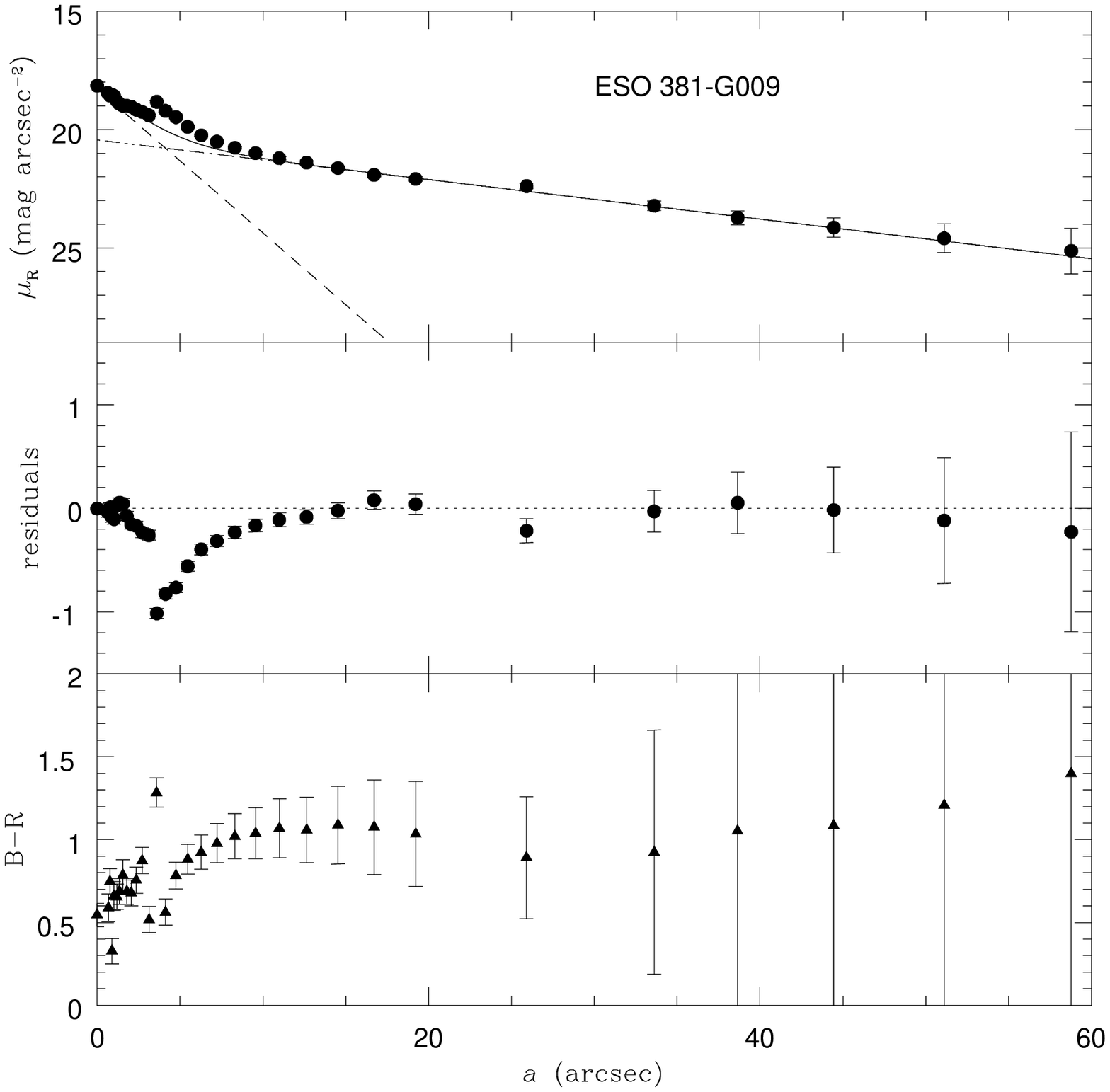}
\caption{From top to bottom: fit to the $R$-band surface brightness profile of ESO381-G009 
with an exponential bulge (dashed line) and an exponential disk (dash-dotted line) component; 
fit residuals, and \br\ radial profile.\label{eso_col}}
\end{figure}

A bulge-disk decomposition of the surface brightness profiles of both galaxies in the 
three passbands was performed without a PSF-deconvolution, assuming an exponential law 
for the two components \citep{and98,as94}:
\begin{equation}
\Sigma_{\rm bulge}(r)=5.36 B_e \exp\left[-1.68\frac{r}{r_e}\right]
\label{expbulge}
\end{equation}
\begin{equation}
\Sigma_{\rm disk}(r)=D_0 \exp\left(\frac{r}{h_D}\right),
\label{disk}
\end{equation}
where $\Sigma_{\rm bulge}(r)$ and $\Sigma_{\rm disk}(r)$ are the
surface intensities of the bulge and disk components (in mag arcsec$^{-2}$)
at a radius $r$, $B_e$ is the surface intensity at the effective radius $r_e$ 
(i.e. the radius containing half of the bulge light), $D_0$ is the central
surface intensity of the disk and $h_D$ is the disk scale-length.
The structural parameters averaged on the three passbands, the bulge-to-total light ratios $B/T$,
and the total magnitudes integrated to infinity are listed in Table\ref{phot}.
Magnitudes corrected for Galactic extinction are reported as well.
The B and V magnitudes measured for Tol1238-364 are in agreement with \citet{hunt99}.
The B/T ratios are typical of disk-dominated galaxies. Let us note that 
the existence of a real bulge in ESO381-G009 is uncertain. Actually the central 
part of the galaxy consists of a structure composed by two knots, the brightest of
which was chosen as the nucleus, although it is not located in the geometrical center 
and we lack information about the kinematics of the system.
In order to investigate the color gradient \br, we built a new $B$ surface brightness profile
by applying the same ellipses fitted to the R-band isophotes.
The \br\ profile of Tol1238-364 (Fig.~\ref{tol_col}) shows a red color for the bar, 
which is probably dominated by an old stellar population, blue spiral arms, and becomes red
again in the outermost part of the galaxy, where the disk might be affected by dust extinction.
In ESO381-G009, instead, the inner part of the bar exhibits a blue color (Fig.~\ref{eso_col}), 
probably as an effect of
the presence of central star-forming regions in agreement with the bright knots visible in the 
H$\alpha$ image. The profile reddens in the outer part of the bar and becomes blue again in 
correspondence of the spiral arms. 

\placetable{phot}

The continuum-subtracted H$\alpha$ image shows the presence of bright emitting regions
distributed all over Tol1238-364, but more concentrated
toward its nucleus and in the north-eastern spiral arm facing the companion galaxy (Figs.~\ref{images}
and \ref{contour}),
and confirms the structures already revealed by the analysis of the broad-band images for ESO381-G009. 
Additionally, a sort of plume departing from the northern side of ESO381-G009 is visible in all
images, as well as an extension of star forming regions (better visible in the H$\alpha$ image) on
the south, outside the ``ring'', especially in direction of the companion galaxy.
All the above features are also emphasized as blue regions in the \br\ image (Fig.~\ref{cmap}). 

\placefigure{tol_col}
\placefigure{eso_col}

\subsection{Spectroscopic Classification of the Emission-Line Regions}

\begin{figure}
\plotone{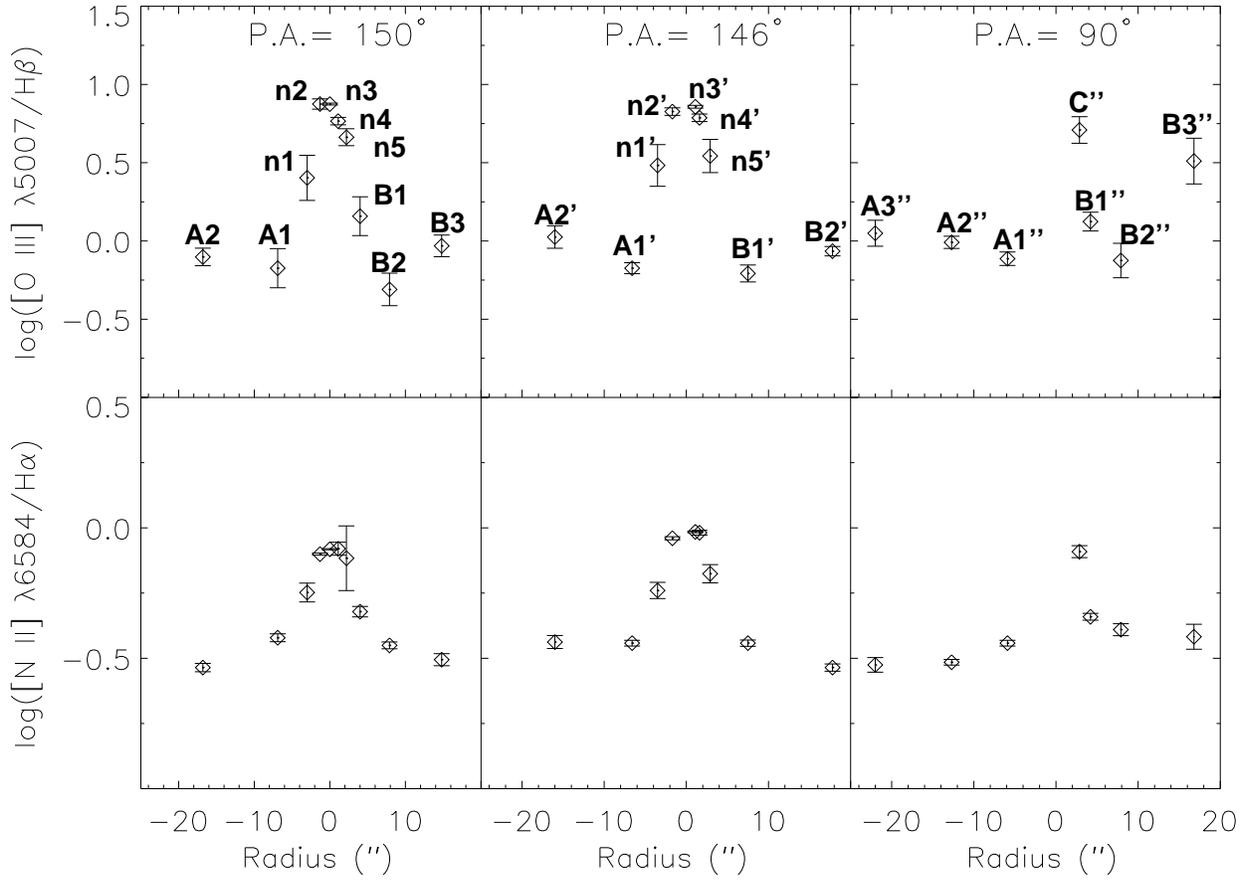}
\caption{Trend of the ionization degree with 
the radial distance, represented by the
emission-line ratios [\ion{O}{3}] $\lambda$5007/H$\beta$ and [\ion{N}{2}] 
$\lambda$6583/H$\alpha$, for Tol1238-364. Positive radii indicate
regions East to the nucleus. \label{tol_iondeg}} 
\end{figure}

We analyze in this section the spectroscopic characteristics of the two 
galaxies separately:

\emph{Tol1238-364}. The trend of the ionization degree along the direction of 
the slits is shown in Fig.~\ref{tol_iondeg}, where the emission-line ratios [\ion{O}{3}] 
$\lambda$5007/H$\beta$, and [\ion{N}{2}] $\lambda$6583/H$\alpha$
are plotted against the radial distance
to the nucleus for the regions identified as explained in \S~2.
All the diagrams show a peak in correspondence of the nucleus as 
expected in the case of
photoionization by a source with a power-law spectrum, and a decreasing ionization degree 
moving away from it. Nevertheless the outermost regions exhibit
an increasing trend of the [\ion{O}{3}]/H$\beta$ ratio and an exactly opposite trend 
of the 
[\ion{N}{2}]/H$\alpha$ ratio. This symmetric behavior is typical of \ion{H}{2}-like regions,
however we stress that this trend is observed in only two bins per slit position.
The rise of the ionization degree is more evident at P.A. 90$^{\circ}$ for the
regions at the East side of the nucleus and could be partly caused by an 
underestimate of
H$\beta$, difficult to measure in spectra with relatively
low SNR, and partly an effect of the presence of a hot interstellar 
medium.

\placefigure{tol_iondeg}

\begin{figure}
\plotone{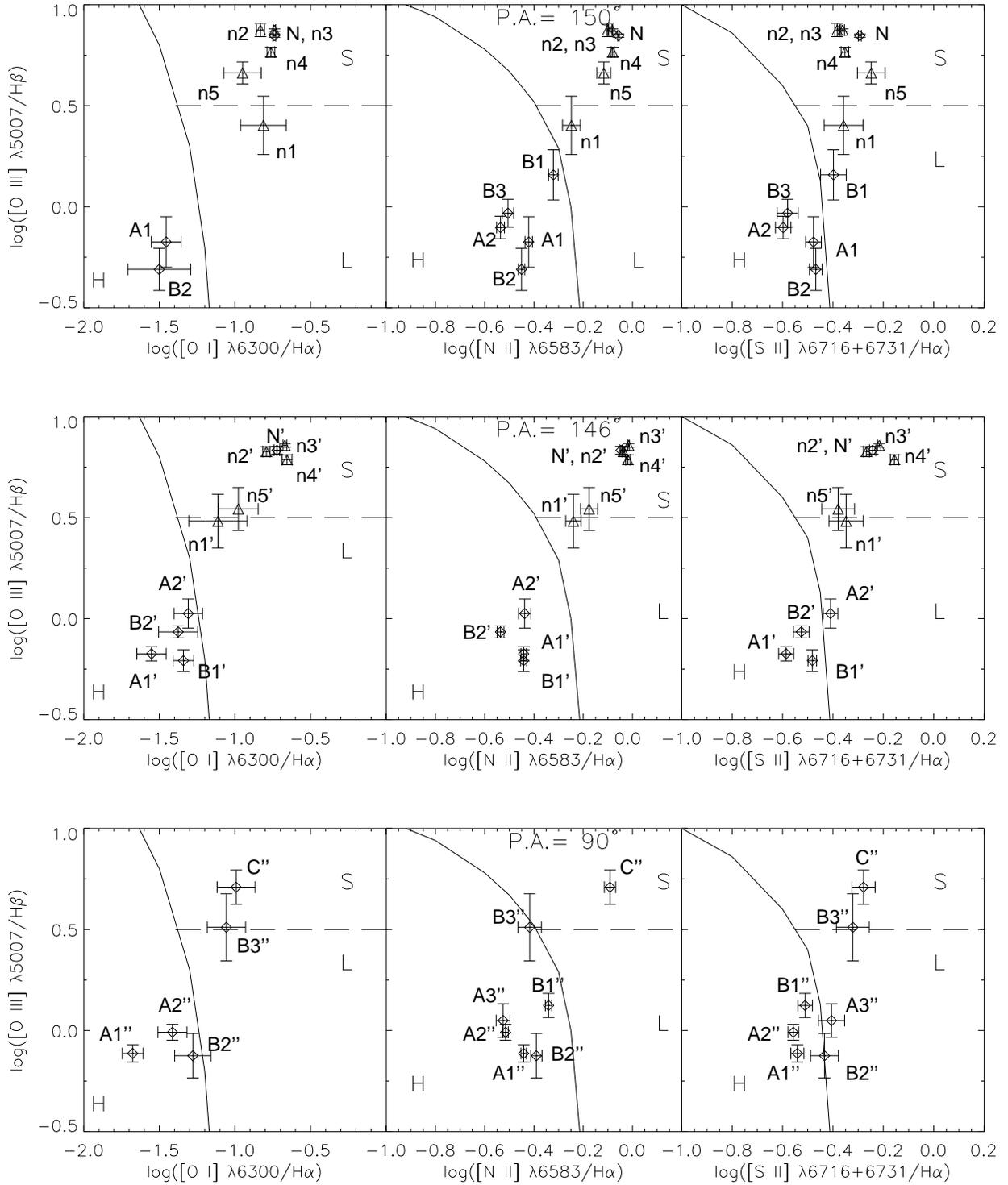}
\caption{Veilleux - Osterbrock (VO) diagnostic
diagrams for all the regions extracted from the
spectra of Tol1238-364. S, L, and H indicate Seyfert, Liner, and \ion{H}{2} areas of 
the diagrams; the solid line sets the empirical separation between AGN 
and \ion{H}{2}-like regions, while the dashed line divides Seyferts from
Liners. Diamonds indicate the extranuclear regions and the central regions at the
three P.A.; triangles indicate the nucleus and circumnuclear regions extracted from
the central kiloparsec.  \label{tol_vo}}
\end{figure}

\begin{figure}
\plotone{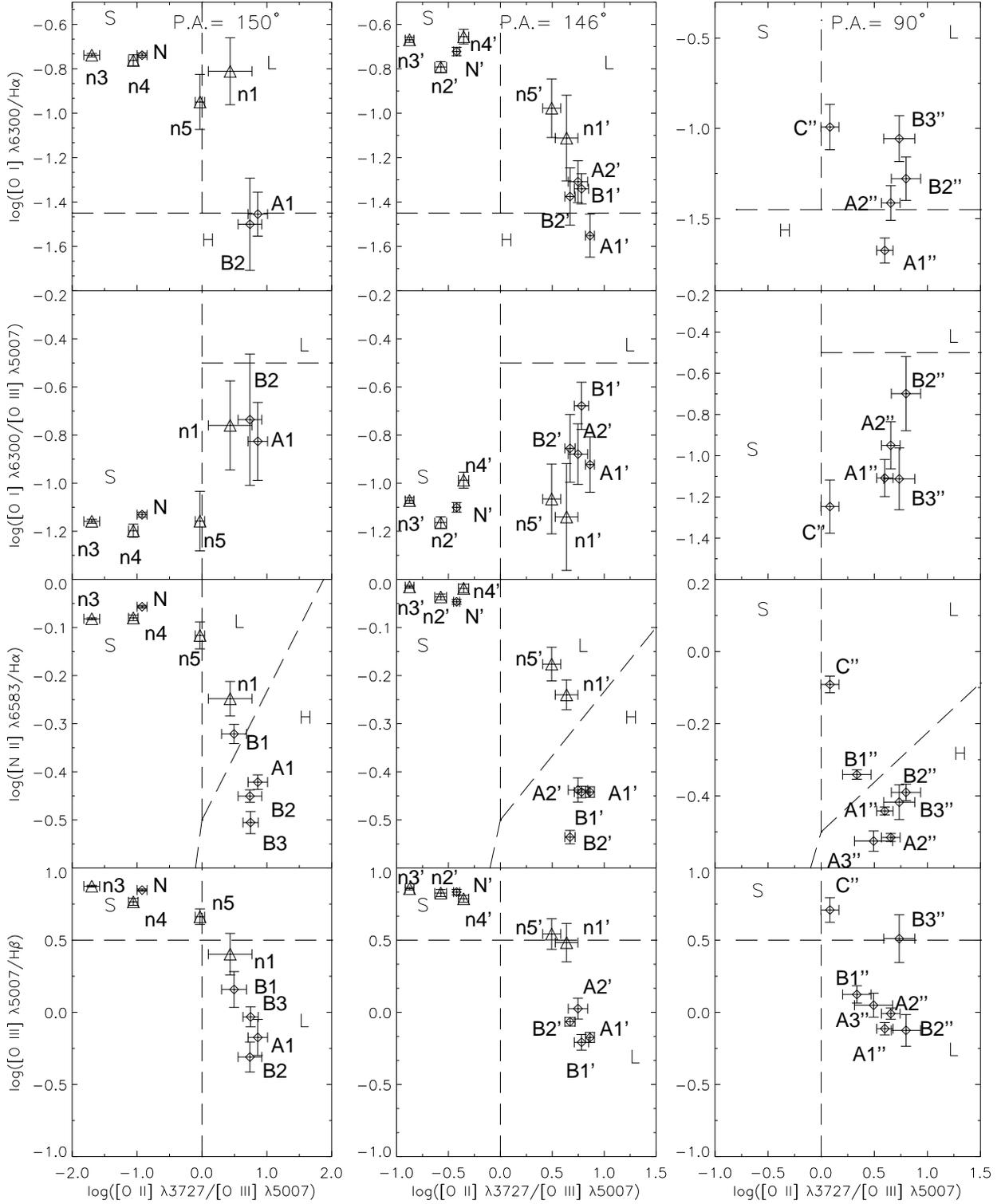}
\caption{[OII] $\lambda$3727 - based diagnostic
diagrams for all the regions of Tol1238-364. Symbols and S, L, and H are as in
Fig.~\ref{tol_vo}; the corresponding areas of the diagrams are separated by
dashed lines. \label{tol_diag}} 
\end{figure}

In order to investigate the nature of the emission line regions and of their 
ionizing sources we used the intensities
listed in the Appendix (Tables A1 through A6) in order to build the
classic diagnostic diagrams (Figs.~\ref{tol_vo} and \ref{tol_diag})
[\ion{O}{3}] $\lambda$5007/H$\beta$ vs [\ion{O}{1}] $\lambda$6300/H$\alpha$, [\ion{N}{2}] 
$\lambda$6583/H$\alpha$, [\ion{S}{2}] $\lambda$6716+6731/H$\alpha$ 
\citep[VO diagrams]{vo87}, 
and [\ion{O}{1}] $\lambda$6300/H$\alpha$, [\ion{O}{1}] $\lambda$6300/[\ion{O}{3}] 
$\lambda$5007, 
[\ion{N}{2}] $\lambda$6583/H$\alpha$, [\ion{O}{3}] $\lambda$5007/H$\beta$ vs
[\ion{O}{2}] $\lambda$3727/[\ion{O}{3}] $\lambda$5007 \citep{bpt81,sf90,hs83}.

A careful examination of the diagnostic diagrams allowed us to investigate in some detail 
the nature of the circumnuclear 
and extranuclear regions (throughout this paper we call ``circumnuclear 
regions''  the regions within 1 kpc from the nucleus, assuming 
H$_0$ = 75 km s$^{-1}$ Mpc$^{-1}$). 
The nuclear and circumnuclear regions n2, n3, n4, n2', n3', and n4' extracted from the 
central portions of the bidimensional spectra have line-ratios characteristic of Seyfert 
galaxies, 
while the adjacent regions n1, n5, n1', and n5', as well as region C'' at P.A. = 90\degr, 
gradually move towards the LINER area of the diagrams, probably indicating a transition 
from a dominating non-thermal photoionizing source to a dominating thermal source.

Along P.A. = 150$^{\circ}$ a 
progressive transition from Seyfert-like to \ion{H}{2}-like properties is evident when 
going out from the 
nucleus. The region A1 on the west side of the nucleus shows typical \ion{H}{2}-like 
features, while B1 on the opposite side has transition properties between LINERs and
\ion{H}{2} regions.
At P.A. = 146$^{\circ}$ the extranuclear regions all fall in the \ion{H}{2}-like
area of the diagrams,
although A2' approaches the LINER region in the [\ion{O}{1}]/H$\alpha$ diagram 
and enters it in the [\ion{S}{2}]/H$\alpha$ one.  
A similar behavior is observed in the regions B2'' and A3'' at P.A. = 90$^{\circ}$;
the region B3'' has an anomalously high [\ion{O}{3}]/H$\beta$ ratio,
which locates it at the border between Seyferts and LINERs. However, this could
be a consequence of an underestimate\footnote{The correction for underlying stellar absorption
was particularly difficult for this spectrum, therefore the measured H$\beta$ flux is uncertain.}  
of H$\beta$ and the region is most probably to be
classified as LINER type.
In contrast with \citet{braetal98} we do not 
observe any marked E/W asymmetry in the distribution of the emission line ratios.
The discrepancy among different diagnostic 
diagrams could be caused by an enhancement of [\ion{O}{1}] and [\ion{S}{2}] lines 
consequent to ionization
by starburst-driven shocks, as suggested by \citet{braetal98}. 
However, some of the anomalous regions are not well centered
on any knot of emission in the H$\alpha$ image (Fig.~\ref{contour}), and have
low density of star formation rate 
(see last column of Table~\ref{tab2}; details on the content of the table
are given in \S 3.4). This could indicate that these regions are
weakly ionized by a low number of young stars.
Actually, ``faint emission'' regions are found to have relatively high values of
[\ion{S}{2}]/H$\alpha$ ($\sim$ 0.3 -- 0.5) and [\ion{O}{1}]/H$\alpha$ ($\sim$ 0.04 -- 0.1) 
because of their low ionization degree \citep{smetal93}. 

\placefigure{tol_vo}
\placefigure{tol_diag}

\begin{figure}
\plotone{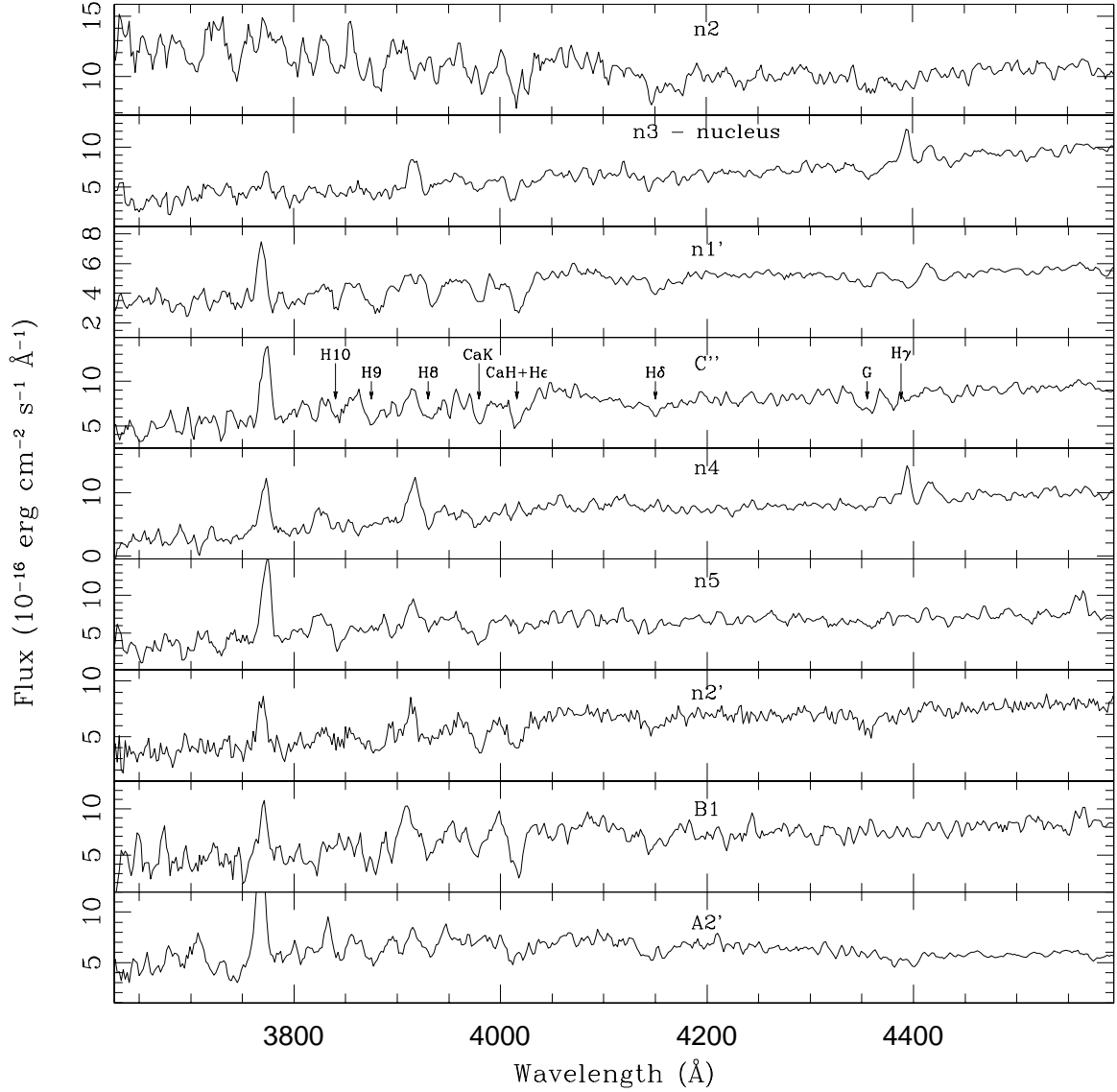}
\caption{Enlargement of the blue range of a few circumnuclear regions and two extranuclear regions of
Tol1238-364 showing different behaviors of metal and Balmer absorption lines, which suggest different
underlying stellar populations (see text for details). 
\label{spectra_zoom}}
\end{figure}

A qualitative examination of the spectra of the individual extranuclear and circumnuclear regions
(Figs.~\ref{tolspec} and \ref{nucspec}) gives further insights into their nature.
In particular, we notice that the spectra of the outermost regions tend to be
dominated by young (or sometimes very young) stellar populations, as suggested by the weakness
of metal absorption lines, the strength of Balmer absorption lines, the high strength of the 
CaH+H$\epsilon$ doublet (2.3 \AA\ $\lesssim$ EW $\lesssim$ 9.5 \AA) relative to the CaK absorption line
(0.5 \AA\ $\lesssim$ EW $\lesssim$ 5.2 \AA), and the blue continuum usually 
peaked around 4000 \AA\ (e.g. region A2' in Fig.\ref{spectra_zoom}). These spectra appear to be 
dominated by a population of A-type (or
F-type) stars. In two of the outermost regions, B3 and B2', we observe a very blue continuum
without detectable metal absorption lines and very weak Balmer absorption lines, indicative of
a dominating population of O-B stars. Toward inner radii, the extranuclear regions show a 
slightly redder continuum, peaked around 4500 \AA, progressively deeper metal absorption lines,
but still very strong Balmer absorption lines indicating important contributions to the spectrum
from both a relatively young and an old (G to K-type) stellar population (e.g. region B1 in
Fig.\ref{spectra_zoom}).
The most important contribution from old stellar populations is found in the circumnuclear
regions, which exhibit a red continuum, strong metal lines and comparably strong Balmer 
absorption lines (see e.g. regions n5, n2' in Fig.\ref{spectra_zoom}). 
The CaK (3.5 \AA\ $\lesssim$ EW $\lesssim$ 7.7 \AA) absorption line becomes stronger 
than the CaH+H$\epsilon$ doublet (1.6 \AA\ $\lesssim$ EW $\lesssim$ 4.5 \AA), 
as typical of G-type or later stellar populations \citep[e.g.][]{ro84,lr96}, but not 
in all the regions within
the central kiloparsec. In particular this is not the case for regions n1', C'', and especially the 
nucleus n3 (EW(CaK) $\sim$ 3 \AA, EW(CaH+H$\epsilon$) $\sim$ 5 \AA), 
where an A-type population probably gives a non-negligible contribution to the spectrum, and 
region n2, where a red continuum is overlapped with a component rising toward blue wavelengths
(Fig.\ref{spectra_zoom}).
These last spectra give evidence of the presence of (relatively) recent nuclear and circumnuclear
starbursts. The general trend described above, with old population contributions becoming 
progressively more important with decreasing radius and young populations dominating the 
outermost regions (at radii of about 3-5 kpc), is in agreement with the stellar population gradients
observed by \citet{raietal02}. 
However, these authors (their Fig.~3) find that in the nucleus the CaK absorption line has equivalent width 
significantly larger than CaH+H$\epsilon$. This discrepancy with our results could be a consequence 
of a different spatial sampling and size of the extracted regions.
While the spatial scale of their spectra is 0$^{\prime\prime}$.7 pixel$^{-1}$ and they  
extracted one-dimensional spectra in windows of 2$^{\prime\prime}$.1 in the nuclear regions,
our spatial scale is 0$^{\prime\prime}$.336 pixel$^{-1}$ and the good seeing conditions allowed us
to extract nuclear regions $\sim$ 1$^{\prime\prime}$ in size along the slit.
Rebinning our spectra to match the spatial sampling of Raimann et al. and extracting regions of similar size,
we verified that a 1-pixel offcenter (i.e. ~0$^{\prime\prime}$.7) is sufficient to obtain a spectrum of the
nucleus dominated by an old stellar population, and thus reproduce the findings of \citet{raietal02}.
Indeed, the distribution of stellar populations in the circumnuclear regions appears very inhomogeneous
with strong variations within a few arcseconds, which makes the analysis very sensitive to the extraction 
method.
\placefigure{spectra_zoom}

Since in \ion{H}{2} regions the equivalent widths of Balmer emission lines are proportional 
to the ratio of ionizing photons to visible continuum photons from the embedded stars \citep{kkb89},
we can obtain additional information on the population of the \ion{H}{2} regions from the
observed values of EW(H$\alpha$) listed in Table~\ref{tab2}.
The H$\alpha$ EWs of the extranuclear regions are at (or even below) the lower limit of the distribution 
found by \citet{kkb89} for disk \ion{H}{2} regions (defined as the regions located at a distance $>$ 1 kpc 
from the nucleus) of spiral and irregular galaxies. These low emission-line equivalent widths could
indicate that the contribution of the young stellar associations to the optical continuum is rather small
and might suggest a significant contribution to the continuum by older stellar populations or a
prolonged period of star formation over the disk. 

\begin{figure}
\plotone{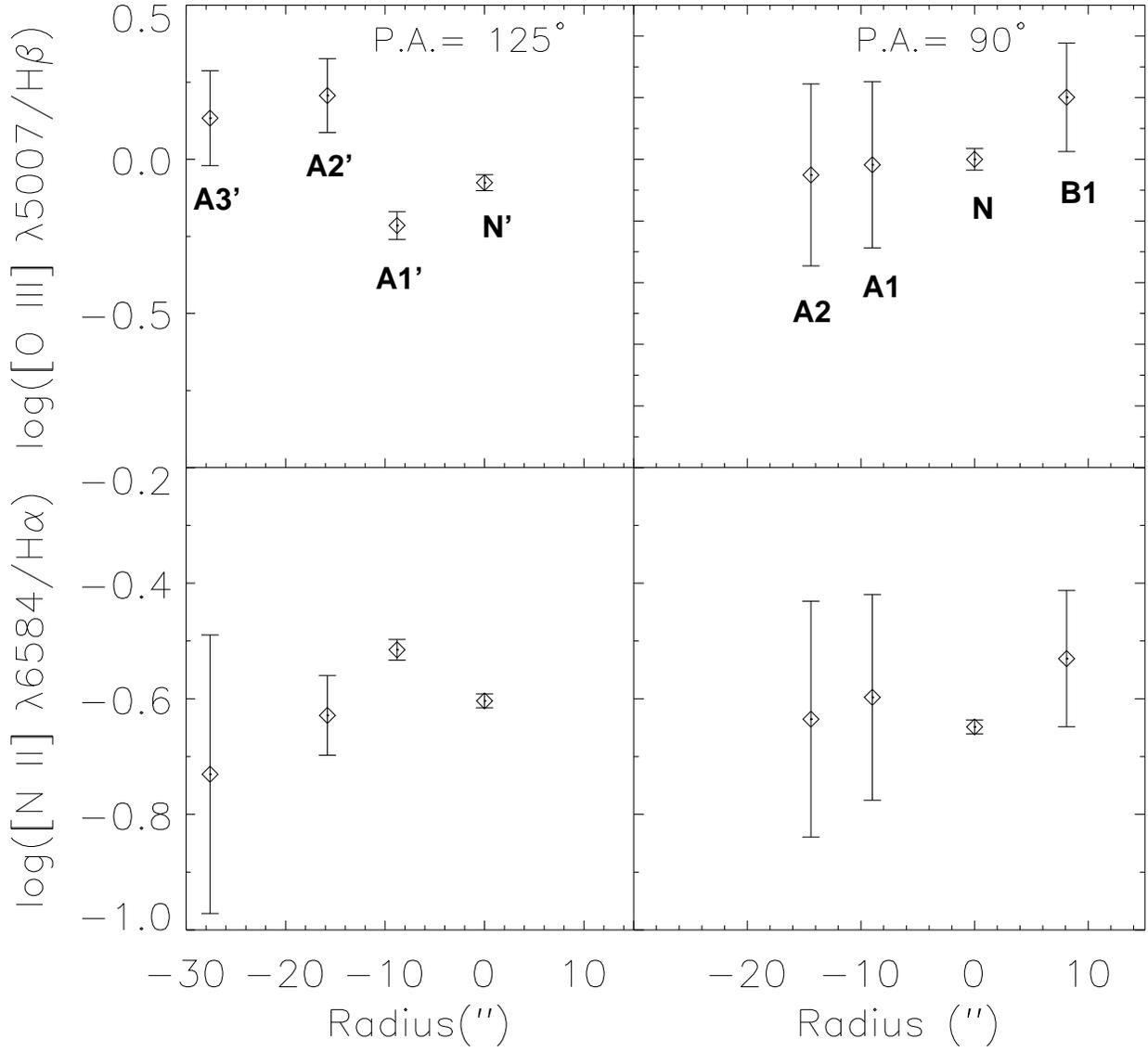}
\caption{Same as Fig.~\ref{tol_iondeg}, but for ESO381-G009. 
\label{eso_iondeg}}
\end{figure}

\begin{figure}
\plotone{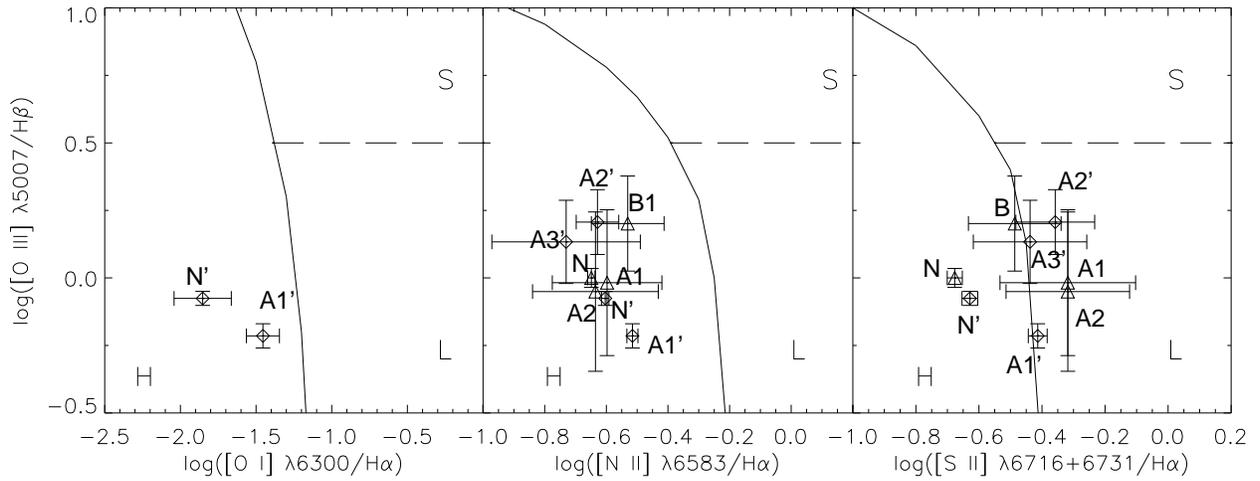}
\caption{VO diagnostic diagrams for ESO381-G009. Diamonds and triangles indicate regions 
extracted from the spectra at P.A. = 125$^{\circ}$ and P.A. = 90$^{\circ}$, respectively.
\label{eso_vo}}
\end{figure}

\emph{ESO381-G009.} The ionization-sensitive line ratios at several 
positions along the slits and the diagnostic diagrams have been plotted for
the companion of Tol1238-364 as well (emission line fluxes are given
in Tables A6 and A7). As expected, the [\ion{O}{3}]/H$\beta$ and 
[\ion{N}{2}]/H$\alpha$
(Fig.~\ref{eso_iondeg}) have symmetric trends typical of \ion{H}{2}-like regions, except for the 
regions A3', A4', A5',
which exhibit the same trend in both the line ratios. 
Actually, also for this galaxy not all the line ratios can be explained by
pure photoionization from thermal radiation: the shift of several points toward
the LINER area in the [\ion{S}{2}]/H$\alpha$ diagnostic 
diagram (Fig.~\ref{eso_vo}) could be caused in some cases by the faintness of the 
emission regions (A2, B1), but
in other cases (A1, A1', A2', A3') they are probably an indication of shock
heating as additional ionization mechanism \citep{ds95}.

\begin{figure}
\plotone{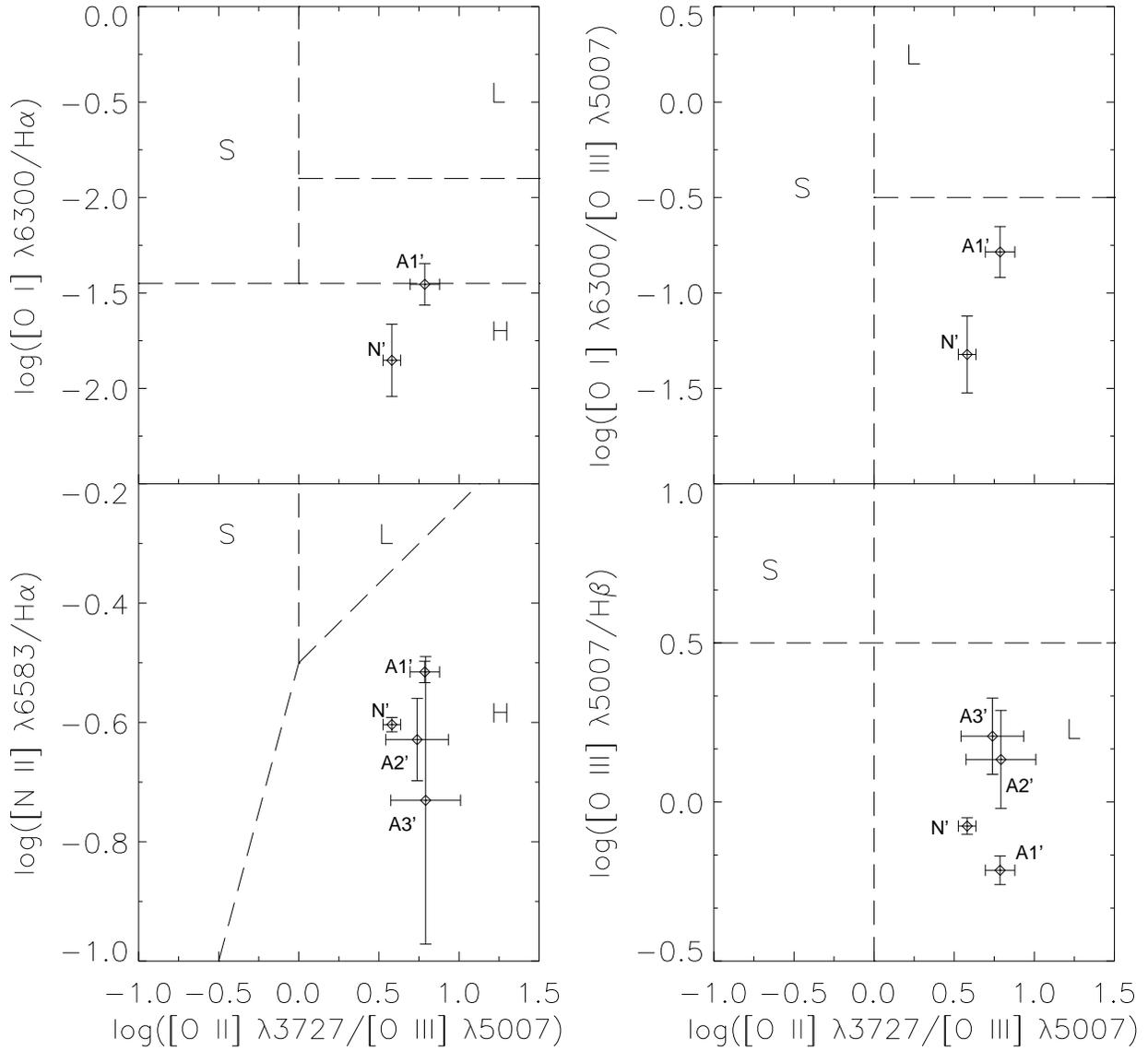}
\figcaption{Same as Fig.~\ref{tol_diag}, but for ESO381-G009 
at P. A. = 125$^{\circ}$. \label{eso_diag}}
\end{figure}

\placefigure{eso_iondeg}
\placefigure{eso_vo}
\placefigure{eso_diag}

The spectra of all the regions show typical properties of young stellar populations
with a blue continuum. Metal absorption lines are detectable only in the nucleus (N and N')
together with deep Balmer absorptions. However, the H$\alpha$ equivalent width measured
in the spectra are rather low (see Table~\ref{tab3}), ranging from $\sim$ 15 to $\sim$ 83 \AA.

\subsection{Photoionization Models}

The code CLOUDY 90 \citep{feetal98} has been used to
calculate photoionization models, which have been compared to observational 
emission-line flux ratios, in order to evaluate the physical parameters of
extranuclear and circumnuclear regions.
CLOUDY is designed to simulate emission-line regions
ranging from intergalactic medium to the BLR of quasars. Assuming diluted gas,
heated and ionized by the radiation field of a central object, and
simultaneously solving the equations of statistical and thermal equilibrium, 
CLOUDY can predict the physical conditions of the gas and its
resulting emission line spectrum.

Simple models can be obtained by specifying as input parameters 
the continuum emitted by the source, a set of assumed chemical abundances,
the total hydrogen density N$\rm_H$, 
and the ionization parameter U = Q$\rm _H$/(4$\pi$r$^2$N$\rm_H$c), 
where Q$\rm _H$ is the number of ionizing photons and r is the 
distance between the source and the inner side 
of the gas cloud assumed to have a plane parallel geometry.

We derived a first estimate of the physical parameters and abundances of
the ionized gas in the extranuclear regions by comparison with empirical diagrams
\citep{mg91,dtt02}.
In particular the R$_{23}$ ratio, defined as $\log[($ [\ion{O}{2}] $\lambda$ 3727 $+$
[\ion{O}{3}] $\lambda$ 4959,5007$)$/H$\beta]$, was compared with the model grid in the plane
[\ion{O}{3}]/[\ion{O}{2}] $vs$ R$_{23}$ in Fig.~10 of \citet{mg91}.
From this comparison the ionization parameter was found in the range
$-$3.5 $\lesssim$ $\log$U $\lesssim$ $-$3.0.
The metal abundances were derived by means of the N2 = [\ion{N}{2}] $\lambda$ 6584/H$\alpha$
calibrator \citep[][their Fig.~1]{dtt02} by referring to the model track for $\log$U = $-$3.
They were found in the ranges\footnote{[12$+\log$(O/H)]$_{\odot}$ = 8.9} 
8.2 $\lesssim$ 12$+\log$(O/H) $\lesssim$ 8.6 and 
8.2 $\lesssim$ 12$+\log$(O/H) $\lesssim$ 8.45 for Tol1238-364 and ESO381-G009, respectively.
The choice not to directly calculate the abundances was motivated by the fact that the 
metallicity is very sensitive to the gas temperature,
whose determination requires the measurement of the [\ion{O}{3}] $\lambda$4363 line.
Some authors \citep{raietal00} found a ratio [\ion{O}{3}] $\lambda$ 4363/H$\beta$ 
$\lesssim$ 0.02 in \ion{H}{2} and starburst galaxies, 
which implies electronic temperatures lower than 10$^4$ K.
Unfortunately, the low SNR of our spectra in the blue range
prevented the detection of this line at intensity levels lower than 0.1
relative to H$\beta$ in all the extranuclear regions.

\begin{figure}
\plottwo{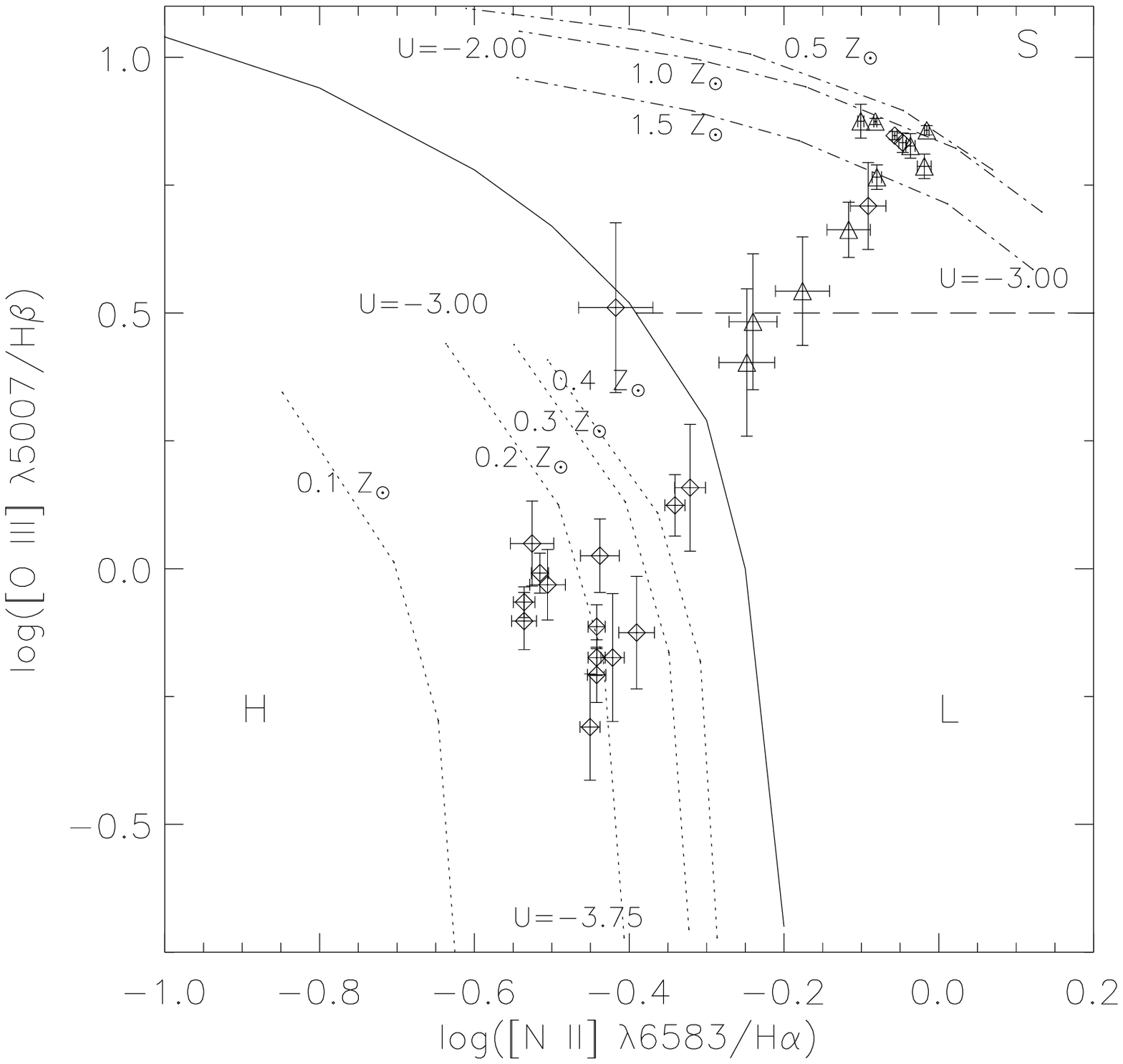}{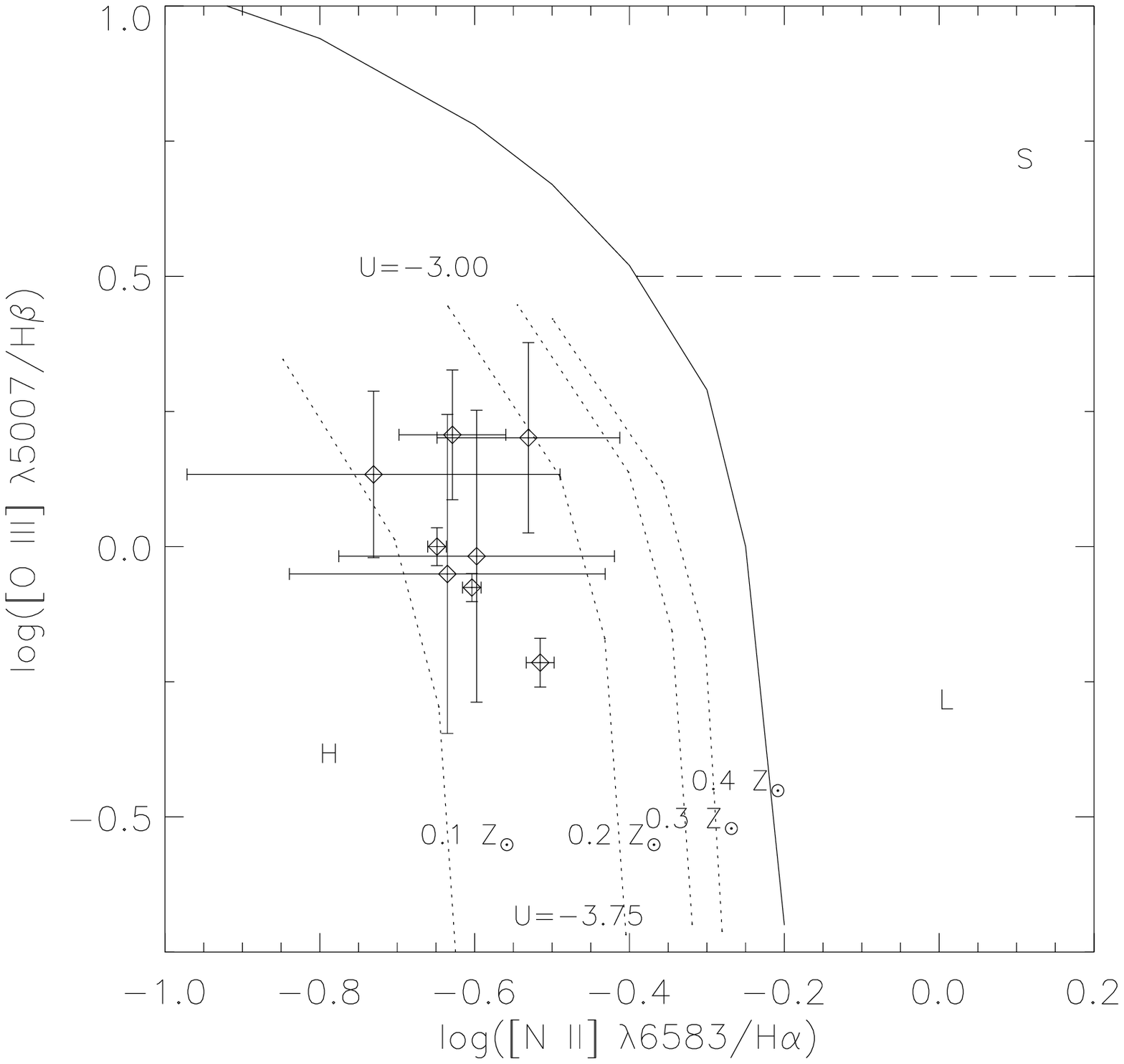}
\caption{Grid of photoionization models for \ion{H}{2}
regions calculated with the code CLOUDY using the input parameters 
described in the text and four different values of metallicity (dotted lines)
Z = 0.1, 0.2, 0.3, and 0.4 Z$_{\odot}$. 
The grid is represented onto VO-diagnostic diagrams together with the 
observed logarithmic line ratios for the \ion{H}{2} regions of Tol1238-364 (\emph{left})
and ESO381-G009 (\emph{right}). 
The photoionization parameter $\log$U
varies from  $-$3.75 to $-$3.0 with a step of 0.25 from bottom to top
along the model lines. The observational points are mainly concentrated 
around the line at Z = 0.2 Z$_{\odot}$ for Tol1238-364 and between 0.1 and 0.2 Z$_{\odot}$
for ESO381-G009. Circumnuclear regions of Tol1238-364 within a radius of 0.5 kpc are compared
with power-law ionization models (dot-dashed lines) for three values of metallicity, 0.5, 1.0, and 
1.5 Z$_{\odot}$. $\log$U changes from $-$3.00 to $-$2.00 with a step of 0.25. The observational 
points appear in agreement with solar or supersolar metallicities. 
\label{N2models}}
\end{figure}

Models were obtained assuming as continuum Mihalas non-LTE stellar atmospheres
\citep{mih72} with T = 40000 K and as N$\rm _H$
a typical value derived from the [\ion{S}{2}]6716/6731 ratios (N$\rm _H$ $\sim$ 50 cm$^{-3}$),
while the ionization parameter 
was left free to vary in the range $-$4 $<$ $\log$U $<$ $-$3 with step 0.25 dex,
and the metal abundances
were varied in the range 0.1Z$_{\odot}$ $\leq$ Z $\leq$ 0.4Z$_{\odot}$ with step 0.1.
The resulting grid of models (dotted lines) is plotted onto the [\ion{N}{2}]/H$\alpha$ 
diagnostic diagram in Fig.~\ref{N2models} overlapped with the observational points.
The models that appear to better reproduce the observed line ratios for most of 
the extranuclear regions of Tol1238-364
are those with $-$3.50$\lesssim\log$U$\lesssim$ $-$3.25 and abundances slightly 
changing around  0.2Z$_{\odot}$. The same models seem to be in agreement also
with ESO381-G009 flux ratios, but with a larger spread in metallicity between
0.1Z$_{\odot}$ and 0.2Z$_{\odot}$ and with higher ionization parameters, distributed around 
$-$3.25.
 
In the case of the circumnuclear regions of Tol1238-364, which are dominated by 
the non-thermal ionization (\S 3.2), the models were calculated by assuming
as continuum a power law with spectral index  
$\alpha$ = $-$1.8 (F$_{\nu}\,\propto \, \nu^{\alpha}$), N$\rm _H$ = 10$^3$ cm$^{-3}$ 
(as derived from the [\ion{S}{2}]6716/6731 ratios),
and $-$3 $<$ $\log$U $<$ $-$2.
Models with different power law index ($-$1.5 and $-$2.0) have also been attempted but
they significantly deviate from the observed values.
In these regions the electronic temperature of the gas can be estimate since 
[\ion{O}{3}] $\lambda$4363 is sufficiently strong to be measured. Nevertheless,
the usual formulas \citep[e.g.][]{paetal92,itl94} for the direct calculation of metal 
abundances are calibrated with
typical \ion{H}{2} regions and not with active nuclei, therefore they cannot be
applied in this case.
Trying to model the N2 calibrator as a function of metallicity and ionization parameter 
for a power-law ionizing continuum with index $-$1.5 we have obtained theoretical tracks,
which are far below the observed [\ion{N}{2}]/H$\alpha$ ratios. This is an indication that
there is a significant overabundance of nitrogen. Indeed, theoretical tracks obtained 
assuming a triple nitrogen abundance approach much more the data in the range of the
expected ionization parameter for an AGN. For the above reason we constructed a grid of
models with metallicities Z = 0.5, 1.0, and 1.5Z$_{\odot}$ and nitrogen abundances higher by
a factor of 3. The models with $-$2.75 $<$ $\log$U $<$ $-$2.50 and solar or even supersolar
metallicity (Fig.~\ref{N2models}, dash-dot lines) reproduce in a reasonable way the nuclear and
circumnuclear regions within a radius of $\sim$ 0.5 kpc from the nucleus.
The comparison of these models with the observed [\ion{S}{2}] emission-line fluxes reveals
also a possible overabundance (by a factor of $\sim$ 2) of sulfur in the circumnuclear regions
of Tol1238-36.4. Similar cases of nitrogen and sulfur overabundances have been frequently observed
in Seyfert 2 galaxies \citep{sp90} and are probably a consequence of circumnuclear starburst activity.
However, our observed flux-ratios show also a good agreement with the two-component 
models by \citet{mrv02}, which combine the effects of shocks with AGN photoionization.
In particular, the strong [\ion{O}{3}] $\lambda4363$ and [\ion{O}{1}] $\lambda$6300 measured in the 
circumnuclear regions of Tol1238-364 are well reproduced by these models, 
whereas pure photoionization models predict too low values for these emission lines.
Therefore we cannot exclude as alternative explanation to our flux-ratios, the presence of 
shocks as an additional source of ionization.

Regions n1, n5, n1', n5', and C'' are not reproduced by any of the above models.
As we have already pointed out in \S~3.2, these regions are located in a transition area of the
diagnostic diagrams. Their properties can be explained with hybrid models involving
 different proportions of mixed thermal and non-thermal ionization, as explained in \citet{rhr98}. 
  
\placefigure{N2models}

\subsection{Star Formation and Energy Budget}

The H$\alpha$ line intensity of all emitting regions extracted from 
the spectra of Tol1238-364 and ESO381-G009 were converted into 
luminosities (Table~\ref{tab2} and \ref{tab3}) using the distances 47.9 Mpc and 44.3 Mpc, 
respectively, derived from the mean redshift values measured along 
the slits for each of the two galaxies.
These luminosities were used to calculate the corresponding star 
formation rates (except for the pure Seyfert-like regions) adopting 
the relation of \citet{hg86}:

\begin{equation}
\rm SFR = 7.07 \times 10^{-42} L_{H\alpha} ~~ M_{\odot}~ yr^{-1}
\end{equation}

The values shown in Tables~\ref{tab2} and \ref{tab3} for Tol1238-364 and ESO381-G009
range from about 0.001 to 0.15 M$_{\odot}$ yr$^{-1}$,
but much more significant are the SFR densities, which are 10 to 100 times 
higher than in normal spirals \citep{elm98}.
The mean value of the SFR density derived considering all \ion{H}{2}-like regions 
identified
along the slits up to a distance of 28$^{\prime\prime}$ from the nucleus is 
1.60 $\times$ 10$^{-7}$ M$_{\odot}$ yr$^{-1}$ pc$^{-2}$ in Tol1238-364.

The total flux measured on the calibrated H$\alpha$ image within a 
radius of 37$^{\prime\prime}$ is F(H$\alpha$+[\ion{N}{2}]) = 3.83$\times$10$^{-12}$ ergs s$^{-1}$
cm$^{-2}$. 
The percentage of [NII] emission detected by the
H$\alpha$ interference filter has been evaluated taking the average of the 
[NII]/H$\alpha$ ratios measured in the extranuclear spectra ($<$[NII]6548+6583/H$\alpha$$>$ = 0.48),
although in the central kpc the [NII] lines contribute to the flux in the H$\alpha$ image by
a higher percentage. 
Galactic and internal extinction corrections were applied.
For the last, the mean extinction value derived from the spectra, 
E(B-V) = 0.4, was used.
No attempt was made to correct for the contribution from the active nucleus. 
The resulting corrected H$\alpha$ luminosity is
L$_{H\alpha}$ = 1.44$\times$10$^{42}$ ergs s$^{-1}$,
which yields a total SFR = 10.22 M$_{\odot}$ yr$^{-1}$ and a SFR density of
4.41 $\times$ 10$^{-8}$ M$_{\odot}$ yr$^{-1}$ pc$^{-2}$.

In a similar way, mean SFR densities of 1.1 $\times$ 10$^{-7}$ and 1.62 
$\times$ 10$^{-8}$ 
M$_{\odot}$ yr$^{-1}$ pc$^{-2}$ were 
obtained from the spectra and the H$\alpha$ image of ESO381-G009.
The total luminosity and SFR evaluated from the image within a radius of 41$^{\prime\prime}$
are 5.58$\times$10$^{41}$ ergs s$^{-1}$ and 3.95 M$_{\odot}$ yr$^{-1}$, respectively.
In this case the average observed ratio 
[\ion{N}{2}]6548+6583/H$\alpha$ = 0.33 and the mean internal extinction value
E(B-V) = 0.59 were used to correct the flux F(H$\alpha$+[\ion{N}{2}]) = 1.0$\times$10$^{-12}$ 
ergs s$^{-1}$ cm$^{-2}$ measured in the image.
In both cases the SFR densities derived from the spectra are considerably larger 
than the corresponding value derived from the H$\alpha$ images.
This apparent discrepancy is a natural consequence of the orientation of the 
slit, preferentially located along alignments of knotty structures on the galaxies.
A comparison with the work of \citet{bu87} shows that both galaxies have SFR densities
approaching the upper value found for a sample of interacting spiral galaxies (6.1$\times$10$^{-11}$ 
$\lesssim$ SFRD $\lesssim$ 7.6$\times$10$^{-8}$ M$_{\odot}$ yr$^{-1}$ pc$^{-2}$) and clearly higher
than values found in a sample of isolated spirals (2$\times$10$^{-10}$ $\lesssim$ SFRD $\lesssim$ 
2$\times$10$^{-8}$ M$_{\odot}$ yr$^{-1}$ pc$^{-2}$).
Low values of \ion{H}{1} depletion time scales, $\log$(M$_{\rm HI}$)/SFR) = 8.5 - 9 yr, are also observed.
This is expected in case of interacting systems, according to the distribution shown in Figure~11
of \citet{bu87}.

\begin{figure}
\plottwo{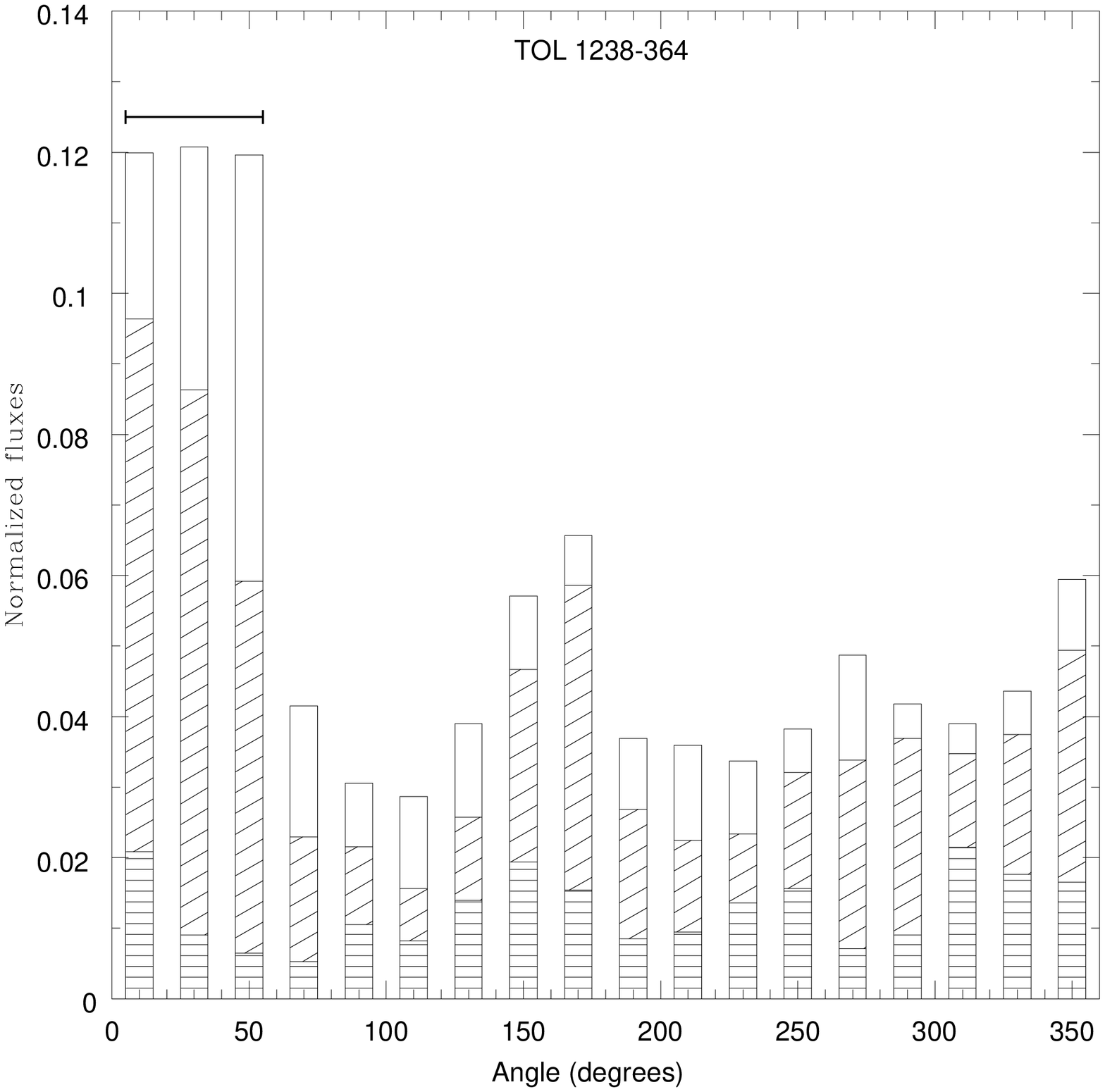}{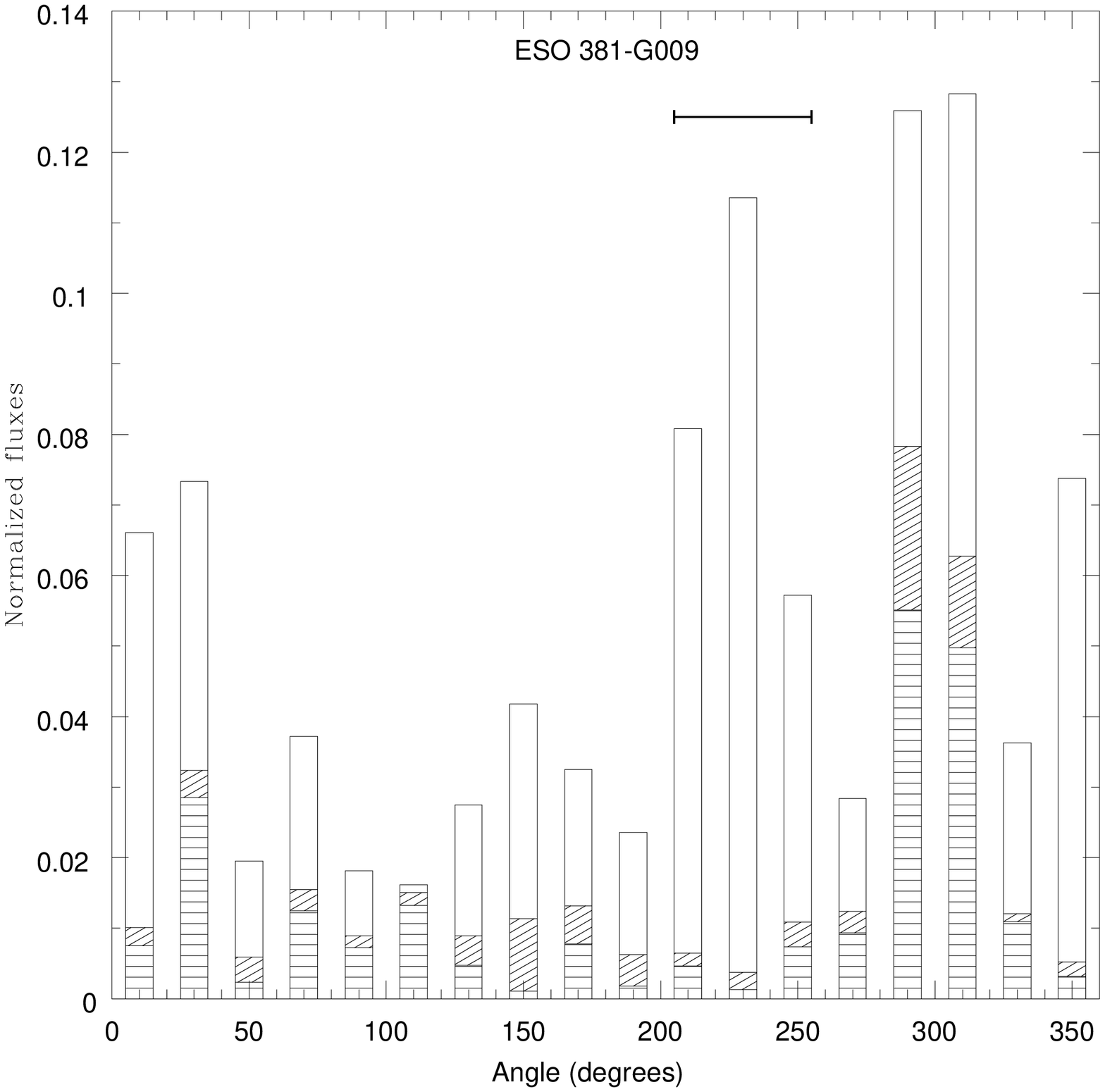}
\caption{Angular and radial distribution of H$\alpha$ fluxes normalized
to the total H$\alpha$ flux (excluding the nucleus) of Tol1238-364 (\emph{left})
and ESO381-G009 (\emph{right}). Fluxes are summed within circular sectors with aperture
20\degr. Horizontal dashed bars indicate fluxes within a radius of 2 kpc, inclined dashed bars
indicate fluxes between 2 and 4 kpc, and empty bars indicate fluxes at radii $>$ 4 kpc.
The horizontal bars indicate the range of angles in which each galaxy faces the companion. \label{Ha_histog}}
\end{figure}

As already noticed in \S~3.1 the H$\alpha$ emitting regions of Tol1238-364 appear more
concentrated in the spiral arm facing the companion galaxy. Also in ESO381-G009 a 
concentration of \ion{H}{2} regions seem to be present in direction of Tol1238-364 (Fig.~\ref{cmap}).
In order to better investigate the distribution of the emitting regions in the two galaxies,
we measured the H$\alpha$ fluxes in 18 circular sectors with aperture 20\degr\ and radii 2, 4, and  
$\sim$ 10 kpc, excluding the nuclei. In each galaxy, the fluxes were normalized to the total galaxy 
flux (excluding the nucleus). The angular distribution of the normalized fluxes is represented with
bar histograms in Fig.~\ref{Ha_histog}. Bars with horizontal dashes indicate fluxes within 2 kpc, 
bars with inclined dashes indicate fluxes between 2 and 4 kpc, and empty bars indicate fluxes at
radii $>$ 4kpc. The range of angles in which each galaxy faces the companion are marked with 
horizontal bars. The enhancement of H$\alpha$ flux at these positions is evident, especially at the 
outer radii for ESO381-G009 and both at intermediate and outer radii for the Seyfert galaxy.
A second peak in the angular flux distribution of ESO381-G009
is found around P.A. 300\degr\ in correspondence of the NW end of its bar. 

\placefigure{Ha_histog}

SFRs in the active galaxy have been calculated for all the
extranuclear regions and those circumnuclear regions whose diagnostic 
emission-line ratios suggest a mixture of thermal and non-thermal 
ionizing radiation. In the latter case the SFR values quoted in Table~\ref{tab2} 
are upper limits (marked with a colon), since the fractional contribution
of the active nucleus to the H$\alpha$ luminosity is unknown. 

However, some simple energy budget considerations could be done
in order to verify the capability of the Seyfert's non-thermal source 
to ionize the regions outside the nucleus. 
The number of ionizing photons necessary to produce the observed H$\alpha$
luminosity, 
\begin{equation}
{\rm Q}_{ion} = 7.3 \times 10^{11} \, {\rm L_{H\alpha}} \,\,\,{\rm
photons \,\, s^{-1}}
\end{equation}
\citep{ken98}, was evaluated in every region 
and compared (Table~\ref{tab2}) with the number of nuclear ionizing photons  which 
in principle could reach the considered region (Q$^{\prime}_{nuc}$).

Q$^{\prime}_{nuc}$ is actually a fraction of the total number
of ionizing photons produced per second by the central source (Q$_{nuc}$). 
In fact it is diluted by the covering factor $\Omega$/4$\pi$, which depends on the size and 
distance of each region to the nucleus 
($\Omega$ is the solid angle under which the considered region is seen 
from the nucleus). Obviously a precise measure of this  
factor cannot be achieved, since we see only the projected sizes and 
distances of the regions.

The value of Q$^{\prime}_{nuc}$ is given by
\begin{equation}
{\rm Q}^{\prime}_{nuc} = {\rm Q}_{nuc}\frac{\Omega}{4\pi} .
\end{equation}

Because of the impossibility to estimate the covering factor of the very 
central region n3, the value of Q$_{nuc}$ has been evaluated in  
indirect way by selecting among all the circumnuclear regions those 
exhibiting a clear Seyfert nature and similar line ratios
according to the diagnostic diagrams, namely n2, n4, n2', n3' and n4'.
The Q$_{ion}$ of such regions was multiplied by 4$\pi$/$\Omega$
and the average of the resulting values was assumed to be the actual 
number of nuclear ionizing photons 
\begin{equation}
{\rm Q}_{nuc} = <{\rm Q}_{ion}\frac{4\pi}{\Omega}>,
\end{equation}
included in Table~\ref{tab2} in correspondence of the central subregion n3.
This Q$_{nuc}$ was then diluted to obtain Q$^{\prime}_{nuc}$ of every
region.

\placetable{tab2}

In the region-by-region comparison of Q$_{ion}$ and Q$^{\prime}_{nuc}$ 
there was no need to take into account the filling factor, that is 
the fraction of the total volume occupied by the gas. This factor is believed 
to be very low, typically few 0.01 \citep{du90}, and is generally assumed to 
affect in the same measure both the Narrow Line Region and the 
extranuclear regions.

This comparison revealed that Q$_{ion}>$Q$^{\prime}_{nuc}$ for all the 
regions located at distances R$\gtrsim$1 kpc from n3  
indicating that their ionization is 
not dominated by the active nucleus. This result is in 
agreement with the diagnostic diagrams, which show a clear thermal 
nature of the ionizing source in these regions. 
Within the central kiloparsec Q$_{ion}$$\lesssim$Q$^{\prime}_{nuc}$,
except for the region n2.
In fact the nuclear ionization is clearly dominating 
in the regions n4, n2', n3' and n4', which lie inside 
the Seyfert area of the diagnostic diagrams. Instead it is likely mixed 
in different percentages to the ionization from thermal sources in the 
regions n5, n5' and C'', which lie in the Seyfert area, but slightly 
displaced toward the Liners, and in the regions n1' and n1, 
which occupy the Seyfert-Liner transition region and the Liner area, 
respectively. According to the diagnostic diagrams (Fig.~\ref{tol_vo}),
also the circumnuclear region n2 is dominated by the nuclear non-thermal
radiation, however it exhibits an opposite behavior
(Q$_{ion}>$Q$^{\prime}_{nuc}$) with respect to the other circumnuclear
regions. In particular its high observed Q$_{ion}$ would require a 
number of nuclear ionizing photons a factor ~3.4 higher than the estimated one.
One could argue that Q$_{nuc}$, as determined above, is still 
underestimated due to dust absorption and
that only in direction of n2 the real amount of nuclear ionizing
radiation can be estimated. In such a circumstance we would expect 
to observe a similarly high Q$_{ion}$ also in the immediately adjacent
n1 region. Since this is not the case, we rule out this hypothesis
and confirm instead the presence of a circumnuclear starburst inside 
n2 as already suggested in \S~3.2 on the basis of the analysis of its spectral properties. 
This idea is also supported by the estimated value of internal
extinction in n2, A$_V$ $\simeq$ 3.3 mag, much higher than 
in the surrounding zones.

The number of ionizing photons produced in every region was 
calculated also 
for ESO381-G009 (Table~\ref{tab3}), and  expressed in equivalent number of O5 
stars, N(O5),
assuming that each O5 star emits $\sim$ 5 $\times$ 10$^{49}$ ionizing photons 
s$^{-1}$ \citep{ost89}.
The obtained values range from the upper limit for normal \ion{H}{2} regions 
ionized by clusters or associations of OB stars to values found for 
``giant'' and ``supergiant'' \ion{H}{2} regions \citep{keh89}. 
The starbursts with the highest 
values of SFR are located in the nucleus of the galaxy.

\placetable{tab3}

\section{FIR EMISSION}

The connection between the enhanced far-infrared emission and galaxy 
interactions 
has been established in many works \citep[e.g.][]{twd88,boetal00}.
In order to study the infrared activity of each member
of the pair, position and flux calibrated 
raw IRAS data were extracted from the IRAS database server of the
Space Research Organisation of the Netherlands (SRON). The program 
GIPSY (Groningen Image Processing 
System, Assendorp et al. 1995), was used to create low resolution 
co-added maps at 12, 25, 60 and 100 $\mu m$ centered on Tol1238-364.

\begin{figure}
\plotone{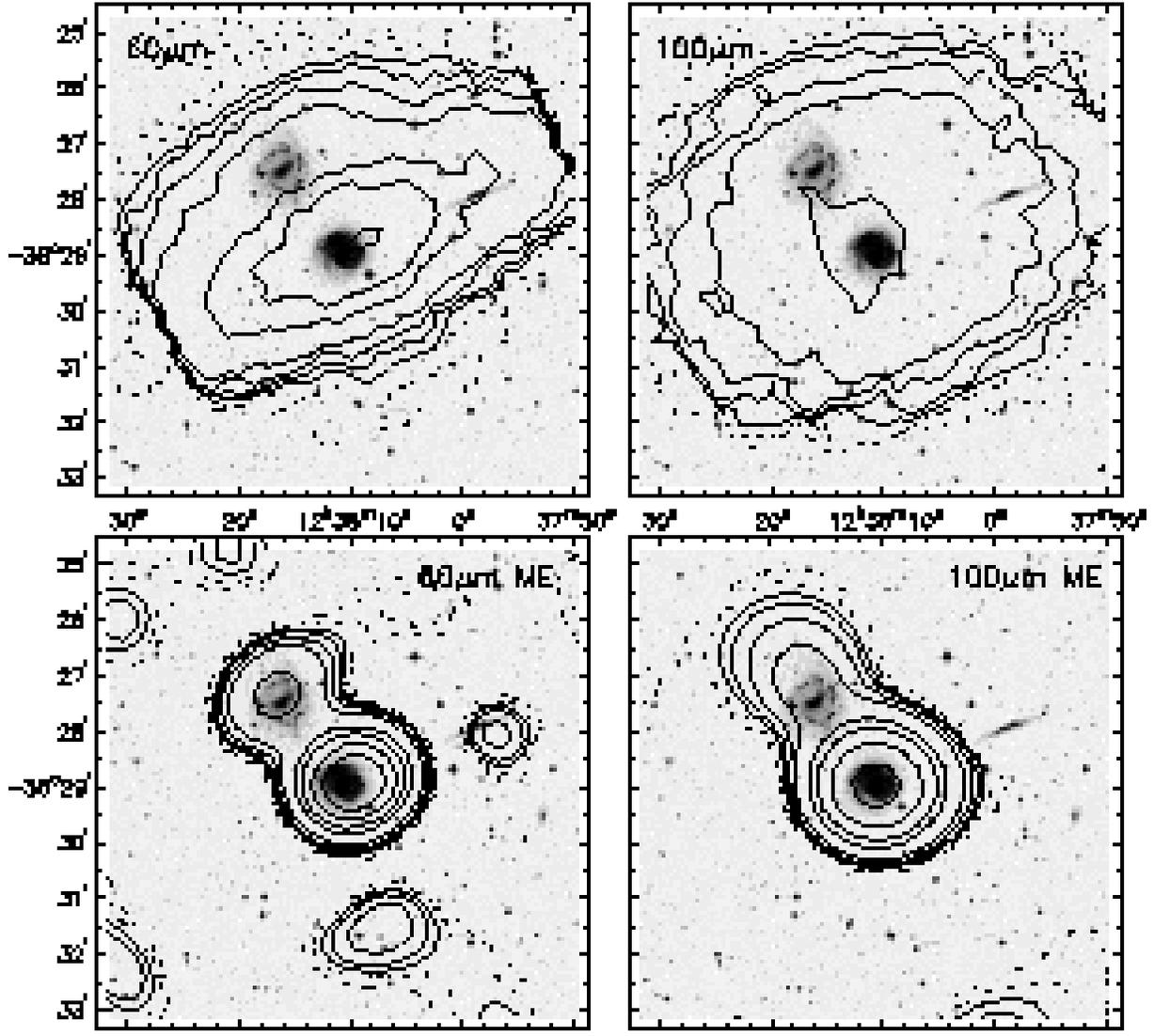}
\caption{IRAS co-added contour-maps of the
galaxy pair at 60 and 100 $\mu m$ (top) and the corresponding high 
resolution HIRAS contour-maps (bottom), overimposed to the 
Digitized Sky Survey optical image; 1 and 2$\sigma$ (dashed lines), 
and 3, 4, 6, 10, 30, 50, 100, 200$\sigma$
 (solid lines) contour levels are drawn. Right Ascension and Declination 
(1950.0) are displayed along {\emph x} and {\emph y} axes. \label{IRmap}}
\end{figure}

The overlap of these maps on the corresponding Digitized Sky Survey optical 
image allowed to show
the fainter emission of ESO381-G009 at 12 and 25 $\mu m$, compared with the 
active galaxy. At 60 and 100 $\mu m$ both galaxies 
seemed to be embedded in 
the same emission, because of the larger size of the detectors at 
those wavelengths \citep{asetal95}.
To remove the ``confusion'' and evaluate the contributions of the two objects,
a higher resolution (by a factor of $\sim$5) was achieved (Fig.~\ref{IRmap}) by 
applying the program HIRAS, which drives the MEMSYS5 maximum entropy 
imaging algorithm \citep{bkk94}. 

\placefigure{IRmap}

The result showed ESO381-G009 as an emitting source clearly separated 
from Tol1238-364 in all the four bands.
The sum of the flux densities of the two galaxies, measured by means of the 
GIPSY task FLUX (Table~\ref{tab4}), was compared with previously published values, for 
which only a unique emitting source was  
considered. In particular our results are
in good agreement both with the Point Source Catalog, apart from 
a slightly higher emission at 100 $\mu m$ in our measurement, and with the 
values of \citet{rms93} obtained using
the ADDSCAN procedure of the Infrared Processing and Analysis Center (IPAC). 
 
\begin{figure}
\plotone{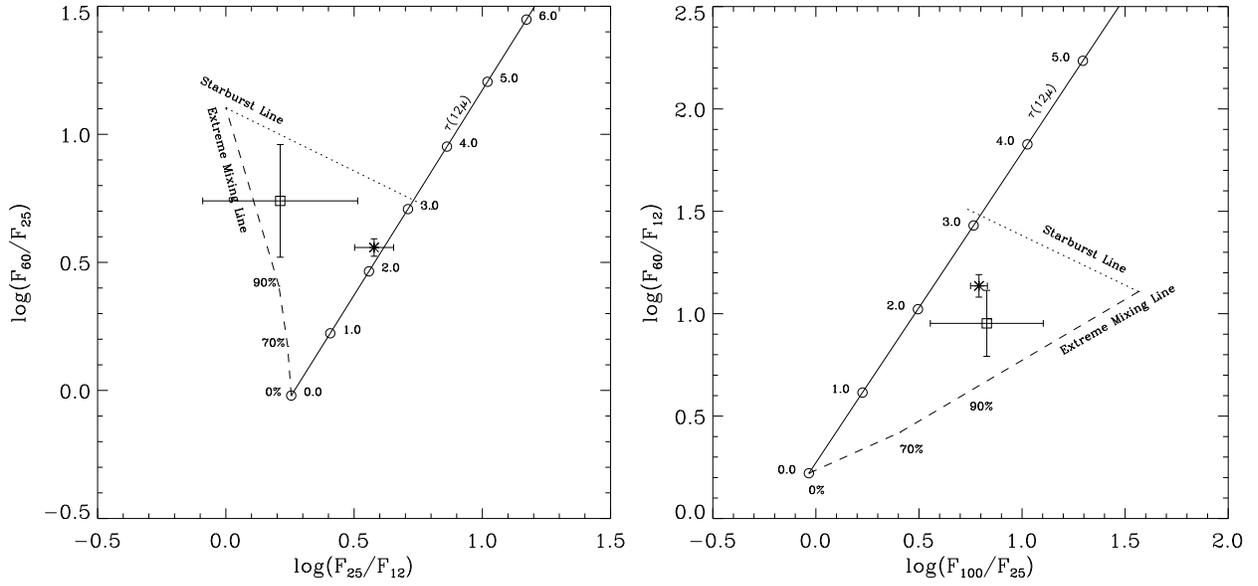}
\caption{Infrared color-color diagrams, similar 
to the ones in \citet{doetal98}, but with 
the two galaxies represented as distinct sources: ESO381-G009 (\emph{square}) 
clearly appears as a ''warm'' starburst, while Tol1238-364
(\emph{star}) is dominated by a moderately obscured AGN. \label{IRcol}}
\end{figure}

Our flux ratios were compared with the infrared color-color 
diagrams log(F$_{25}$/F$_{12}$) $vs$ log(F$_{60}$/F$_{25}$) and 
log(F$_{100}$/F$_{25}$) $vs$ log(F$_{60}$/F$_{12}$), 
which can indicate whether the FIR emission is dominated by a 
dust-extinguished 
active nucleus, a pure starburst or a mixture of these two components 
\citep{doetal98}.  We found that the FIR emission of
Tol1238-364 is dominated by a moderately obscured AGN, while
ESO381-G009 has an expected ``warm'' starburst nature (Fig.~\ref{IRcol}).
Further confirmations came from the analysis of 
the spectral indices $\alpha_{100,60}$ and $\alpha_{60,25}$, and mostly of the 
60$\mu m$ curvature ($\alpha_{60,25} - \alpha_{100,60}$),  
whose values $-$0.425 and $-$1.545 are typical of a Seyfert 2 and an \ion{H}{2} galaxy 
respectively \citep{mns85}.

\placefigure{IRcol}

The total fluxes between 42.5 and 122.5 $\mu m$ have been calculated following Helou et al. (1985):

\begin{equation}
\rm FIR = 1.26\times10^{-11}(2.58F_{60}+F_{100})  ~~~ erg~ cm^{-2}~ sec^{-1} 
\end{equation}

and converted into luminosities, using the distances given in \S~3.4.
The resulting values (Table~\ref{tab4}) indicate that Tol1238-364 is $\sim$ 10 times 
brighter than ESO381-G009, a clear effect
of the presence of an active nucleus.
In fact the dusty torus is expected to contribute notably to the far-IR 
luminosity \citep{smw92,geetal98,luetal98}, which 
is generally produced
by interstellar dust heated by the UV radiation field of young and hot stars.
Considering a single temperature component and a $\lambda^{-1}$ emissivity law
\citep{yoetal89}, a major content of warm dust was revealed in Tol1238-364
(Table~\ref{tab4}), likely concentrated 
into the bar and in correspondence of the giant \ion{H}{2} regions populating its 
spiral arms.
However, the star formation rate, derived from the FIR luminosities
following the relation given by \citet{huetal86}, is high in both
galaxies as expected to occur in an interacting system.
The overall SFR estimated from L$_{FIR}$ (SFR$_{L_{FIR}}$ $\sim$ 15.9 M$_{\odot}$ yr$^{-1}$)
appears somewhat higher than that estimated from L$_{H\alpha}$ (SFR$_{L_{H\alpha}}$ $\sim$ 10.2
M$_{\odot}$ yr$^{-1}$). 
In fact, the far infrared 
luminosity generally includes the contribution of the cirrus component, 
diffuse dust heated by the starlight radiation field, whose effect is to 
give an overestimate of the star formation rate, up to a factor 2
\citep{elm98}. Furthermore, in the case of the Seyfert galaxy the dust 
heated by the nuclear non-thermal radiation contributes to the total FIR 
emission in a percentage which cannot be determined.
On the opposite, for ESO381-G009, SFR$_{L_{H\alpha}}$ is a factor $\sim$ 2 lower than 
SFR$_{L_{FIR}}$. A possible reason could be the error related to the ``deblending''
of the two IRAS sources.

\placetable{tab4}

\section{DISCUSSION AND CONCLUSIONS}

We have analyzed the physical properties of the galaxy pair Tol1238-364 and
ESO381-G009, which belongs to a triple system together with ESO381-G006, 
on the basis of optical imaging and long-slit spectroscopy
in order to point out possible effects of interaction.

The triple system was analyzed by \citet{bw01} as well, who suggested that also
a fourth galaxy, ESO381-G014, might belong to this small galaxy group, because of its
detection in radio observations at a radial velocity of 3304 km s$^{-1}$.
For the supposed four-members group they reported a median radial velocity of 3285 km s$^{-1}$,
a median projected velocity dispersion of 11 km s$^{-1}$, a median projected galaxy-galaxy 
separation of 110 kpc, and a median crossing time of order 10 Gyr. 
Since they did not detect strong signs of galaxy interactions, like a common envelope of
neutral gas or prominent tidal tails, they speculated that the triple system might be
much looser than the projected density implies.
ESO381-G014 has been indicated as a member of the group also in the catalog 
of groups of nearby optical galaxies (NOGG)\footnote{In the NOGG catalog, which contains only galaxies
brighter than B = 14 mag, the group is erroneously indicated as a quartet, because the
galaxy ESO381-G009 is counted twice, with the alternative names PGC 042519 and PGC 097487.} 
compiled by \citet{giur00}. However, its projected separation from the triplet
(we assume as average coordinates of the triplet $\alpha$(J2000) = 12$^h$40$^m$50$^s$.7, $\delta$(J2000) = 
$-$36\degr44$^{\prime}$32$^{\prime\prime}$) is $\sim$ 42$^{\prime}$, corresponding to $\sim$ 0.5 Mpc,
assuming H$_0$ = 75 km s$^{-1}$ Mpc$^{-1}$), i.e. $\sim$33 times the diameter
of Tol1238-364 as estimated at the 25 mag arcsec$^{-2}$ B isophote. Instead, the centers of
the galaxies in the triplet are all encompassed by a minimum circle of radius 1$^{\prime}$.87
(25 kpc) and have a median projected separation of
2$^{\prime}$.63 (i.e. $\sim$ 35 kpc). While we have no pieces of evidence other than the concordant
radial velocity to support either exclude the membership of ESO381-G014 to the group, we consider it
unlikely for this galaxy to have a strong gravitational influence on the members of the triplet.

The radial velocity of the galaxies pose them at the same distance of the Centaurus cluster
(Abell 3526, $\alpha$(J2000.0) = 12$^h$48$^m$48$^s$.7, $\delta$(J2000.0) = 
-41\degr18$^{\prime}$44$^{\prime\prime}$). However, they are at a projected distance of $\approx$ 3.5 Mpc
from the cluster center and well separated from the main cluster galaxy concentration,
therefore it is unclear whether they can be considered cluster members. In any case, 
at such a separation the influence of the Centaurus cluster on the triplet, if any, is most
likely negligible with respect to the mutual influence of the galaxies within the triplet itself,
provided this is really a tight system.
We have actually observed properties in Tol1238-364 and ESO381-G009 (the tightest pair within
the triple system, at least in projection on the sky) that are consistent with an interaction
scenario, as we discuss in the following.

Both galaxies show a prominent bar, and ESO381-G009 also a ring of \ion{H}{2} regions. 
Theoretical N-body simulations show that close encounters between galaxies can lead to
bar formation \citep[e.g.][]{n87,gca90}, although bars can form in isolated galaxies, as well,
as a consequence of disk instabilities.
Indeed, studies of the relative frequency of bars among various Hubble types and in different environments 
have demonstrated that galaxies
in close binary systems and some groups have a greater occurrence of bars than field galaxies \citep{eeb90}.
This result is somewhat controversial, in fact it was recently found \citep{vb02} that the bar frequency
does not appear to depend on the galaxy environment. 
Nevertheless, we think that the presence of a bar in at least two out of three members of a galaxy system,
like in our case (since ESO381-G006 is edge-on we cannot establish whether it is barred or not), 
strongly favors the hypothesis that the mutual interaction has played a role in the bar formation.
Further support to this hypothesis is given by the detection of star formation along the bars,
which indicate that the bars are still relatively young and actively funneling gas toward the center
of the galaxies.

The two galaxies show evidences of only slight morphological distortions (see \S 3.1).
Specifically, we observe a twist of the inner isophotes, most likely caused by the bars, and
an elongation and change of the P.A. of the outermost isophotes of both galaxies toward one 
another. Additionally, the decentering degree of the most external isophotes with respect to the
luminosity center of Tol1238-364 is higher than the maximum value (5\%) found by \citet{mm99} for
isolated galaxies, indicating a significant asymmetry of the outskirts of the disk likely caused by
interaction processes. On the contrary, ESO381-G009 does not exhibit such an asymmetry of the outermost
isophotes, but its H$\alpha$ image shows an extension of \ion{H}{2} regions,
also detectable in the broad band images, on the southern side of the disk, outside the ring-like
structure, facing the Seyfert galaxy. 
Additionally, the optical images show a plume on its northern side. 
This plume appears more extended
in the contour maps of the \ion{H}{1} 21 cm line \citep{bpj00, bw01}. 
No other tidal features are visible in the optical images, but
there are signs of an \ion{H}{1} bridge between 
Tol1238-364 and ESO381-G006 \citep{bpj00}. However,
since the detection of this bridge is limited to one velocity channel,
the possibility that this is a spurious feature cannot be excluded.
Most of the neutral hydrogen remains located in the discs of the galaxies.

We note that the formation of prominent tidal tails is not ubiquitous in interacting galaxies,
but depends on the geometry of the encounters, being particularly favored in prograde interactions,
as shown by numerical simulations \citep[see, e.g., the seminal work of][]{tt72}.
In a recent work, Barton Gillespie, Geller, \& Kenyon (2003) identified within a sample of
close galaxy pairs a number of galaxies exhibiting triggered star formation but not long tidal tails.
The color profiles of these galaxies show blue dips in their centers, in analogy with the 
\br\ profile we obtained for the non-active galaxy ESO381-G009.

In fact, we could not identify a real bulge structure in this galaxy, but two knots with high SFR,
the brightest of which we adopted as nucleus. This kind of structure appears in agreement with 
the gas infall and centrally concentrated star formation predicted by the numerical simulations of 
\citet{mh96} for bulgeless galaxies in the early stages of interactions. 

Tol1238-364 hosts a typical Seyfert 2 nucleus, with a hidden BLR.  
The non-thermal ionization produced by the central power-law source is confined within the first
kiloparsec and no evidence of ionization cones is found along the three studied
directions. Considerations
on the energy budget (\S~3.4) reveal the presence of at least one circumnuclear starburst,
a hypothesis that finds support in the observed spectral features.
Besides the central starbursts, a large number of \ion{H}{2} 
regions are found in both galaxies with a major 
concentration in the zone of their minimal mutual distance, further supporting the
interaction scenario. In ESO381-G009 the \ion{H}{2} regions are mainly located 
along the bar and in the ring, while in Tol1238-364 they are diffused all over the disk.
The SFR and the density of SFR are strongly enhanced with respect to normal and/or isolated
spirals, and higher in Tol1238-364 than in its companion (see \S~3.4).
This enhancement could be a combined effect of the mutual interactions of the galaxies, 
as shown by the comparison between the star formation properties of interacting and isolated
galaxies \citep[see e.g.][]{bu87,ketal87,kvs92},
and of the perturbation of the whole galactic disk
by the stellar bar \citep{a99}, although the relation between the global 
star formation and the presence of a bar in a galaxy is still debated.
Also, numerical simulations demonstrated that both galaxy interactions \citep[e.g.][]{bh91,bh96,mh96} 
and bars or, in general, non-axisymmetric features \citep[as summarized e.g. in][]{com01} can produce
torques and gas radial inflows toward the center, inducing nuclear starbursts and/or fueling a
central AGN.  

The \ion{H}{2} regions of both galaxies,whose 
excitation is in general the effect of photoionization from hot, young stars
although in some cases the contribution of starburst
driven shocks cannot be excluded, exhibit subsolar metallicities.
However, the Seyfert nucleus shows higher (solar or even supersolar) metallicity, 
with traces of overabundance of nitrogen and sulphur, which could be related to 
the presence of circumnuclear starburst.

The IRAS infrared colors confirmed the dual starburst - AGN nature of 
Tol1238-364 and
the simple starburst nature of ESO381-G009, but the infrared emission of the 
Seyfert appears dominated by the active nucleus and shows higher FIR luminosity
and dust content.

Based on the result of their analysis, \citet{bgk03} suggested that blue central
colors, moderate EW(H$\alpha$), and small velocity separations in galaxy pairs,
are indicative of galaxies that have undergone a close pass and than moved apart,
while their triggered burst of star formation ages. Accordingly,
numerical simulations by \citet{mh94}, showed that
two interacting disk/halo galaxies remain relatively unperturbed and show no 
increase of star forming activity until they reach the perigalacticon.
Furthermore, an enhanced star formation activity is also induced in the disks of the galaxies before 
they reach the widest separation after their first encounter.
The observational properties we have outlined in the present work suggest that this
stage of the evolution, i.e. the phase subsequent to a first close passage, might be
the case for the galaxy pair under study.
The lack of really prominent morphological distortions might indicate that the galaxy separation
is actually larger than appearing in projection on the sky.

In conclusion, although some of the observed properties, taken individually, can be
found also in isolated galaxies, when considered all together they provide,
in our opinion, a significant indication of an undergoing interaction between two 
gas-rich galaxies. From our data we cannot establish which role is played by
ESO381-G006 in the interaction.
On the basis of a comparison with the findings in other interacting pairs \citep[e.g.][]{bgk03}
and the results of galaxy encounter simulations \citep{mh94,mh96}, we suggest that the 
galaxy pair has already undergone a first close passage, which determined the onset of
bursts of star formation both in the disks and in the central regions, caused moderate
morphological distortions, and, possibly, disk instabilities leading to bar formation.
Nevertheless, a clear connection of the Seyfert activity with the interaction cannot be
demonstrated and remains at a speculative level.

\acknowledgments

We are very grateful to	the anonymous referee for precious comments which highly improved
the presentation of the paper.
ST acknowledges support from the University of Padova through a grant
which has funded part of this work. JV acknowledges support provided
by the Estonian Science Foundation, grant 4702.

\appendix
\section{Measured Emission Line Fluxes of Tol1238-364 and ESO381-G009}

\placetable{tabA1}
\placetable{tabA2}
\placetable{tabA3}
\placetable{tabA4}
\placetable{tabA5}
\placetable{tabA6}

\clearpage

\clearpage

\clearpage

\begin{deluxetable}{ccccc}
\tablecaption{Galaxy Properties \label{literature}}
\tablecolumns{5}
\tablehead{
\colhead{Object} &\colhead{$\alpha$(J2000.0)} & \colhead{$\delta$(J2000.0)} &
\colhead{Radial Velocity\tablenotemark{a}} &\colhead{M$_B$\tablenotemark{b}} \\
 &( h m s)&\colhead{($\circ$ $\prime$ $\prime\prime$)} & \colhead{(km s$^{-1}$)}   & \colhead{(mag)} 
}
\startdata
Tol1238-364  & 12 40 52.9& $-$36 45 22 & 3282 & $-$20.2 \\
ESO 381-G009 & 12 40 58.4& $-$36 43 55 & 3288 & $-$19.3 \\
ESO 381-G006 & 12 40 40.8& $-$36 44 20 & 3101 & $-$17.2 \\
\enddata
\tablecomments{Units of right ascension are hours, minutes, and seconds, and units of declination are degrees,
arcminutes, and arcseconds.}
\tablenotetext{a}{Obtained from the \ion{H}{1} 21 cm line \citet{bw01}.}
\tablenotetext{b}{Converted from apparent magnitudes given in the NASA Extragalactic Database.}
\end{deluxetable}

\begin{deluxetable}{ccccccccc}
\tablecaption{Observation Summary \label{tab1}}
\rotate
\tablewidth{0pt}
\tabletypesize{\footnotesize}
\tablecolumns{9}
\tablehead{
\colhead{Object} &\colhead{Date} & \colhead{U.T.} &
\colhead{Exp. Time} &\colhead{P.A.} & \colhead{Spectral Range} & \colhead{Filter} & \colhead{$\mu_{lim}$\tablenotemark{a}} &
\colhead{m$_{lim}$\tablenotemark{b}}\\
 & &\colhead{(hh:mm)} & \colhead{(sec)}   && \colhead{(\AA)} & & \colhead{(mag arcsec$^{-2}$)} & \colhead{(mag)}}
\startdata
\multicolumn{9}{c}{Spectroscopy}\\
Tol1238-364         & 1995 Apr. 1& 09:23&900  & 150$^{\circ}$ & 3850 - 7950 & \nodata& \nodata&\nodata\\
'' & 1995 Apr. 1& 09:47&1200 &  '' & 3200 - 6030 &\nodata & \nodata & \nodata \\
''          &  1995 Apr. 3 & 05:55&900  & 90$^{\circ}$  & 3850 - 7950 & \nodata & \nodata &\nodata \\
''          & 1995 Apr. 3 & 06:47&900  & ''  & 3200 - 6030 & \nodata & \nodata &\nodata \\
''          & 1995 Apr.  4& 06:28&1800 &  146$^{\circ}$  & 3850 - 7950 & \nodata &\nodata  &\nodata \\
''          &  1995 Apr. 4& 07:04&2400 &  '' & 3200 - 6030 & \nodata &\nodata  &\nodata \\
\\
 ESO381-G009 &  1995 Apr. 3 & 06:23&900  & 90$^{\circ}$    & 3850 - 7950 & \nodata & \nodata &\nodata \\
''          & 1995 Apr. 3  & 07:59&900  & 125$^{\circ}$& 3850 - 7950 & \nodata & \nodata &\nodata \\
'' &1995 Apr. 3  & 08:16& 900  &   '' & 3200 - 6030 & \nodata & \nodata &\nodata \\
\multicolumn{9}{c}{Imaging}\\
         & 1995 Apr. 1 & 07:49&600 & \nodata& \nodata& B & 23.79 & 22.79 \\
         & 1995 Apr. 1 & 08:02&300 & \nodata& \nodata& V & 23.96 & 23.13 \\
        & 1995 Apr. 1 & 08:10& 300 &\nodata& \nodata& R & 24.11 & 23.12 \\
	    & 1995 Apr. 1 & 07:05&900 & \nodata& \nodata& H$\alpha$\tablenotemark{b} & 3.56$\times$10$^{-16}$ &
	    8.47$\times$10$^{-16}$\\
	    & 1995 Apr. 1 & 07:23&900 & \nodata& \nodata& H$\alpha$-cont. & \nodata & \nodata\\
\tablenotetext{a}{Limit surface brightnesses are evaluated at a 3$\sigma$ level above the background, and the limit
magnitudes are evaluated assuming a 3$\sigma$ signal within the PSF radius corresponding to 80\% of the total flux.}
\tablenotetext{b}{For the H$\alpha$ image we give the brightness and flux limit in units of ergs cm$^{-2}$
s$^{-1}$.}
\enddata
\end{deluxetable}

\begin{deluxetable}{ccccccc}
\tablecaption{Photometric Results \label{phot}}
\tablewidth{0pt}
\tablehead{\colhead{Object} & \colhead{B} & \colhead{V}& \colhead{R} & \colhead{$<r_e>$}& \colhead{$<h_D>$} &
\colhead{$<B/T>$} \\
 & \colhead{(mag)} &\colhead{(mag)} &\colhead{(mag)} & \colhead{(arcsec)} & \colhead{(arcsec)} & 
}
\startdata
Tol1238-364 & 12.92 & 12.45 & 11.98 & 2.2 & 7.5 & 0.2 \\
            & 12.69 & 12.28 & 11.84 & & & \\
ESO381-G009 & 13.71 & 13.22 & 12.74 & 3.0 & 12.5 & 0.12 \\
            & 13.48 & 13.05 & 12.60 & & & \\
\enddata
\tablecomments{For each galaxy, observed magnitudes are listed on the first row and 
magnitudes corrected for Galactic extinction on the second row. Extinction values
in V and R were derived from A$_B$ \citet{bh82} following \citet{ccm89} and assuming a visual selective 
extinction R$_V$ = 3.1.}
\end{deluxetable}

\clearpage

\begin{deluxetable}{cccccccc}
\tablecaption{Tol1238-364: Ionizing Photons and SFR \label{tab2}}
\tablewidth{0pt}
\tabletypesize{\footnotesize}
\tablehead{
\colhead{Id.} & \colhead{R\tablenotemark{a}} & \colhead{L(H$\alpha$)} & 
\colhead{Q$_{ion}$} & \colhead{Q$^{\prime}_{nuc}$} & EW(H$\alpha$)&
\colhead{SFR} & \colhead{SFR/pc$^2$}\\
& \colhead{(kpc)} & \colhead{($\times$10$^{40}$ erg s$^{-1}$)}&
 \colhead{(phot. s$^{-1}$)} & \colhead{(phot. s$^{-1}$)} & (\AA) &
\colhead{(M$_{\odot}$ yr$^{-1}$)} & 
\colhead{(M$_{\odot}$ yr$^{-1}$ pc$^{-2}$)}
}
\startdata
\multicolumn{8}{c}{P.A. = 150$^{\circ}$}\\
A2 & $-$3.90 & 1.57 & 1.15E+52 & 2.20E+51 & 94.93 & 0.111 & 7.47E$-$8\\*
A1 & $-$1.60 & 2.31 & 1.69E+52 & 5.74E+51 & 31.31 & 0.164 & 2.50E$-$7\\*
n1 & $-$0.70 & 0.68 & 4.98E+51 & 8.28E+51 & 9.39  & 0.048: & 2.65E$-$7:\\*
n2 & $-$0.30 &21.21 & 1.55E+53 & 4.53E+52 & 53.96 & \nodata & \nodata \\*
n3 & 0.0     &19.45 & 1.42E+53 & {\bf 2.82E+53} & 106.62& \nodata & \nodata \\*
n4 & 0.26    & 0.89 & 6.49E+51 & 3.61E+52 & 50.57 & \nodata & \nodata \\*
n5 & 0.51    & 0.22 & 1.59E+51 & 1.22E+52 & 12.77 & 0.015: & 1.06E$-$7:\\*
B1 & 0.93    & 1.39 & 1.01E+52 & 6.59E+51 & 16.70 & 0.098 & 3.86E$-$7\\*
B2 & 1.83    & 1.45 & 1.06E+52 & 3.91E+51 & 46.95 & 0.103 & 1.77E$-$7\\*
B3 & 3.43    & 1.14 & 8.32E+51 & 1.75E+51 & 87.16 & 0.081 & 8.87E$-$8\\
\multicolumn{8}{c}{P.A. = 146$^{\circ}$}\\
A2'& $-$3.71 & 0.92 & 6.73E+51 & 2.48E+51 & 36.28 & 0.065 & 4.27E$-$8\\*
A1'& $-$1.53 & 1.35 & 9.85E+51 & 4.84E+51 & 43.68 & 0.095 & 1.88E$-$7\\*
n1'& $-$0.81 & 0.97 & 7.06E+51 & 7.40E+51 &  9.42 & 0.068: & 3.13E$-$7:\\*
n2'& $-$0.39 & 2.43 & 1.77E+52 & 2.68E+52 & 25.38 & \nodata & \nodata \\*
n3'& 0.26    & 2.69 & 1.96E+52 & 3.61E+52 & 49.09 & \nodata & \nodata \\*
n4'& 0.37    & 0.64 & 4.69E+51 & 2.36E+52 & 17.83 & \nodata & \nodata \\*
n5'& 0.67    & 0.77 & 5.61E+51 & 9.04E+51 &  9.86 & 0.054: & 2.99E$-$7:\\*
B1'& 1.74    & 2.05 & 1.50E+52 & 6.19E+51 & 35.73 & 0.145 & 1.73E$-$7\\*
B2'& 4.13    & 0.86 & 6.27E+51 & 1.86E+51 & 98.28 & 0.061 & 4.29E$-$8\\
\multicolumn{8}{c}{P.A. = 90$^{\circ}$}\\
A3''& $-$5.10 & 0.38 & 2.76E+51 & 1.07E+51 & 64.92 & 0.027 & 2.16E$-$8\\*
A2''& $-$2.95 & 1.61 & 1.17E+52 & 2.06E+51  & 94.16 & 0.114 & 1.42E$-$7\\*
A1''& $-$1.37 & 2.19 & 1.60E+52 & 9.12E+51 & 57.72 & 0.155 & 2.03E$-$7\\*
C'' & 0.66    & 0.83 & 6.05E+51 & 1.88E+52 & 11.71 & 0.059: & 1.61E$-$7:\\*
B1''& 0.97    & 2.15 & 1.57E+52 & 7.80E+51 & 24.21 & 0.152 & 4.65E$-$7\\*
B2''& 1.83    & 0.76 & 5.58E+51 & 3.88E+51 & 27.01 & 0.054 & 9.30E$-$8\\*
B3''& 3.90    & 1.07 & 7.80E+51 & 2.11E+51 & 28.00 & 0.076 & 5.33E$-$8\\
\enddata
\tablenotetext{a}{Distance in kpc to the center of the galaxy. Negative distances
indicate regions West or North-West of the nucleus.}
\tablecomments{Q$^{\prime}_{nuc}$ values are calculated taking as 
reference the value given in bold-face in correspondence of the
central region n3, and obtained as explained in the text.
SFR and SFRD values marked with a colon are upper limits.}
\end{deluxetable}

\clearpage

\begin{deluxetable}{cccccccc}
\tablecaption{ESO381-G009: Ionizing Photons and SFR \label{tab3}}
\tabletypesize{\footnotesize}
\tablewidth{0pt}
\tablehead{
\colhead{Id.} & \colhead{R\tablenotemark{a}} & \colhead{L(H$\alpha$)} & 
\colhead{Q$_{ion}$} & \colhead{N(O5)} & EW(H$\alpha$) &\colhead{SFR} & 
\colhead{SFR/pc$^2$}\\
 &\colhead{(kpc)} & \colhead{($\times$10$^{40}$ erg s$^{-1}$)}&
 \colhead{(phot. s$^{-1}$)} & & (\AA) & \colhead{(M$_{\odot}$ yr$^{-1}$)} & 
\colhead{(M$_{\odot}$ yr$^{-1}$ pc$^{-2}$)} 
}
\startdata
\multicolumn{8}{c}{P.A. = 90$^{\circ}$}\\
A2   & $-$3.10 & 0.23 & 1.65E+51 & 33  & 28.41 & 0.016 & 9.36E-9\\*
A1   & $-$1.94 & 1.43 & 1.04E+52 & 208 & 15.47 & 0.101 & 1.01E-7\\*
N    & 0.0     & 5.52 & 4.03E+52 & 805 & 76.07 & 0.390 & 4.49E-7\\*
B1   & 1.74    & 0.37 & 2.72E+51 & 54  & 32.60 & 0.026 & 1.76E-8\\
\multicolumn{8}{c}{P.A. = 125$^{\circ}$}\\
A3'  & $-$5.93 & 0.44 & 3.24E+51 & 65  & 82.61 & 0.031 & 2.77E-8\\*
A2'  & $-$3.40 & 0.40 & 2.91E+51 & 58  & 45.92 & 0.028 & 3.60E-8\\*
A1'  & $-$1.89 & 1.28 & 9.35E+51 & 187 & 58.24 & 0.091 & 1.09E-7\\*
N'   & 0.0     & 3.53 & 2.58E+52 & 516 & 66.08 & 0.250 & 2.68E-7\\*
B1'  & 19.35   & 0.09 & 6.80E+50 & 14  & 16.83 & 0.007 & 2.59E-9\\
\enddata
\tablenotetext{a}{Distance in kpc to the center of the galaxy. Negative distances
indicate regions West of the nucleus.}
\end{deluxetable}
  
\clearpage

\begin{deluxetable}{ccc}
\tablecaption{FIR Data and Dust Content \label{tab4}}
\tablewidth{0pt}
\tablehead{
\colhead{Quantity} & \colhead{Tol1238-364} & \colhead{ESO381-G009}}
\startdata
F$_{12}$ (Jy) &  0.580 $\pm$ 0.065  & 0.125 $\pm$ 0.035 \\
F$_{25}$ (Jy) &  2.198 $\pm$ 0.140  & 0.204 $\pm$ 0.085 \\
F$_{60}$ (Jy) &  7.942 $\pm$ 0.130  & 1.122 $\pm$ 0.102 \\
F$_{100}$(Jy) & 13.591 $\pm$ 0.431  & 1.378 $\pm$ 0.296 \\
&&\\
log(F$_{25}$/F$_{12}$)  & 0.578 $\pm$ 0.076 & 0.212 $\pm$ 0.302 \\
log(F$_{60}$/F$_{25}$)  & 0.558 $\pm$ 0.034 & 0.740 $\pm$ 0.220 \\
log(F$_{100}$/F$_{25}$) & 0.791 $\pm$ 0.041 & 0.829 $\pm$ 0.274 \\
log(F$_{60}$/F$_{12}$)  & 1.136 $\pm$ 0.055 & 0.953 $\pm$ 0.161 \\
&&\\
$\alpha_{60,25}$  & -1.476 $\pm$ 0.090 & -1.947 $\pm$ 0.580 \\
$\alpha_{100,60}$ & -1.051 $\pm$ 0.094 & -0.402 $\pm$ 0.598 \\
&&\\
T$_{\rm dust}$ (K)           &       39.5         &    45.1 \\
M$_{\rm dust}$ (M$_{\odot}$)  & 5.5$\times$10$^6$  & 3.0$\times$10$^5$ \\
&&\\
L$_{\rm FIR}$ (L$_{\odot}$) & 3.04$\times$10$^{10}$ & 3.24$\times$10$^{9}$ \\
SFR (M$_{\odot}$ yr$^{-1}$)    & 15.86                 & 1.69     \\
\enddata
\end{deluxetable}

\clearpage

\begin{deluxetable}{lccccccc}
\tabletypesize{\scriptsize}
\tablewidth{0pt}
\tablenum{A1}
\tablecaption{Tol1238-364 - P.A. = 150$^{\circ}$: Flux Ratios \label{tabA1}}
\tablehead{
\colhead{Line} & \colhead{$\lambda$}& \colhead{A1}& \colhead{A2}&
\colhead{N}& \colhead{B1}& \colhead{B2}& \colhead{B3}\\
 & & \colhead{$-$9.7 -- $-$3.7} &  \colhead{$-$23.5 -- $-$9.7} &
\colhead{$-$3.7 -- 3.0} & \colhead{3.0 -- 5.4} & \colhead{5.4 --
10.8} & \colhead{10.8 -- 19.2}}
\startdata
$[$OII$]$	& 3727 & 3.05  $\pm$ 0.98&\nodata & 0.55  $\pm$ 0.10& 2.35  $\pm$ 0.96& 2.00  $\pm$ 0.70& 4.14  $\pm$ 1.12\\            
                &      &  4.86 &  \nodata&  0.84 &  4.50 &  2.69 & 5.22 \\                              
$[$NeIII$]$ 	& 3869 & \nodata &  \nodata & 0.39  $\pm$ 0.14& 2.16  $\pm$ 1.36& \nodata & \nodata\\                                       
                & &  \nodata& \nodata& 0.56 & 3.81 & \nodata&  \nodata\\                                    
H$\gamma$ 	& 4340 & \nodata & \nodata & 0.12  $\pm$ 0.02& \nodata &  \nodata & \nodata  \\                                                
                & & \nodata& \nodata& 0.15 & \nodata& \nodata&  \nodata\\                                     
$[$OIII$]$ 	& 4363 & \nodata &  \nodata & 0.11  $\pm$ 0.04&\nodata & \nodata& \nodata \\                                                
                & &  \nodata& \nodata& 0.14 &  \nodata&  \nodata& \nodata\\                                     
HeII          	& 4686 &  \nodata &  \nodata & 0.13  $\pm$ 0.03&  \nodata  &  \nodata &  \nodata \\                                                
                & &  \nodata&  \nodata& 0.14& \nodata&  \nodata&  \nodata\\                                     
$[$OIII$]$ 	& 4959 &  \nodata & 0.24  $\pm$ 0.06& 2.41  $\pm$ 0.07& 0.39  $\pm$ 0.23& 0.22  $\pm$ 0.09& 0.38  $\pm$ 0.11\\            
                & &  \nodata& 0.24 &  2.34 &  0.37 &  0.22 &  0.37 \\                                 
$[$OIII$]$ 	& 5007 & 0.71  $\pm$ 0.20& 0.81  $\pm$ 0.11& 7.38  $\pm$ 0.15& 1.55  $\pm$ 0.45& 0.51  $\pm$ 0.12& 0.95  $\pm$ 0.15\\   
                & &  0.67&  0.79 &  7.03& 1.44 & 0.49 &  0.93 \\                                
$[$FeVII$]$     & 5158 & \nodata & \nodata & 0.08  $\pm$ 0.03& \nodata & \nodata&  \nodata \\                                                
                & & \nodata& \nodata& 0.07 & \nodata&  \nodata& \nodata\\                                     
$[$FeVI$]$      & 5177 & \nodata & \nodata & 0.11  $\pm$ 0.03&  \nodata &  \nodata & \nodata \\                                                
                & &  \nodata&  \nodata&  0.10 &\nodata& \nodata& \nodata\\                                     
$[$NI$]$ 	& 5199 & \nodata & \nodata & 0.18  $\pm$ 0.03& \nodata &  \nodata & \nodata \\                                                
                & &  \nodata&  \nodata& 0.16 & \nodata&\nodata& \nodata\\                                     
$[$CaV$]$       & 5309 & \nodata & \nodata& 0.13  $\pm$ 0.03& \nodata  & \nodata & \nodata \\                                                
                & & \nodata& \nodata& 0.11& \nodata& \nodata& \nodata\\                                     
HeI          	& 5876 & 0.16  $\pm$ 0.07& 0.16  $\pm$ 0.04& 0.18  $\pm$ 0.02& 0.37  $\pm$ 0.16& 0.09  $\pm$ 0.04& 0.38  $\pm$ 0.08\\   
                & & 0.11 & 0.13&  0.13 &  0.22 &  0.07 &  0.32 \\                                
$[$OI$]$ 	& 6300 & 0.17  $\pm$ 0.06& \nodata & 0.81  $\pm$ 0.03&  \nodata & 0.12  $\pm$ 0.07&\nodata  \\                              
                & &0.10 & \nodata& 0.52 & \nodata& 0.09& \nodata\\                                   
$[$OI$]$ 	& 6364 & \nodata &  \nodata & 0.32  $\pm$ 0.03& \nodata &  \nodata & \nodata \\                                                
                & & \nodata& \nodata&  0.20 & \nodata& \nodata& \nodata\\                                     
$[$NII$]$ 	& 6548 & 0.55  $\pm$ 0.12& 0.27  $\pm$ 0.05& 1.27  $\pm$ 0.04& 1.43  $\pm$ 0.29& 0.30  $\pm$ 0.06& 0.30  $\pm$ 0.07\\   
                & & 0.32 &  0.20 &  0.77 &  0.66 & 0.21 &  0.23 \\                                
 H$\alpha$ 	& 6563 & 4.97  $\pm$ 0.70& 3.89  $\pm$ 0.27& 4.72  $\pm$ 0.09& 6.20  $\pm$ 0.93& 4.06  $\pm$ 0.37& 3.76  $\pm$ 0.34\\   
                & & 2.85&  2.85 &  2.85&  2.85 &  2.85 &  2.85 \\                                
$[$NII$]$ 	& 6583 & 1.90  $\pm$ 0.30& 1.14  $\pm$ 0.10& 4.16  $\pm$ 0.08& 2.98  $\pm$ 0.51& 1.45  $\pm$ 0.15& 1.18  $\pm$ 0.14\\   
                & &  1.08 &  0.83 &  2.50&  1.36 & 1.01 &  0.89 \\                                
$[$SII$]$ 	& 6717 & 1.06  $\pm$ 0.19& 0.61  $\pm$ 0.07& 1.17  $\pm$ 0.04& 1.57  $\pm$ 0.36& 0.85  $\pm$ 0.10& 0.67  $\pm$ 0.10\\   
                & &  0.59 & 0.44 & 0.68 & 0.69 &  0.58&   0.50 \\                                
$[$SII$]$ 	& 6724 & 1.73  $\pm$ 0.33& 1.01  $\pm$ 0.12& 2.48  $\pm$ 0.07& 2.61  $\pm$ 0.63& 1.42  $\pm$ 0.18& 1.01  $\pm$ 0.16\\   
                & &  0.95 &  0.72 & 1.45&  1.14 &  0.97&  0.75 \\                                
$[$SII$]$ 	& 6731 & 0.67  $\pm$ 0.14& 0.40  $\pm$ 0.06& 1.31  $\pm$ 0.04& 1.04  $\pm$ 0.28& 0.57  $\pm$ 0.08& 0.35  $\pm$ 0.06\\   
                & & 0.37 & 0.29 &  0.76 & 0.45&  0.39 & 0.26\\                                
\multicolumn{2}{l}{Hbeta 4861 = 1.00} \\                                                                 
\multicolumn{2}{l} {F(H$\alpha$) ($10^{-15}$ erg/s/cm$^2$)}  &                                            
27.2 $\pm$ 0.3& 30.4 $\pm$ 0.3& 238.8 $\pm$ 0.0& 10.4 $\pm$ 0.1& 25.7 $\pm$ 0.3& 23.6 $\pm$ 0.2\\        
&&  84.0& 57.2&  664.8&  50.4& 52.8&  41.4\\                             
\multicolumn{2}{l} {F(H$\beta$) ($10^{-15}$ erg/s/cm$^2$)}  &                                             
5.5 $\pm$ 0.7& 7.8 $\pm$ 0.5& 50.6 $\pm$ 1.0& 1.7 $\pm$ 0.2& 6.3 $\pm$ 0.5& 6.3 $\pm$ 0.5\\              
&&  29.5& 20.1& 233.2& 17.7& 18.5&  14.5\\                             
\multicolumn{2}{l}{c\tablenotemark{a}}&  0.73 &  0.41 &  0.66 &  1.02 &  0.47 &  0.36 \\                                  
\multicolumn{2}{l}{E(B-V)}&  0.51 &  0.28 &  0.46 &  0.71 &  0.32 &  0.25 \\   
\enddata
\tablenotetext{a}{Extinction coefficient}                          
\tablecomments{
In this and the following tables, bold face numbers
indicate flux ratios and absolute fluxes corrected for internal
reddening. Below the label of every region, the distance in arcsec of
the region's extremes to the intensity peak along the slit are
indicated; positive distances are in East direction.}
\end{deluxetable}
\clearpage

\begin{deluxetable}{lcccccc}
\tabletypesize{\scriptsize}
\tablewidth{0pt}
\tablenum{A2}
\tablecaption{Tol1238-364 - P.A. = 146$^{\circ}$: Flux Ratios \label{tabA2}}
\tablehead{
\colhead{Line} & \colhead{$\lambda$}& \colhead{A1'}& \colhead{A2'}&
\colhead{N'}& \colhead{B1'}& \colhead{B2'} \\
 & & \colhead{$-$8.7 -- $-$4.0}&  \colhead{$-$22.8 -- $-$8.7}&  \colhead{$-$4.0
-- 3.7}&  \colhead{3.7 -- 11.4}&  \colhead{11.4 -- 24.5}}
\startdata
$[$OII$]$ 	& 3727 & 3.74  $\pm$ 0.30& 4.14  $\pm$ 0.87& 1.61  $\pm$ 0.16& 2.74  $\pm$ 0.36& 3.62  $\pm$ 0.40\\ 
    & & 4.89 &  5.92 &  2.58 & 3.75 &  4.03 \\ 
$[$NeIII$]$ 	& 3869 & \nodata&  \nodata & 0.82  $\pm$ 0.26&  \nodata & \nodata  \\ 
   & & \nodata& \nodata& 1.24&  \nodata&  \nodata\\ 
H$\gamma$ 	& 4340 &  \nodata &  \nodata &  \nodata & \nodata & 0.48  $\pm$ 0.09\\ 
   & & \nodata& \nodata& \nodata&  \nodata&  0.51 \\ 
$[$OIII$]$ 	& 4363 & \nodata & \nodata & 0.17  $\pm$ 0.07& \nodata & \nodata\\ 
   & &  \nodata&  \nodata&  0.21 & \nodata&  \nodata\\ 
HeII 	& 4686 & \nodata &\nodata & \nodata &\nodata & \nodata \\ 
   & &  \nodata& \nodata& \nodata& \nodata& \nodata\\ 
$[$OIII$]$ 	& 4959 & 0.22  $\pm$ 0.04& 0.37  $\pm$ 0.12& 2.27  $\pm$ 0.14& 0.23  $\pm$ 0.06& 0.30  $\pm$ 0.04\\ 
   & & 0.22&  0.36 &  2.20 &  0.22 &  0.30 \\ 
$[$OIII$]$ 	& 5007 & 0.69  $\pm$ 0.06& 1.10  $\pm$ 0.19& 7.19  $\pm$ 0.36& 0.64  $\pm$ 0.08& 0.87  $\pm$ 0.06\\ 
  & &  0.67 &  1.06 &6.81 &  0.62 & 0.86 \\ 
$[$NI$]$ 	& 5199 & \nodata & \nodata & 0.15  $\pm$ 0.05& \nodata & \nodata\\ 
   & & \nodata& \nodata& 0.13 & \nodata& \nodata\\ 
HeI 	& 5876 & 0.16  $\pm$ 0.03& \nodata & 0.25  $\pm$ 0.04& 0.15  $\pm$ 0.03& 0.12  $\pm$ 0.03\\ 
   & &  0.13 & \nodata&  0.17 &  0.12 &  0.11\\ 
$[$OI$]$ 	& 6300 & 0.10  $\pm$ 0.03& 0.21  $\pm$ 0.06& 0.88  $\pm$ 0.07& 0.18  $\pm$ 0.04& 0.13  $\pm$ 0.04\\ 
   & &  0.08 & 0.14 &  0.54 &  0.13 & 0.12 \\ 
$[$OI$]$ 	& 6364 & \nodata & \nodata & 0.30  $\pm$ 0.05&  \nodata & \nodata \\ 
  & &  \nodata&  \nodata& 0.18 & \nodata& \nodata\\ 
$[$NII$]$ 	& 6548 & 0.48  $\pm$ 0.04& 0.40  $\pm$ 0.10& 1.51  $\pm$ 0.09& 0.53  $\pm$ 0.06& 0.37  $\pm$ 0.04\\ 
   & &  0.35&  0.26 & 0.86 & 0.36&  0.33 \\ 
H$\alpha$ 	& 6563 & 3.93  $\pm$ 0.16& 4.37  $\pm$ 0.44& 5.00  $\pm$ 0.25& 4.15  $\pm$ 0.25& 3.24  $\pm$ 0.13\\ 
   & & 2.85&  2.85 &  2.85 & 2.85 & 2.85 \\ 
$[$NII$]$ 	& 6583 & 1.43  $\pm$ 0.07& 1.60  $\pm$ 0.19& 4.52  $\pm$ 0.23& 1.51  $\pm$ 0.11& 0.95  $\pm$ 0.05\\ 
   & &  1.03 &  1.04 & 2.56&  1.03 &  0.83 \\ 
$[$SII$]$ 	& 6717 & 0.56  $\pm$ 0.05& 1.09  $\pm$ 0.13& 1.48  $\pm$ 0.09& 0.85  $\pm$ 0.07& 0.61  $\pm$ 0.05\\ 
   & &  0.40 &  0.69 &  0.81 &  0.57 &  0.53 \\ 
$[$SII$]$ 	& 6724 & 1.05  $\pm$ 0.09& 1.76  $\pm$ 0.23& 2.97  $\pm$ 0.18& 1.41  $\pm$ 0.11& 0.97  $\pm$ 0.09\\ 
  & &  0.74 & 1.11&  1.63 &  0.94 & 0.85\\ 
$[$SII$]$ 	& 6731 & 0.49  $\pm$ 0.05& 0.68  $\pm$ 0.10& 1.49  $\pm$ 0.09& 0.56  $\pm$ 0.05& 0.37  $\pm$ 0.04\\ 
  & &  0.35 &  0.43 & 0.82 & 0.37&  0.32\\ 
\multicolumn{2}{l}{Hbeta 4861 = 1.00} \\ 
\multicolumn{2}{l} {F(H$\alpha$) ($10^{-15}$ erg/s/cm$^2$)} & 
25.5 $\pm$ 0.3& 14.1 $\pm$ 0.3& 89.9 $\pm$ 0.9& 34.7 $\pm$ 0.3& 24.1 $\pm$ 0.2\\ 
&& 49.0&  33.5&  281.2&  74.4& 31.2\\ 
\multicolumn{2}{l} {F(H$\beta$) ($10^{-15}$ erg/s/cm$^2$)} &
6.5 $\pm$ 0.2& 3.2 $\pm$ 0.3& 18.0 $\pm$ 0.7& 8.4 $\pm$ 0.4& 7.4 $\pm$ 0.2\\ 
&& 17.2&  11.7& 98.7&  26.1& 11.0\\ 
\multicolumn{2}{l}{c\tablenotemark{a}}&  0.42 &  0.56 &  0.74 &  0.49 &  0.17 \\ 
\multicolumn{2}{l}{E(B$-$V)}&  0.29 &  0.39 &  0.51 &  0.34 &  0.12 \\ 
\enddata
\tablenotetext{a}{Extinction coefficient.}
\end{deluxetable}

\clearpage
\begin{deluxetable}{lcccccccc}
\tabletypesize{\scriptsize}
\tablewidth{0pt}
\tablenum{A3}
\tablecaption{Tol1238-364 - P.A. = 90$^{\circ}$: Flux Ratios \label{tabA3}}
\tablehead{
\colhead{Line} & \colhead{$\lambda$}& \colhead{A1''}& \colhead{A2''}&
\colhead{A3''}& \colhead{C''}& \colhead{B1''}& \colhead{B2''}&
\colhead{B3''\tablenotemark{a}} \\
 & & \colhead{$-$9.4 -- $-$2.3}& \colhead{$-$16.8 -- $-$9.4}& \colhead{$-$28.2 -- 
$-$16.8}& \colhead{$-$2.3 -- 1.1}& \colhead{1.1 -- 4.1}&
\colhead{4.1 -- 9.5}& \colhead{9.5 -- 22.6}}
\startdata
$[$OII$]$ 	& 3727 & 2.15  $\pm$ 0.34& 3.01  $\pm$ 0.57& 3.08  $\pm$ 1.48& 4.05  $\pm$ 1.34& 1.73  $\pm$ 0.54& 3.40  $\pm$ 0.92& 9.37  $\pm$ 4.69\\ 
   & & 3.06 &4.43 & 3.49&  6.19 &  2.88 &  4.72 & 17.60 \\ 
$[$NeIII$]$ 	& 3869 & 0.64  $\pm$ 0.35& \nodata & \nodata& 1.83  $\pm$ 1.15& \nodata & \nodata &  \nodata \\
 H$\gamma$ 	& 4340 & \nodata & \nodata & \nodata &  \nodata& \nodata & \nodata & \nodata \\ 
    & & \nodata&\nodata&\nodata& \nodata& \nodata&\nodata& \nodata\\ 
$[$OIII$]$ 	& 4363 & \nodata &\nodata & \nodata&\nodata & \nodata & \nodata &\nodata \\ 
    & &\nodata& \nodata&  \nodata& \nodata& \nodata& \nodata& \nodata\\ 
HeII 	& 4686 & \nodata & \nodata & \nodata & \nodata & \nodata & \nodata & \nodata \\ 
   & & \nodata&\nodata& \nodata& \nodata& \nodata& \nodata& \nodata\\ 
$[$OIII$]$ 	& 4959 & 0.25  $\pm$ 0.05& 0.34  $\pm$ 0.06& 0.48  $\pm$ 0.17& 1.66  $\pm$ 0.46& 0.46  $\pm$ 0.13& 0.24  $\pm$ 0.04& 0.99 $\pm$ 0.50\\ 
  & & 0.24 & 0.33 & 0.48 & 1.61&  0.44 &  0.23 &  0.95 \\ 
$[$OIII$]$ 	& 5007 & 0.80  $\pm$ 0.08& 1.02  $\pm$ 0.09& 1.14  $\pm$ 0.22& 5.38  $\pm$ 1.08& 1.41  $\pm$ 0.20& 0.78  $\pm$ 0.20& 3.48 $\pm$ 1.15\\ 
  & &  0.77& 0.98& 1.12&  5.12 &  1.33&  0.75 & 3.24 \\ 
$[$NI$]$ 	& 5199 &\nodata  & \nodata&\nodata &\nodata  & \nodata & \nodata & \nodata\\ 
   & & \nodata&\nodata& \nodata& \nodata& \nodata&\nodata& \nodata\\ 
HeI 	& 5876 & 0.14  $\pm$ 0.04& 0.15  $\pm$ 0.05& \nodata & 0.40  $\pm$ 0.24& 0.22  $\pm$ 0.05&\nodata& \nodata \\ 
   & & 0.11 & 0.11& \nodata& 0.29 &0.15& \nodata&\nodata\\ 
$[$OI$]$ 	& 6300 & 0.09  $\pm$ 0.02& 0.17  $\pm$ 0.04&\nodata& 0.46  $\pm$ 0.20&\nodata& 0.21  $\pm$ 0.08& 0.48  $\pm$ 0.24\\ 
  & &  0.06&  0.11 &  \nodata& 0.29 &\nodata&  0.15 & 0.25\\ 
$[$OI$]$ 	& 6364 & \nodata &\nodata& \nodata& \nodata& \nodata& \nodata & \nodata \\ 
   & & \nodata&\nodata&\nodata& \nodata&\nodata&\nodata& \nodata\\ 
$[$NII$]$ 	& 6548 & 0.34  $\pm$ 0.04& 0.37  $\pm$ 0.04& 0.30  $\pm$ 0.09& 1.76  $\pm$ 0.40& 0.57  $\pm$ 0.09& 0.32  $\pm$ 0.10& 0.33 $\pm$ 0.19\\ 
  & & 0.22& 0.23&  0.26 & 1.06 &  0.31 &0.22 & 0.16 \\ 
H$\alpha$ 	& 6563 & 4.34  $\pm$ 0.22& 4.52  $\pm$ 0.23& 3.31  $\pm$ 0.46& 4.73  $\pm$ 0.90& 5.25  $\pm$ 0.42& 4.22  $\pm$ 0.51& 6.06 $\pm$ 1.70\\ 
  & &2.85 &2.85&  2.85 &  2.85&  2.85 &  2.85 & 2.85\\ 
$[$NII$]$ 	& 6583 & 1.58  $\pm$ 0.09& 1.38  $\pm$ 0.08& 0.99  $\pm$ 0.18& 3.85  $\pm$ 0.77& 2.41  $\pm$ 0.22& 1.73  $\pm$ 0.24& 2.33 $\pm$ 0.77\\ 
  & & 1.03 & 0.87& 0.85& 2.31&  1.30& 1.16&  1.09\\ 
$[$SII$]$ 	& 6717 & 0.79  $\pm$ 0.06& 0.74  $\pm$ 0.06& 0.77  $\pm$ 0.17& 1.33  $\pm$ 0.33& 0.98  $\pm$ 0.12& 0.94  $\pm$ 0.18& 1.79 $\pm$ 0.63\\ 
  & & 0.50 &0.45 & 0.66 &  0.77 & 0.51& 0.62&  0.80 \\ 
$[$SII$]$ 	& 6724 & 1.29  $\pm$ 0.12& 1.30  $\pm$ 0.10& 1.31  $\pm$ 0.30& 2.58  $\pm$ 0.65& 1.70  $\pm$ 0.22& 1.60  $\pm$ 0.34& 3.05 $\pm$ 1.13\\ 
  & & 0.82 & 0.79 &  1.12 &  1.50 & 0.88 &  1.05 &  1.36\\ 
$[$SII$]$ 	& 6731 & 0.50  $\pm$ 0.06& 0.56  $\pm$ 0.05& 0.54  $\pm$ 0.14& 1.25  $\pm$ 0.31& 0.72  $\pm$ 0.10& 0.66  $\pm$ 0.15& 1.26 $\pm$ 0.49\\ 
   & & 0.32& 0.34& 0.46&  0.73 &0.37& 0.43&  0.56 \\ 
\multicolumn{2}{l}{Hbeta 4861 = 1.00} \\ 
\multicolumn{2}{l} {F(H$\alpha$) ($10^{-15}$ erg s$^{-1}$ cm$^{-2}$)} &
33.8 $\pm$ 0.3& 22.9 $\pm$ 0.2& 10.1 $\pm$ 0.1& 10.8 $\pm$ 0.2& 22.6 $\pm$ 0.2& 12.5 $\pm$ 0.3& 8.4 $\pm$ 0.3\\ 
&& 79.5& 58.4& 13.7&  30.1& 78.2& 27.8& 38.8\\ 
\multicolumn{2}{l} {F(H$\beta$) ($10^{-15}$ erg s$^{-1}$ cm$^{-2}$)} &
7.8 $\pm$ 0.3& 5.1 $\pm$ 0.2& 3.1 $\pm$ 0.4& 2.3 $\pm$ 0.4& 4.3 $\pm$ 0.3& 3.0 $\pm$ 0.3& 1.4 $\pm$ 0.3\\ 
&&27.9& 20.5& 4.8& 10.6&27.5& 9.8& 13.6\\ 
\multicolumn{2}{l}{c\tablenotemark{b}}&  0.55 &  0.61 &  0.20 &  0.67 &  0.80 &  0.52 &  0.99 \\ 
\multicolumn{2}{l}{E(B$-$V)}&  0.39 &  0.42 &  0.14 &  0.46 &  0.56 &  0.36 &  0.69 \\ 
\enddata
\tablenotetext{a}{The emission-line ratios for region B3'' are somewhat uncertain because of difficulties in
appropriately correcting H$\beta$ for the underlying stellar absorption. This additional source of uncertainty
is not contained in the errors quoted in the table.}
\tablenotetext{b}{Extinction coefficient.}
\end{deluxetable}

\clearpage

\begin{deluxetable}{lcccccccccccc}
\tabletypesize{\tiny}
\tablewidth{0pt}
\rotate
\tablenum{A4}
\tablecaption{Tol1238-364: Flux Ratios of the Subnuclear Regions\label{tabA4}}
\tablehead{
 & & \multicolumn{5}{c}{P.A. = 150$^{\circ}$} & &\multicolumn{5}{c}{P.A. = 146$^{\circ}$}\\
\colhead{Line} & \colhead{$\lambda$}& \colhead{n1}& \colhead{n2}&
\colhead{n3}& \colhead{n4}& \colhead{n5}& &\colhead{n1'}& \colhead{n2'}& \colhead{n3'}& 
\colhead{n4'}& \colhead{n5'}\\
 & & \colhead{$-$3.7 -- $-$2.0} &  \colhead{$-$2.0 -- $-$0.3} & \colhead{$-$0.3 -- 0.7}&
\colhead{0.7 -- 1.7} & \colhead{1.7 -- 3.0} & &\colhead{$-$4.0 -- $-$2.0} & 
\colhead{$-$2.0 -- $-$0.3}& \colhead{$-$0.3 -- 0.7}& \colhead{0.7 -- 2.0} &\colhead{2.0 -- 3.7}}
\startdata
$[$OII$]$ 	& 3727 & 4.00  $\pm$ 3.56& \nodata & 0.08  $\pm$ 0.02& 0.51  $\pm$ 0.10& 4.18  $\pm$ 1.00&& 6.14  $\pm$ 2.27& 1.05  $\pm$ 0.18& 0.68  $\pm$ 0.07& 2.15  $\pm$ 0.32& 5.58  $\pm$ 1.79\\               
                & & 6.87& \nodata& 0.15 &0.51&  4.28 &&13.20 &1.79&  0.96 & 2.71 &10.89\\                                                    
$[$NeIII$]$ 	& 3869 &\nodata& \nodata& 0.22  $\pm$ 0.13& 0.52  $\pm$ 0.13& 1.59  $\pm$ 0.48&& \nodata& 0.74  $\pm$ 0.19& 0.44  $\pm$0.08& 1.14  $\pm$ 0.35&\nodata\\                                          
                & &\nodata&\nodata& 0.38 &0.52& 1.62&& \nodata& 1.18 &0.60 & 1.40 & \nodata\\                                                       
H$\gamma$ 	& 4340 &\nodata&\nodata& 0.16  $\pm$ 0.02& 0.30  $\pm$ 0.05&\nodata&&\nodata& \nodata& 0.21  $\pm$ 0.03& 0.12  $\pm$ 0.04&\nodata \\                                                            
                & & \nodata&\nodata&0.22& 0.30& \nodata&&\nodata&\nodata& 0.25 & 0.13& \nodata\\                                                         
$[$OIII$]$ 	& 4363 &\nodata & \nodata& 0.11  $\pm$ 0.03& 0.19  $\pm$ 0.05& \nodata && 1.27  $\pm$ 0.61& 0.22  $\pm$ 0.13& 0.12 $\pm$0.03& 0.25  $\pm$ 0.11&\nodata\\
HeII 	        & 4686 &\nodata&\nodata & 0.12  $\pm$ 0.03& 0.12  $\pm$ 0.04&\nodata&& \nodata & \nodata& 0.14  $\pm$ 0.04& 0.13  $\pm$0.06&\nodata\\                                                            
                & &  \nodata&\nodata& 0.14&  0.12 &\nodata&&\nodata&\nodata&0.15& 0.14& \nodata\\                                                         
$[$OIII$]$ 	& 4959 & 1.07  $\pm$ 0.59& 2.57  $\pm$ 0.26& 2.62  $\pm$ 0.05& 1.96  $\pm$ 0.14& 1.58  $\pm$ 0.28&& 0.75  $\pm$ 0.41& 2.30  $\pm$ 0.16& 2.39  $\pm$ 0.07& 2.06  $\pm$ 0.16& 1.19  $\pm$ 0.45\\      
                & & 1.03 & 2.40&  2.51 & 1.96& 1.58&&  0.71&  2.21 & 2.33&  2.03 & 1.13\\                                                   
$[$OIII$]$ 	& 5007 & 2.69  $\pm$ 0.89& 8.38  $\pm$ 0.67& 8.05  $\pm$ 0.08& 5.83  $\pm$ 0.29& 4.61  $\pm$ 0.60&& 3.32  $\pm$ 1.00& 7.14  $\pm$ 0.43& 7.49  $\pm$ 0.15& 6.29  $\pm$ 0.38& 3.77  $\pm$ 0.90\\      
                & & 2.53 &7.50 & 7.49&  5.83 & 4.60&&  3.04 & 6.71 & 7.20& 6.12& 3.49\\                                                   
$[$FeVII$]$ 	& 5158 &\nodata & \nodata& 0.06  $\pm$ 0.01& 0.08  $\pm$ 0.03&\nodata &&\nodata & \nodata& 0.05  $\pm$ 0.02&\nodata&\nodata\\                                                                     
                & &\nodata&\nodata& 0.05&0.08 & \nodata&&\nodata& \nodata& 0.05& \nodata&\nodata\\                                                          
$[$FeVI$]$ 	& 5177 &\nodata  & \nodata & \nodata & 0.08  $\pm$ 0.03&\nodata &&\nodata &\nodata & \nodata& \nodata & \nodata\\                                                                                       
                & & \nodata& \nodata& \nodata&0.08 & \nodata&& \nodata& \nodata& \nodata& \nodata& \nodata\\                                                            
$[$NI$]$ 	& 5199 & \nodata & 0.26  $\pm$ 0.08& 0.15  $\pm$ 0.02& 0.14  $\pm$ 0.03&\nodata  && \nodata & 0.15  $\pm$ 0.04& 0.16  $\pm$0.02& 0.12  $\pm$ 0.03& \nodata \\                                          
                & & \nodata& 0.20 & 0.13 & 0.14 &\nodata&&\nodata&  0.13 & 0.14 & 0.11 & \nodata\\                                                       
$[$CaV$]$ 	& 5309 &\nodata & \nodata & 0.06  $\pm$ 0.02& 0.07  $\pm$ 0.02& \nodata &&\nodata & \nodata& \nodata & \nodata &\nodata \\                                                                              
                & & \nodata&\nodata&0.05&  0.07 &\nodata&& \nodata&\nodata& \nodata& \nodata& \nodata\\                                                           
$[$NII$]$ 	& 5755 & \nodata &\nodata  & 0.08  $\pm$ 0.02& 0.09  $\pm$ 0.02& \nodata && \nodata &\nodata& \nodata& \nodata& \nodata \\                                                                              
                & & \nodata& \nodata&  0.05 & 0.09 & \nodata&&\nodata&\nodata&\nodata& \nodata&\nodata\\                                                           
HeI 	        & 5876 & \nodata& 0.26  $\pm$ 0.08& 0.21  $\pm$ 0.02& 0.10  $\pm$ 0.01& 0.16  $\pm$ 0.08&& 0.56  $\pm$ 0.63& 0.23  $\pm$0.05& 0.22  $\pm$ 0.02& 0.20  $\pm$ 0.06&\nodata  \\                        
		& & \nodata& 0.12&  0.13 & 0.10 & 0.16 && 0.31 & 0.15& 0.17& 0.17 & \nodata\\                                                     
$[$FeVII$]$     & 6086 &\nodata  &\nodata & 0.11  $\pm$ 0.03&\nodata & \nodata &&\nodata  & \nodata & 0.09  $\pm$ 0.02& \nodata & \nodata\\                                                                                   
                & &  \nodata& \nodata&  0.06 & \nodata& \nodata&& \nodata& \nodata&0.07 & \nodata& \nodata\\                                                                              
$[$OI$]$ 	& 6300 & 0.77  $\pm$ 0.42& 1.15  $\pm$ 0.15& 1.01  $\pm$ 0.03& 0.37  $\pm$ 0.03& 0.33  $\pm$ 0.12&& 0.50  $\pm$ 0.32& 0.81  $\pm$ 0.08& 0.88  $\pm$ 0.04& 0.80  $\pm$ 0.10& 0.61  $\pm$ 0.27\\      
                & &0.44 & 0.42& 0.52&  0.37 & 0.32 && 0.22& 0.46 & 0.61 &  0.63 &  0.30 \\                                                   
$[$OI$]$ 	& 6364 & \nodata& 0.40  $\pm$ 0.10& 0.37  $\pm$ 0.02& 0.15  $\pm$ 0.03&\nodata && \nodata & 0.29  $\pm$ 0.06& 0.29  $\pm$ 0.03& 0.25  $\pm$ 0.07&   ...  \\                                          
                & & \nodata&  0.14 & 0.19&  0.15 &  \nodata&& \nodata& 0.16 & 0.20 &  0.19 &  \nodata\\                                                       
$[$NII$]$ 	& 6548 & 1.06  $\pm$ 0.39& 2.53  $\pm$ 0.23& 1.65  $\pm$ 0.03& 0.51  $\pm$ 0.04& 0.81  $\pm$ 0.16&& 1.36  $\pm$ 0.50& 1.58  $\pm$ 0.13& 1.30  $\pm$ 0.04& 1.10  $\pm$ 0.10& 0.94  $\pm$ 0.36\\      
                & &  0.56 &  0.81 &  0.78 &  0.51 &  0.79 &&  0.55 &  0.84 &  0.86 &  0.84 &  0.42 \\                                                   
H$\alpha$ 	& 6563 & 5.44  $\pm$ 1.36& 9.00  $\pm$ 0.63& 6.06  $\pm$ 0.06& 1.83  $\pm$ 0.09& 2.93  $\pm$ 0.38&& 7.12  $\pm$ 1.71& 5.40  $\pm$ 0.32& 4.32  $\pm$ 0.09& 3.76  $\pm$ 0.23& 6.34  $\pm$ 1.33\\      
                & &  2.85 &  2.85 &  2.85 &  1.83 &  2.85 &&  2.85 &  2.85 & 2.85 &  2.85 &   2.85 \\                                                   
$[$NII$]$ 	& 6583 & 3.10  $\pm$ 0.84& 7.22  $\pm$ 0.58& 5.06  $\pm$ 0.05& 1.59  $\pm$ 0.08& 2.24  $\pm$ 0.31&& 4.13  $\pm$ 1.07& 5.00  $\pm$ 0.30& 4.19  $\pm$ 0.08& 3.61  $\pm$ 0.22& 4.25  $\pm$ 0.98\\      
                & &  1.61 &  2.26 &   2.36 &  1.59 &   2.18 &&  1.64 &  2.62 &   2.75 &   2.73 &   1.90 \\                                                   
$[$SII$]$ 	& 6717 & 1.34  $\pm$ 0.48& 1.92  $\pm$ 0.17& 1.27  $\pm$ 0.03& 0.47  $\pm$ 0.03& 0.92  $\pm$ 0.17&& 1.91  $\pm$ 0.63& 1.55  $\pm$ 0.11& 1.30  $\pm$ 0.04& 1.35  $\pm$ 0.11& 1.66  $\pm$ 0.46\\      
                & & 0.67 &  0.56 &  0.57 &  0.47 &  0.89 &&  0.72&  0.78 &  0.83 &   1.00 &  0.71 \\                                                   
$[$SII$]$ 	& 6724 & 2.49  $\pm$ 0.92& 4.04  $\pm$ 0.36& 2.77  $\pm$ 0.06& 0.95  $\pm$ 0.07& 1.66  $\pm$ 0.33&& 3.40  $\pm$ 1.16& 3.05  $\pm$ 0.21& 2.70  $\pm$ 0.08& 2.67  $\pm$ 0.21& 2.81  $\pm$ 0.84\\      
                & &   1.25 &  1.18 &  1.24 &  0.95 &  1.61 &&  1.28 &  1.54 &  1.73 &   1.99 &   1.19 \\                                                   
$[$SII$]$ 	& 6731 & 1.15  $\pm$ 0.44& 2.12  $\pm$ 0.19& 1.49  $\pm$ 0.03& 0.48  $\pm$ 0.03& 0.75  $\pm$ 0.16&& 1.49  $\pm$ 0.54& 1.51  $\pm$ 0.11& 1.40  $\pm$ 0.04& 1.32  $\pm$ 0.11& 1.14  $\pm$ 0.36\\      
                & &  0.57 &  0.62 &   0.66 &  0.48 &   0.73 &&   0.56 &  0.76 &   0.90 &   0.98 &   0.48 \\                                                   
\multicolumn{2}{l}{Hbeta 4861 = 1.00} \\                                                                                                                                    
\multicolumn{2}{l} {F(H$\alpha$) ($10^{-16}$ erg/s/cm$^2$)}     &                                                                                                            
66.6 $\pm$ 2.0& 746.5 $\pm$ 0.0& 1528.0 $\pm$ 0.0& 322.7 $\pm$ 3.2& 74.7 $\pm$ 2.2&& 54.7 $\pm$ 1.6& 240.9 $\pm$ 2.4& 419.5 $\pm$ 0.0& 132.9 $\pm$ 1.3& 55.1 $\pm$ 1.7\\     
&&  247.5&  7702.0& 7065.0&  322.7& 79.0&&  351.0& 881.4& 975.8&  233.2&  279.1\\                                      
\multicolumn{2}{l} {F(H$\beta$) ($10^{-16}$ erg/s/cm$^2$)}    &                                                                                                              
12.3 $\pm$ 2.7& 83.0 $\pm$ 5.8& 252.2 $\pm$ 2.5& 176.8 $\pm$ 7.1& 25.5 $\pm$ 2.5&& 7.7 $\pm$ 1.6& 44.6 $\pm$ 2.2& 97.2 $\pm$ 1.9& 35.3 $\pm$ 1.8& 8.7 $\pm$ 1.6\\            
&& 86.9&  2703.0& 2479.0&  176.8&  27.7&&  123.1&  309.3& 342.6&  81.8&  97.9\\                                         
\multicolumn{2}{l}{c\tablenotemark{a}}&  0.85 &  1.51 &  0.99 &  0.00 &  0.04 &&  1.20 &  0.84 &  0.55 &  0.36 &  1.05 \\                                                                     
\multicolumn{2}{l}{E(B-V)}&  0.59 &  1.05 &  0.69 &  0.00 &  0.03 &&  0.84 &  0.59 &  0.38 &  0.25 &  0.73 \\                                                                
\enddata
\tablenotetext{a}{Extinction coefficient.}
\end{deluxetable}

\clearpage
\pagestyle{plain}
\begin{deluxetable}{lccccc}
\tablewidth{0pt}
\tabletypesize{\scriptsize}
\tablenum{A5}
\tablecaption{ESO381-G009 - P.A. = 90$^{\circ}$: Flux Ratios \label{tabA5}}
\tablehead{
\colhead{Line} & \colhead{$\lambda$}& \colhead{A1}& \colhead{A2}&
\colhead{N}& \colhead{B1}\\
&& \colhead{$-$14.4 -- $-$3.6} &  \colhead{$-$32.9 -- $-$14.4} &
\colhead{$-$3.6-- 5.8} &  \colhead{5.8 -- 21.8}}
\startdata
$[$OIII$]$ 	& 4959 &  \nodata  &  \nodata & 0.31  $\pm$ 0.06&  \nodata \\ 
  & &  \nodata&  \nodata& 0.30 &  \nodata\\ 
$[$OIII$]$ 	& 5007 & 1.07  $\pm$ 0.66& 0.94  $\pm$ 0.64& 1.06  $\pm$ 0.08& 1.72  $\pm$ 0.71\\ 
  & &  0.96 & 0.89 &   1.00 &  1.59 \\ 
$[$NI$]$ 	& 5199 &  \nodata  &  \nodata &  \nodata  &   \nodata \\ 
  & & \nodata&  \nodata&  \nodata&  \nodata\\ 
HeI 	& 5876 &  \nodata &   \nodata  & 0.20  $\pm$ 0.04&  \nodata  \\ 
  & &  \nodata&  \nodata&0.14 &  \nodata\\ 
$[$OI$]$ 	& 6300 &  \nodata  &  \nodata  &  \nodata  &  \nodata \\ 
  & &  \nodata&  \nodata&  \nodata&  \nodata\\ 
$[$OI$]$ 	& 6364 &   \nodata  &  \nodata &  \nodata &  \nodata  \\ 
& &  \nodata&  \nodata&  \nodata&  \nodata\\ 
$[$NII$]$ 	& 6548 & 0.81  $\pm$ 1.44&   \nodata  & 0.44  $\pm$ 0.04&  \nodata  \\ 
& &   0.26 &  \nodata& 0.25 & \nodata\\ 
H$\alpha$ 	& 6563 & 8.78  $\pm$ 3.86& 4.75  $\pm$ 2.00& 5.01  $\pm$ 0.20& 6.59  $\pm$ 1.58\\ 
 & &  2.85 &  2.85 &  2.85& 2.85 \\ 
$[$NII$]$ 	& 6583 & 2.23  $\pm$ 1.36& 1.11  $\pm$ 0.79& 1.13  $\pm$ 0.07& 1.96  $\pm$ 0.76\\ 
& &  0.72 &  0.66 &  0.64 &  0.84 \\ 
$[$SII$]$ 	& 6717 & 2.23  $\pm$ 1.56& 1.08  $\pm$ 0.79& 0.61  $\pm$ 0.05& 1.09  $\pm$ 0.51\\ 
 & &  0.67 &  0.63&  0.33 &  0.45 \\ 
$[$SII$]$ 	& 6724 & 4.57  $\pm$ 3.15& 2.36  $\pm$ 1.63& 1.10  $\pm$ 0.10& 2.27  $\pm$ 1.04\\ 
 & &  1.37 &   1.37 &  0.60 &   0.93 \\ 
$[$SII$]$ 	& 6731 & 2.35  $\pm$ 1.60& 1.28  $\pm$ 0.84& 0.49  $\pm$ 0.05& 1.17  $\pm$ 0.53\\ 
 & &  0.70 &  0.74 &  0.27 &   0.48\\ 
\multicolumn{2}{l}{Hbeta 4861 = 1.00} \\ 
\multicolumn{2}{l} {F(H$\alpha$) ($10^{-16}$ erg/s/cm$^2$)} &
61.8 $\pm$ 7.4& 34.1 $\pm$ 3.1& 745.5 $\pm$ 0.0& 28.8 $\pm$ 1.7\\ 
&&  606.0&  96.2&  2343.0&  158.1\\ 
\multicolumn{2}{l} {F(H$\beta$) ($10^{-16}$ erg/s/cm$^2$)} &
7.0 $\pm$ 2.3& 7.2 $\pm$ 2.4& 148.8 $\pm$ 6.0& 4.4 $\pm$ 0.8\\ 
&&  212.7&  33.8&  821.9&  55.5\\ 
\multicolumn{2}{l}{c\tablenotemark{a}}&  1.48 &  0.67 &  0.74 &  1.10 \\ 
\multicolumn{2}{l}{E(B$-$V)}&  1.03 &  0.47 &  0.52 &  0.77 \\ 
\enddata
\tablenotetext{a}{Extinction coefficient.}
\end{deluxetable}

\clearpage

\begin{deluxetable}{lcccccc}
\tablewidth{0pt}
\tabletypesize{\scriptsize}
\tablenum{A6}
\tablecaption{ESO381-G009 - P.A.=125$^{\circ}$: Flux Ratios \label{tabA6}}
\tablehead{
\colhead{Line}& \colhead{$\lambda$}& \colhead{A1'}& \colhead{A2'}&
\colhead{A3'}& \colhead{N'} \\
&& \colhead{$-$12.8 -- $-$3.7} & \colhead{$-$21.2 -- $-$12.8} & \colhead{$-$33.3
-- $-$21.2} & \colhead{$-$3.7 -- 6.4}}
\startdata
$[$OII$]$ 	& 3727 & 2.80  $\pm$ 0.53& 5.58  $\pm$ 2.96& 6.08  $\pm$ 3.47& 2.27  $\pm$ 0.30\\ 
& &  3.73 &  8.82 &  8.42 &  3.20 \\ 
$[$NeIII$]$ 	& 3869 & 0.89  $\pm$ 0.59&  \nodata  &   \nodata & 0.81  $\pm$ 0.21\\ 
 & &  1.14&  \nodata&  \nodata&   1.09 \\ 
H$\gamma$ 	& 4340 &   \nodata  &  \nodata  &  \nodata &  \nodata  \\ 
 & &  \nodata&  \nodata &  \nodata&  \nodata\\ 
$[$OIII$]$ 	& 4363 &  \nodata  &  \nodata &   \nodata  &  \nodata  \\ 
 & &  \nodata&  \nodata&  \nodata& \nodata\\ 
HeII 	& 4686 &  \nodata  &   \nodata &  \nodata &   \nodata  \\ 
 & &  \nodata& \nodata&  \nodata& \nodata\\ 
$[$OIII$]$ 	&  4959 & 0.26  $\pm$ 0.06& 0.75  $\pm$ 0.40& 0.47  $\pm$ 0.30& 0.27  $\pm$ 0.03\\ 
& &  0.25 &  0.73 &  0.46 &   0.26 \\ 
$[$OIII$]$ 	& 5007 & 0.63  $\pm$ 0.07& 1.70  $\pm$ 0.48& 1.41  $\pm$ 0.49& 0.87  $\pm$ 0.05\\ 
 & &  0.61 &  1.61 &   1.36 &  0.84 \\ 
$[$NI$]$ 	& 5199 &  \nodata &  \nodata &  \nodata &   \nodata \\ 
& &  \nodata&  \nodata&  \nodata&  \nodata\\ 
HeI 	& 5876 & 0.14  $\pm$ 0.03&  \nodata & 0.23  $\pm$ 0.15& 0.20  $\pm$ 0.02\\ 
 & &   0.11 &  \nodata&  0.18&  0.15 \\ 
$[$OI$]$ 	& 6300 & 0.14  $\pm$ 0.04&  \nodata &  \nodata & 0.06  $\pm$ 0.03\\ 
 & &  0.10 &  \nodata&  \nodata&   0.04 \\ 
$[$OI$]$ 	& 6364 &   \nodata &   \nodata  &  \nodata  & \nodata \\ 
& &  \nodata &  \nodata &  \nodata &  \nodata \\ 
$[$NII$]$ 	& 6548 & 0.32  $\pm$ 0.05& 0.24  $\pm$ 0.19& 0.23  $\pm$ 0.41& 0.38  $\pm$ 0.03\\ 
 & &  0.23 &   0.14 &  0.16 &   0.25 \\ 
H$\alpha$ 	& 6563 & 4.01  $\pm$ 0.20& 4.93  $\pm$ 1.04& 4.21  $\pm$ 1.26& 4.30  $\pm$ 0.17\\ 
& &  2.85 &  2.85 &   2.85 &  2.85 \\ 
$[$NII$]$ 	& 6583 & 1.23  $\pm$ 0.09& 1.16  $\pm$ 0.36& 0.78  $\pm$ 0.53& 1.08  $\pm$ 0.05\\ 
 & &  0.87&  0.67 &   0.53 &  0.71\\ 
$[$SII$]$ 	& 6717 & 0.90  $\pm$ 0.08& 1.35  $\pm$ 0.54& 0.91  $\pm$ 0.46& 0.57  $\pm$ 0.05\\ 
 & &   0.63 &   0.75&  0.60 &  0.37 \\ 
$[$SII$]$ 	& 6724 & 1.59  $\pm$ 0.16& 2.25  $\pm$ 0.99& 1.58  $\pm$ 0.85& 1.04  $\pm$ 0.08\\ 
& &  1.10 &   1.25 &  1.04&  0.67 \\ 
$[$SII$]$ 	& 6731 & 0.68  $\pm$ 0.07& 0.90  $\pm$ 0.45& 0.68  $\pm$ 0.40& 0.47  $\pm$ 0.04\\ 
 & &  0.47 &  0.50 &  0.45 &  0.30 \\ 
\multicolumn{2}{l}{Hbeta 4861 = 1.00} \\ 
\multicolumn{2}{l} {F(H$\alpha$) ($10^{-15}$ erg/s/cm$^2$)} &
 27.2 $\pm$ 0.3& 5.6 $\pm$ 0.2& 8.5 $\pm$ 0.8& 65.1 $\pm$ 0.7\\ 
 && 54.4& 17.0&  18.8&  150.1\\ 
 \multicolumn{2}{l} {F(H$\beta$) ($10^{-15}$ erg/s/cm$^2$)} &
 6.8 $\pm$ 0.3& 1.1 $\pm$ 0.2& 2.0 $\pm$ 0.4& 15.1 $\pm$ 0.5\\ 
 &&  19.1&  5.9&  6.6&  52.6\\ 
 \multicolumn{2}{l}{c\tablenotemark{a}}&  0.45 &  0.72 &  0.51 &  0.54 \\ 
 \multicolumn{2}{l}{E(B-V)}&  0.31 &  0.50 &  0.36 &  0.38 \\ 
\enddata
\tablenotetext{a}{Extinction coefficient.}
\end{deluxetable}

\end{document}